\documentclass{bmvc2k}
\title{Region-of-Interest Based \\ Neural Video Compression}
\addauthor{Yura Perugachi-Diaz}{y.m.perugachidiaz@vu.nl}{$*$1,2}
\addauthor{Guillaume Sauti\`ere}{gsautie@qti.qualcomm.com}{2}
\addauthor{Davide Abati}{dabati@qti.qualcomm.com}{2}
\addauthor{Yang Yang}{yyangy@qti.qualcomm.com}{3}
\addauthor{Amirhossein Habibian}{habibian@qti.qualcomm.com}{2}
\addauthor{Taco Cohen}{tacos@qti.qualcomm.com}{2}
\addinstitution{
    Vrije Universiteit Amsterdam
}
\addinstitution{
 Qualcomm AI Research, Qualcomm Technologies Netherlands B.V.
}
\addinstitution{
 Qualcomm AI Research, Qualcomm Technologies, Inc.
}
\runninghead{Perugachi Diaz et al}{ROI-Based Neural Video Compression}
\usepackage{graphicx}
\usepackage{amsmath}
\usepackage{amssymb}
\usepackage{booktabs}
\usepackage{amsmath}
\usepackage{dsfont}
\usepackage{rotating}
\usepackage{array}
\usepackage{multirow}
\usepackage{listings}
\usepackage{wrapfig}
\usepackage[symbol]{footmisc}
\usepackage{xcolor}

\def\eg{\emph{e.g}\bmvaOneDot}

\def\etal{\emph{et al}\bmvaOneDot}
\def\ie{\emph{i.e.}\bmvaOneDot}

\newcommand{\round}[1]{\ensuremath{\lfloor#1\rceil}}

\begin{document}
\maketitle
\setcounter{footnote}{1}
\footnotetext{Work completed during internship at Qualcomm Technologies Netherlands B.V.}
\setcounter{footnote}{2}
\footnotetext{Qualcomm AI Research is an initiative of Qualcomm Technologies, Inc. and/or its subsidiaries.}
\vspace{-0.7cm}
\begin{abstract}
Humans do not perceive all parts of a scene with the same resolution, but rather focus
on few regions of interest (ROIs).
Traditional Object-Based codecs take advantage of this biological intuition, and are capable of non-uniform allocation of bits in favor of salient regions, at the expense of increased distortion the remaining areas: such a strategy allows a boost in perceptual quality under low rate constraints.
Recently, several neural codecs have been introduced for video compression, yet they operate uniformly over all spatial locations, lacking the capability of ROI-based processing.
In this paper, we introduce two models for ROI-based neural video coding.
First, we propose an implicit model that is fed with a binary ROI mask and it is trained by de-emphasizing the distortion of the background.
Secondly, we design an explicit latent scaling method, that allows control over the quantization binwidth for different spatial regions of latent variables, conditioned on the ROI mask.
By extensive experiments, we show that our methods outperform all our baselines in terms of Rate-Distortion performance in the ROI.
Moreover, they can generalize to different datasets and ROI specifications at inference time.
Finally, they do not require expensive pixel-level annotations during training, as synthetic ROI masks can be used with little to no degradation in performance.
To the best of our knowledge, our proposals are the first solutions that integrate ROI-based capabilities into neural video compression models.
\end{abstract}
\section{Introduction}
The most common approach in neural lossy video compression is to rely on variational autoencoders to minimize the expected rate-distortion (R-D) objective, $D + \beta R$~\cite{habibian2019video, minnen2018joint, ssf, elfvc, lin2020mlvc}.
Although this approach has proven to be successful, a model trained to minimize the expected rate-distortion tradeoff uniformly over all pixels may allocate too few bits to salient regions of a specific video.
This clashes with the model of the human visual system, which is space-variant and has the highest spatial resolution at the the foveation point~\cite{wandell1995foundations,itti2004automatic}.
Exploiting this phenomenon, \eg by encoding Regions-Of-Interest (ROIs) with higher fidelity, can significantly contribute to the subjective quality under a low bitrate regime.
The key idea of traditional \emph{ROI-based codecs}~\cite{sikora1995shape, li2000shape, chen2006dynamic, han2008object, li2018learning, cai2020endtoend, xia2020object} is to allocate different bitrate budgets for objects or regions of interest, and therefore to allow for non-uniform reconstruction qualities.
For instance, traditional codecs like JPEG2000~\cite{skodras2001jpeg} and MPEG-4~\cite{vetro1999mpeg4} were used as basis to build object-based coding methods~\cite{han2008object, chen2006dynamic}.
However, these ideas lacked widespread adoption due to their complexity and to their block-based nature, limiting their capability to deal with arbitrary ROI shapes.

\begin{wrapfigure}{r}{0.5\textwidth}
\begin{center}
\includegraphics[width=0.5\textwidth]{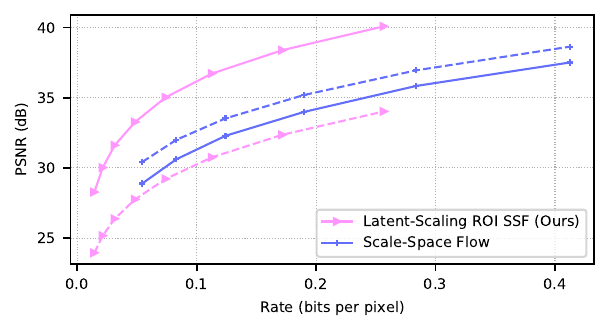}  
\end{center}
\caption{R-D improvements on DAVIS~\cite{davis}, where ROI-PSNR (solid) is higher than non-ROI PSNR (dashed). The improvement is equivalent to 69.3\%  BD-rate gain~\cite{bjntegaard2001calculation}.}
\label{fig:spotlight}
\end{wrapfigure}
More recently, some works have developed ROI-based neural \emph{image} codecs, either by implicitly identifying the ROI as part of the encoding process~\cite{li2018learning, cai2020endtoend}, or by relying on external algorithms for its extraction~\cite{xia2020object}.
Under both approaches the R-D objective can be spatially weighted and, additionally, the latent variables can be masked before the quantization step to reduce their entropy~\cite{habibian2019video, cai2020endtoend, xia2020object}.
Nevertheless, existing neural ROI-based codecs have the following limitations:
\emph{(i)} they only work for images, 
\emph{(ii)} they use intricate masking schemes to spatially control the rate, without exploiting the Gaussian structure of the latent prior distribution and
\emph{(iii)} the encoding operations are tightly coupled with ROI prediction, which makes it hard for the codecs to be adapted to different ROI requirements.

In this paper, we present the first two neural codecs capable of ROI-based compression.
The first \emph{implicit} model is fed with the ROI mask and is trained with an ROI-aware loss, where the distortion of the background is de-emphasized. 
Secondly, the \emph{latent-scaling} model extends the implicit model by exploiting a recent technique originally developed for variable rate coding~\cite{cui2020g,chen2020variable,chen2019neural,plonq}.
We extend its design by introducing an auxiliary autoencoder (AE) being fed with the ROI map, and regressing a gain tensor explicitly controlling the quantization binwidth for different spatial regions.
This can be seen as the continuous equivalent of the masking scheme used in conjunction with scalar quantization~\cite{li2018learning, mentzer2018conditional, cai2020endtoend}.
We describe our solution in the context of a Scale Space Flow (SSF)~\cite{ssf} architecture;
however, we argue that they are in principle compatible with most state-of-the-art models based on hyperpriors~\cite{elfvc,lin2020mlvc,pourreza2021extending}

We show that our methods outperform all our baselines on the DAVIS dataset~\cite{davis} in terms of R-D performance, as measured in PSNR in the ROI (Fig.~\ref{fig:spotlight}).
Moreover, further analyses show that they generalize to any arbitrary ROI which can be specified by the user at inference time and that expensive pixel-dense annotations are not required during training, as synthetic ROI can be used with little to no degradation in performance.
\section{Related work}
\paragraph{Non-uniform coding.}
The literature on spatially variant image encoding mainly focuses on two separate problems: \emph{(i)} how to estimate the ROI and \emph{(ii)} how to exploit it to improve coding.
Most traditional block-based methods~\cite{han2006image, han2008object, chen2006dynamic} fall under the former category, and simply exploit non-uniform coding capabilities of standard codecs such as JPEG2000~\cite{skodras2001jpeg} and MPEG-4~\cite{vetro1999mpeg4}. 
These solutions are limited in their capabilities due to their block-based approach to compression, which hinders the encoding of arbitrarily shaped objects and does not allow for pixel-level bit allocation optimisation~\cite{sikora1995shape, li2000shape}.

In contrast, recent work in neural image coding tackle both the above mention problems and target pixel-level ROI~\cite{li2018learning, cai2020endtoend, duan2019contentaware, xia2020object,agustsson2019extreme, akbari2019dsslic, duan2020jpad}.
Among these, Li~\etal~\cite{li2018learning} and Cai~\etal~\cite{cai2020endtoend} learn the ROI implicitly by spatially masking out the latents before scalar quantization, whereas Xia~\etal~\cite{xia2020object} use the down-scaled output of the DeepLab~\cite{chen2018deeplab} segmentation network to mask out foreground from background, before sending each stream to a separate hyper-codec for quantization.
Similar to these works, our work focuses on how to use a given ROI to enable non-uniform coding, whilst delegating its extraction to some external automatic model such as~\cite{zhao2017pspnet, chen2018deeplab, le2018video, Wang_2018_CVPR, lai2019video, Wang_2019_revisitingVS, wang2021hrnet}.
However, our approach extends extends neural ROI-coding to the case of video inputs.
\paragraph{Neural video compression.}
Compressing videos with neural networks has been an active field of research recently~\cite{wu2018video, habibian2019video, dvc, rippel2019lvc, golinski2020frae, lin2020mlvc, ssf, elfvc, pourreza2021extending,c2f}.
While varying in their choice of architecture and quantization strategy, neural video codecs generally follow the DVC~\cite{dvc} framework where an I-frame codec compresses the first frame and a P-frame codec uses motion estimation and a residual network to model the subsequent ones.
Recently, Agustsson \etal~\cite{ssf} proposed to use a Scale-Space Flow which addresses uncertainties in motion estimation via interpolation through a Gaussian pyramid.
This allows blurring of the warped frames in regions where optical flow prediction is uncertain or ill-posed, like chaotic motions and obstructed objects.
Our work is established in the same SSF framework, and enables ROI-based coding by means of latent scaling~\cite{chen2019neural, chen2020variable, cui2020g, plonq}, a technique originally introduced for variable bitrate coding.
Differently from these works, that scale the latents globally with a single scalar value, we adjust the quantization step size for every spatial location, thus controlling the levels of distortion and entropy in foreground and background regions.
\\\\
In summary, we are the first work to learn ROI coding end-to-end for video inputs (as opposed to images) and extend latent-scaling spatially to be used in an ROI-based context.
Additionally, other works either learn implicitly the ROI using a subnetwork~\cite{cai2020endtoend,duan2019contentaware,li2018learning} or tie themselves to a restricted set of semantic classes~\cite{agustsson2019extreme,akbari2019dsslic,duan2020jpad}, which would require re-training if testing on unseen classes. In contrast, we explicitly take the ROI as input, which provides the user evaluation time flexibility similar to H.264 and H.265 ROI mode.
\section{ROI-based neural video compression}
\label{sec:methodology}
In this section we first present the neural video codec we use as backbone for our work, Scale Space Flow (SSF)~\cite{ssf}.
Next, we extend SSF to be an ROI-based codec by proposing two models: the \emph{Implicit} and \emph{Latent-scaling} ROI SSF.
Lastly, we will describe the optimization for SSF and the ROI-aware methods.

We define a video frame $x_i \in \mathds{R}^{H \times W\times 3}$ at time step $i$, where $H$ and $W$ represent its height and width respectively.
Then, a video sequence is denoted as $\mathbf{x}= \{ x_0, x_1, \hdots, x_{T} \}$, with $T+1$ frames.
The sequence of binary ROI masks corresponding to the video sequence is defined as $\mathbf{s} = \{s_0, s_1, \hdots, s_{T}\}$, where $s_i \in  \{0, 1\}^{H \times W}$.
The neural video codec SSF consists of an I-frame codec and a P-frame codec.
The I-frame codec is a mean-scale hyperprior AE~\cite{minnen2018joint} which encodes a first frame $x_0$ independently to produce a reconstruction $\hat{x}_0$. 
The P-frame codec is comprised of two hyperprior AEs. 
The first, the \emph{P-frame flow hyperprior AE}, estimates a scale-space flow $g_i$ from the previous reconstruction $\hat{x}_{i-1}$ and current frame $x_i$, which is used to warp the previous reconstruction into $\bar{x}_i$. 
The second hyperprior AE, \emph{the P-frame residual hyperprior AE}, encodes the residual $r_i = \bar{x}_i - x_i$. 
The final reconstruction $\hat{x}_i$ is obtained by adding the warped prediction $\bar{x}_i$ and the estimated residual $\hat{r}_i$. 
The latent codes of each hyperprior AE are denoted by $z_0$, $w_i$ and $v_i$
and are rounded to integer values then  entropy coded using the prior parameters estimated by their respective hyper-decoder.
We omit hyper latent codes for ease of exposition, and we refer to~\cite{ssf} for further details.
\begin{figure*}[t]
\centering
\resizebox{\textwidth}{!}{
\begin{tabular}{ccc}
\includegraphics[height=5cm]{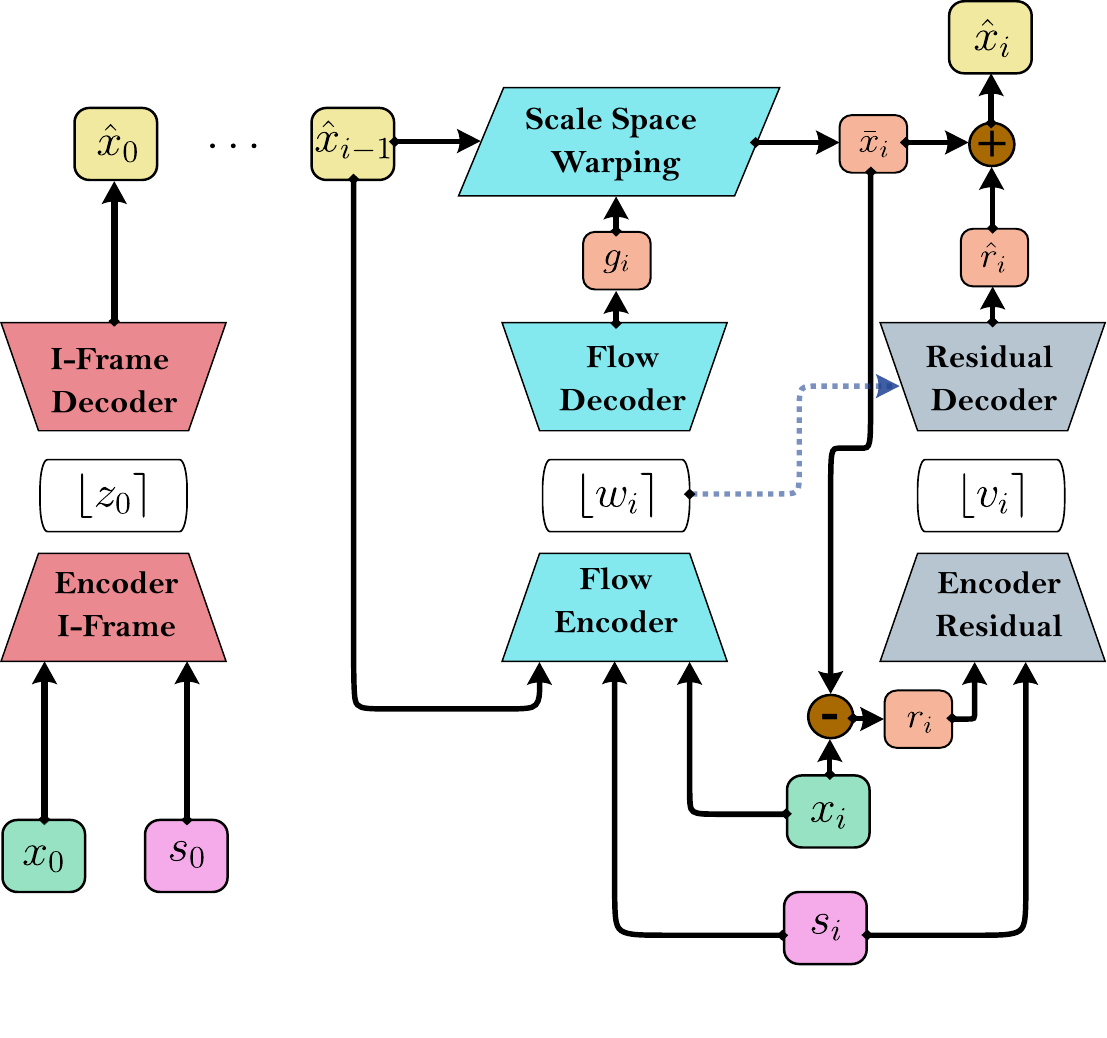}&&
\includegraphics[height=5cm]{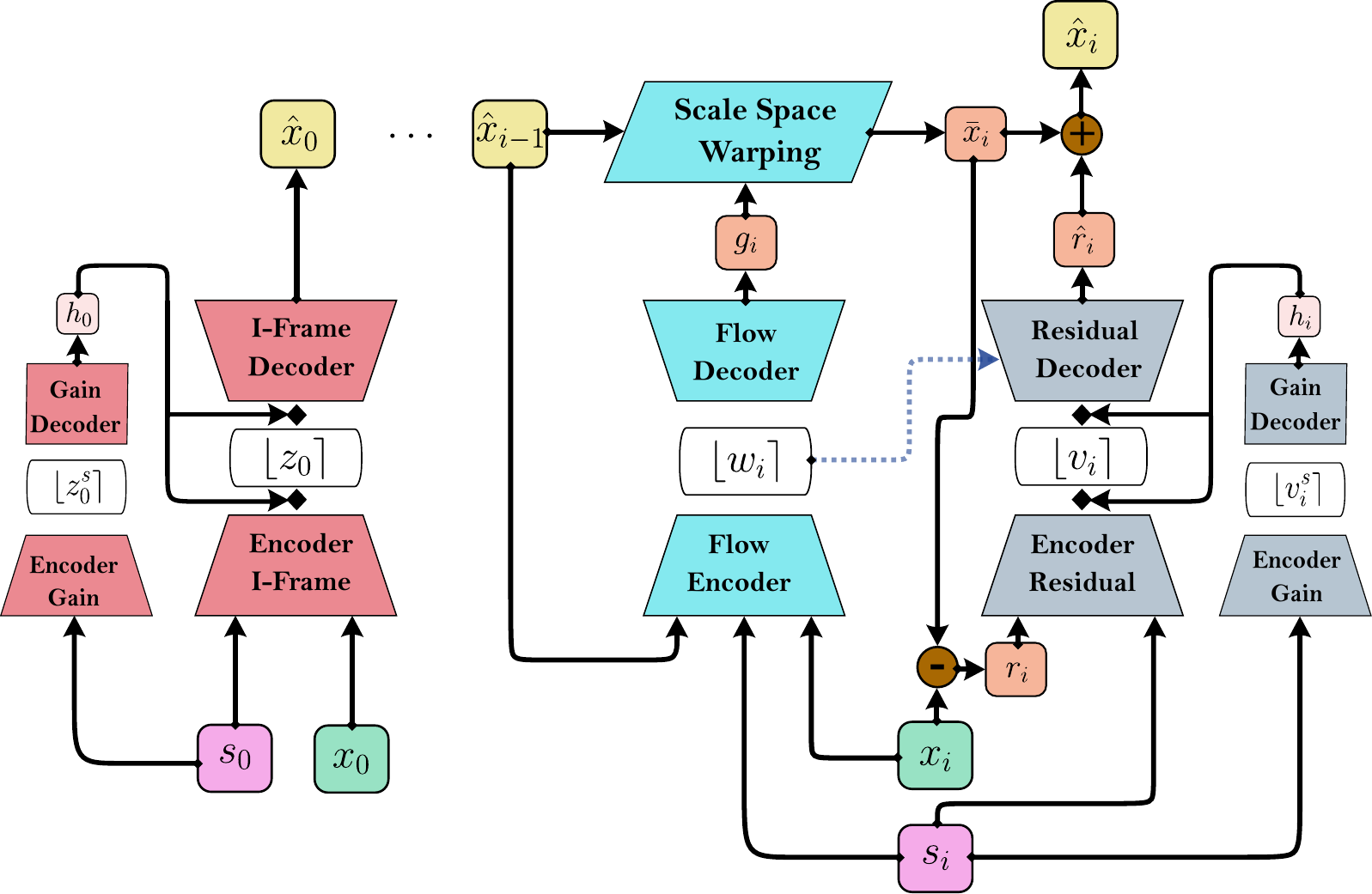}\\
(a) Implicit ROI Scale-Space Flow && (b) Latent-scaling ROI Scale-Space Flow
\end{tabular}}
\caption{Illustration of the proposed ROI-based neural video compression models. Both models learn to utilize an ROI mask $s$. Model (a) feeds a mask along with an image and model (b) utilizes extra hyperpriors to send the ROI mask for latent scaling.}
\vspace{-4mm}
\label{fig:models}
\end{figure*}
\paragraph{Implicit ROI Scale-Space Flow} \label{sec:implicit_ssf}
An immediate extension to SSF to make it ROI-aware is to provide the ROI mask $s_i$ as input to each of the three hyperpriors, see Fig.~\ref{fig:models}a.
Note that the ROI mask is not fed to the decoder, meaning we expect the encoders to implicitly store the relevant ROI information inside the existing latent codes.
Since the decoder does not require the ROI mask, we do not need to transmit a representation of the mask itself. Feeding information of the mask along with the video frame, in combination with the use of an ROI-aware loss, encourages the model to focus on important aspects for the user.
Albeit simple, we show the effectiveness of this approach when paired with an ROI-aware loss in Sec.~\ref{sec:experiments}.
\paragraph{Latent-scaling ROI Scale-Space Flow}\label{sec:latent_scaling_ssf}
Inspired by methods like~\cite{li2018learning, cai2020endtoend} which introduce a mechanism to explicitly control the spatial bit allocation, we adapted a recent technique called latent-scaling~\cite{chen2020variable, cui2020g}. 
Albeit similar in its motivation, it differs from the masking approach of~\cite{cui2020g} by exploiting the Gaussian prior structure of mean-scale hyperprior AE.
The key idea is to apply a scaling factor to the latent which changes the quantization step size, leading to different trade-offs between rate and distortion in ROI and non-ROI areas.
By using ROI-based information to control the scale of latents, the quantization grid can be explicitly adjusted. 
Our model can therefore learn that foreground regions require finer quantization than background regions.
For ease of exposition, we will describe in the next paragraphs latent-scaling for the I-frame hyperprior AE, but the same method is applied to the P-frame residual hyperprior AE.
We do not apply it to the P-frame flow hyperprior AE as initial studies showed the flow code $w_i$ only accounts for a small fraction of the total rate.
For similar reasons, we only apply latent-scaling latents, leaving hyper-latents, which are cheap to encode, unaffected.

We introduce a new hyperprior-like network called \emph{gain hyperprior AE} (see leftmost autoencoder in Fig. \ref{fig:models}b). 
This network encodes the ROI mask $s_0$ into a latent code $z^s_0$, that is decoded to a gain variable $h_0$ which shares the same dimensions as the latent variable $z_0$, both spatially and channel-wise\footnote{previous latent-scaling~\cite{chen2020variable, cui2020g, plonq} work only use channel-wise gain}.
We scale the latent $z_0$ with the inverse of the estimated spatial gain variable $h_0$,
where we restrict $h_0 \geq 1$. 
Such a procedure is akin to making the quantization range larger, depending on the value of $h_0$.
We further denote the mean $\mu$ and scale $\sigma$ as the prior parameters estimated by the I-frame hyper-decoder. 
In the quantization step, we choose to center the scaled latent $z_0 \oslash h_0$ by its prior mean $\mu \oslash h_0$, where $\oslash$ is a elementwise division. 
Next, we apply the rounding operator $\round{\cdot}$ on $(z_0 - \mu) \oslash h_0$ such that the estimated mean $\mu$ learned by the hyper-encoder is on the grid, and then add the offset $\mu \oslash h_0$ back.
The dequantized latent $\hat{z}_0(h_0)$ is obtained by multiplying by $h_0$ after the quantization block. More precisely:
\begin{figure*}[t]
\centering
\resizebox{\textwidth}{!}{
\begin{tabular}{cc}
\includegraphics[height=3cm]{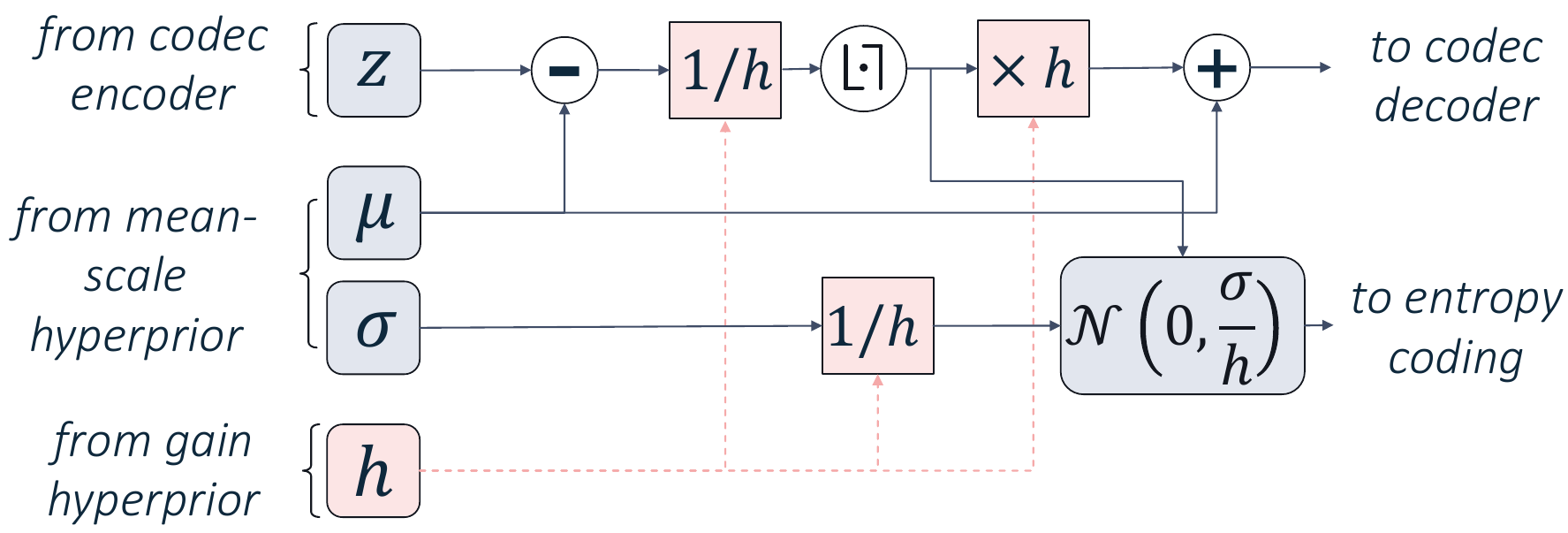}&
\includegraphics[height=3cm]{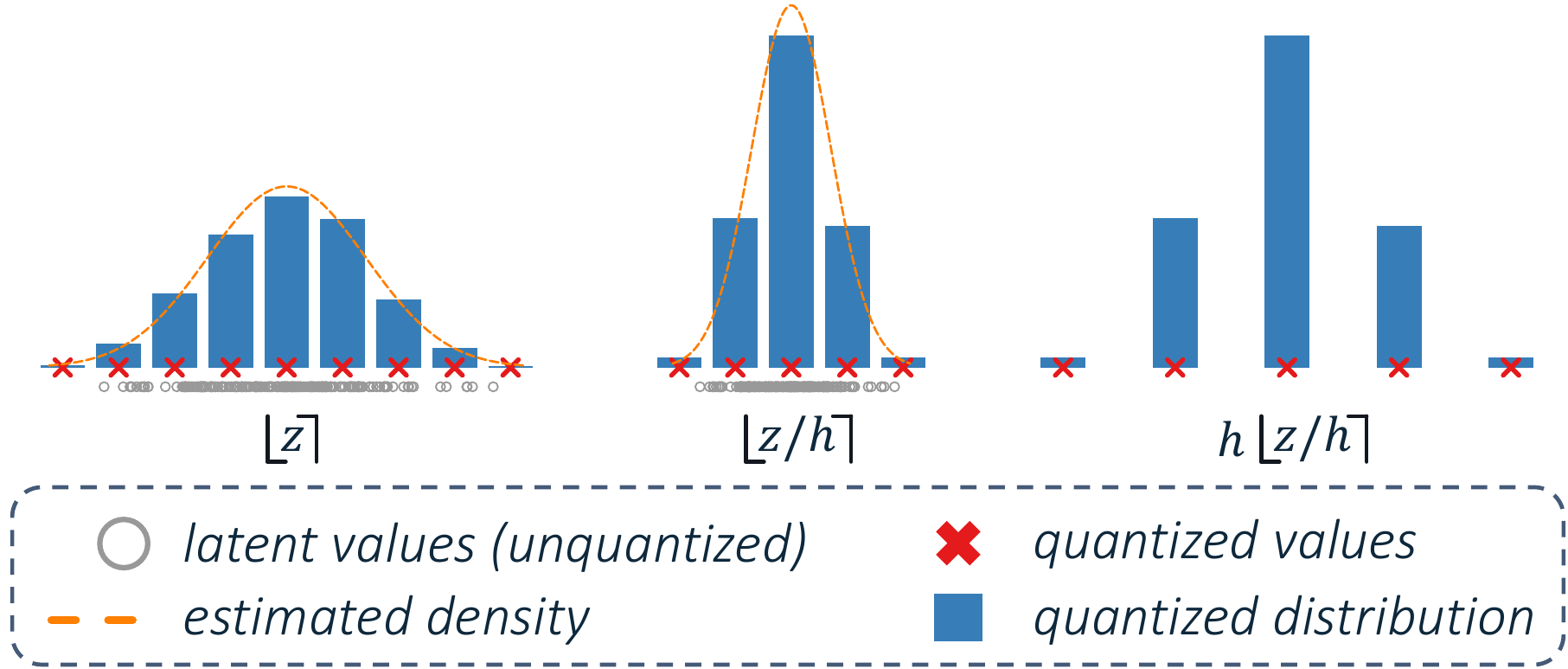}\\
(a) & (b)
\end{tabular}}
\caption{Illustration of the latent scaling mechanism for ROI-based coding. (a) shows how latent-scaling variable $h$ affects latent $z$ and prior parameters $\mu$ and $\sigma$. (b) shows intuition on why scaling the prior is necessary for entropy coding.}
\vspace{-4mm}
\label{fig:latent_scaling}
\end{figure*}

\begin{equation}
    \hat{z}_0(h_0) = \round{(z_0 -  \mu) \oslash h_0} \odot h_0 + \mu,
\end{equation}
where $\odot$ denotes elementwise multiplication.
After the dequantized latent $\hat{z}_0(h_0)$ is obtained, it is passed to the decoder to obtain reconstructed frame $\hat{x}_0$. 
The whole procedure is illustrated in Fig.~\ref{fig:latent_scaling}a.
For rate computation and entropy coding, we use the modified probability $\mathbb{P}$ of $\hat{z}_0(h_0)$ as follows:

\begin{align}
    \mathbb{P}\left(\hat{z}_0(h_0)\right) &= \int_{\hat{z}_0(h_0) - h_0/2}^{\hat{z}_0(h_0) + h_0/2} \mathcal{N}(x - \mu|0, \sigma) dx
\label{eq:change_grid} \\ \label{eq:change_prior}  =& \int_{\hat{z}_0(h_0)/ h_0 - 1/2}^{\hat{z}_0(h_0)/ h_0 + 1/2} \mathcal{N}\left(x - \frac{\mu}{h_0}{\Big|}0, \frac{\sigma}{h_0}\right) dx
\end{align}

As shown in Fig.~\ref{fig:latent_scaling}b and in Eq.~\eqref{eq:change_grid}, latent-scaling can be interpreted as effectively changing the quantization grid~/~binwidth.
In practice, for entropy coding we do not change the quantization grid and round to the integer grid and scale the prior appropriately, as in Fig.~\ref{fig:latent_scaling}a and b (middle plot) and Eq.~\eqref{eq:change_prior}.
As stated above, the same procedure is applied to the P-frame residual latent code $v_t$, as shown in Fig. \ref{fig:models}b.

\paragraph{ROI-aware Rate-Distortion Loss}

We modify the regular R-D loss from SSF to take into account the ROI mask. We sum the rate and distortion for all $T$ frames in the video sequence $\mathbf{x}$ with corresponding ROI masks $\mathbf{s}$:

\begin{equation}
\label{eq:rd_loss}
    \mathcal{L} = \beta \mathcal{L}_{R} + \sum_{i=0}^{T} \mathcal{L}_{D, i},
\end{equation}
where $\beta$ is rate-distortion trade-off variable.  $\mathcal{L}_D$ represents the distortion loss which is a modified mean squared error (MSE) involving the binary ROI mask:
\begin{align} \label{eq:bin_mse}
        \mathcal{L}_{D, i} = \frac{1}{HWC} \sum_{j=1}^{H} \sum_{k=1}^{W} \sum_{l=1}^{C} \left(s_i \odot \epsilon_i + \frac{1}{\gamma} \cdot (1 - s_i) \odot \epsilon_i\right)_{jkl},
\end{align}
where $H, W$ and $C$ denote the image dimensions, $\gamma$ is a penalty hyperparameter for the non-ROI, $ \epsilon_i = (x_i - \hat{x}_i)^2$ is the squared error
and $s_i$ is broadcasted over the channel dimension.
Note that the distortion loss of the original SSF corresponds to the special case where $s_i$ equals one everywhere.
Further, the rate loss $\mathcal{L}_R$ is computed with the estimated cross-entropy $\mathcal{H}(\cdot)$ by the hyperprior of each latent variable present in the model. 
For the implicit ROI SSF the rate loss $\mathcal{L}_{I,R}$ is equal to:

\begin{equation}
\label{eq:rate_implicit}
        \mathcal{L}_{I,R} = \mathcal{H}(z_0) + \sum_{i=1}^{T} \left[ \mathcal{H}(v_{i}) + \mathcal{H}(w_{i}) \right].
\end{equation}
The rate loss $\mathcal{L}_{LS,R}$ of the latent-scaling ROI SSF also includes latent variables $z^s_0$ for the latent scaling of the I-frame hyperprior AE and $v^s_i$ for the latent scaling of the P-frame residual hyperprior AE. 
As such, it is given by:
\begin{equation}
\label{eq:rate_ls}
\begin{split}
        \mathcal{L}_{LS,R} = 
        & \mathcal{H}(z_0^s) + 
        \mathcal{H}(z_0) \\
        & + 
        \sum_{i=1}^{T} \left[ \mathcal{H}(v_{i}^s) + \mathcal{H}(v_{i}) + \mathcal{H}(w_{i}) \right].
\end{split}
\end{equation}
In practice we found that the two extra rate contributions from the ROI masks $\mathcal{H}(z^s_0)$ and $\mathcal{H}(v^s_i)$ are only a small fraction compared to the standard rate components $\mathcal{H}(z_0)$ and $\mathcal{H}(v_i)$ of the model.
Please note that in both Eq.~\ref{eq:rate_implicit} and~\ref{eq:rate_ls} we omit the rate of the hyper latent codes to avoid notational clutter.
\section{Experiments}
\label{sec:experiments}
\paragraph{Datasets.}
As standard video compression benchmarks~\cite{HEVC_dataset,UVG,Xiph} do not come with ROI annotations, we hereby introduce a benchmark for ROI-based codecs, by utilizing publicly available video segmentation datasets and deriving ROI maps from their pixel-level groundtruth labels.
More specifically, we rely on DAVIS~\cite{davis} and Cityscapes~\cite{cityscapes} for training and evaluation of our models.
DAVIS is composed of 90 diverse and short video sequences, for which groundtruth segmentation of salient objects provided.
To create binary ROI masks, we consider all labeled objects as foreground, whereas the rest of the frame is labeled as background.
We use 60 sequences for training and 30 for validation, comprising 4,209 and 1,999 frames respectively.
Cityscapes is composed of 2,120 video sequences from dashcam of vehicles driving around German cities. 
1,885 sequences are used for training and 235 for validation, or 89,248 and 15,000 frames respectively.
As groundtruth segmentation labels are provided only at 1 fps, we extract semantic labels automatically for every frame by running the state of the art segmentation model in~\cite{nvidia_segm}.
The dataset provides a categorization of every pixel into one of 19 classes. 
We select pixels of "\textit{vehicle}", "\textit{road}", "\textit{pedestrian}", "\textit{bicycle}", "\textit{motorcycle}" as belonging to the ROI, and mark other classes as non-ROI.
To reduce compression artifacts, we resize the frames from both datasets to 720p using Pillow~\cite{clark2015pillow}.

As an alternative to ground-truth ROI masks, in some experiments (see Sec.~\ref{sec:results_synthetic_roi})) we rely on synthetic ROI masks generated using Perlin noise~\cite{perlin_noise} (only during training).
The masks contain blobs that evolve continuously over time to cover each of the video frames.
\paragraph{Implementation details.}
We optimize all methods but SSF with the ROI-aware MSE as distortion metric (Eq.~\eqref{eq:bin_mse}), and use $\gamma=30$ as penalty for the non-ROI areas.
Following the training scheme from~\cite{ssf, pourreza2021extending}, all models are warm-started from an SSF pre-trained on the Vimeo-90k dataset~\cite{vimeo90k} for 1M steps, then fine-tuned on the dataset of interest for 300K steps. 
We trained all models at various rate-distortion tradeoffs with $\beta = 2^\alpha \times 10^{-4} : \alpha \in {0,1,...,7}$.
We use Adam optimizer with a learning rate of $10^{-4}$ with batch size 8.
Each example in the batch is comprised of 3 frames (I-P-P), randomly cropped to $256 \times 384$.
The models take about 3 days to train on a single NVIDIA V100 GPU.
We report video quality in terms of PSNR in ROI and non-ROI, where both are first calculated per-frame in the RGB color space, then averaged over all the frames of each video, and finally averaged over all the videos of a dataset. 
The results we report are based on Group-of-Picture of size of 12 for consistency with other neural compression works~\cite{ssf, pourreza2021extending, dvc, lin2020mlvc}.
We refer to the appendix for architecture details, along with information about the computational complexity of the models.
\begin{wrapfigure}{l}{0.5\textwidth}
\begin{center}
\includegraphics[width=0.5\textwidth]{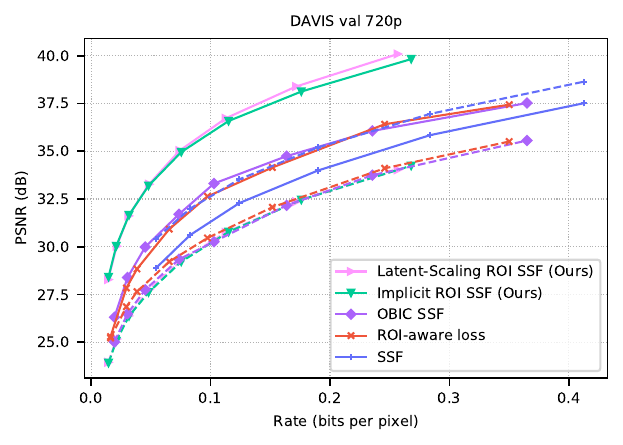}
\end{center}
\caption{All ROI-based neural video compression approaches vs SSF. Solid line denotes ROI PSNR, while dashed non-ROI PSNR.}
\vspace{-10mm}
\label{fig:semantic_models}
\end{wrapfigure}
\paragraph{Compared methods.}
We compare our method to the plain SSF and two further ROI-based baselines. 
The first, dubbed \emph{ROI-aware loss}, consists of SSF trained with our ROI-aware loss as described in Eq.~\eqref{eq:rd_loss}.
While the codec is blind to the ROI, it is expected to implicitly learn it through the training objective, in a similar fashion as the semantic models in Habibian~\etal~\cite{habibian2019video}. 
The second method, dubbed \emph{OBIC SSF}, is based on a recent ROI-based neural image codec~\cite{xia2020object}.
To enable a fair comparison, we train this architecture using our ROI-aware loss, which is slightly different from the formulation in~\cite{xia2020object}.
\paragraph{ROI-based coding}
In Fig.~\ref{fig:semantic_models}, we report the RD-plots of Implicit ROI SSF and Latent-scaling ROI SSF.
We compare our proposed models to the described ROI-aware loss and OBIC SSF baselines, as well as to a plain SSF model that does not involve any ROI-based compression.
\begin{wrapfigure}{r}{0.5\textwidth}
\begin{center}
\includegraphics[width=0.5\textwidth]{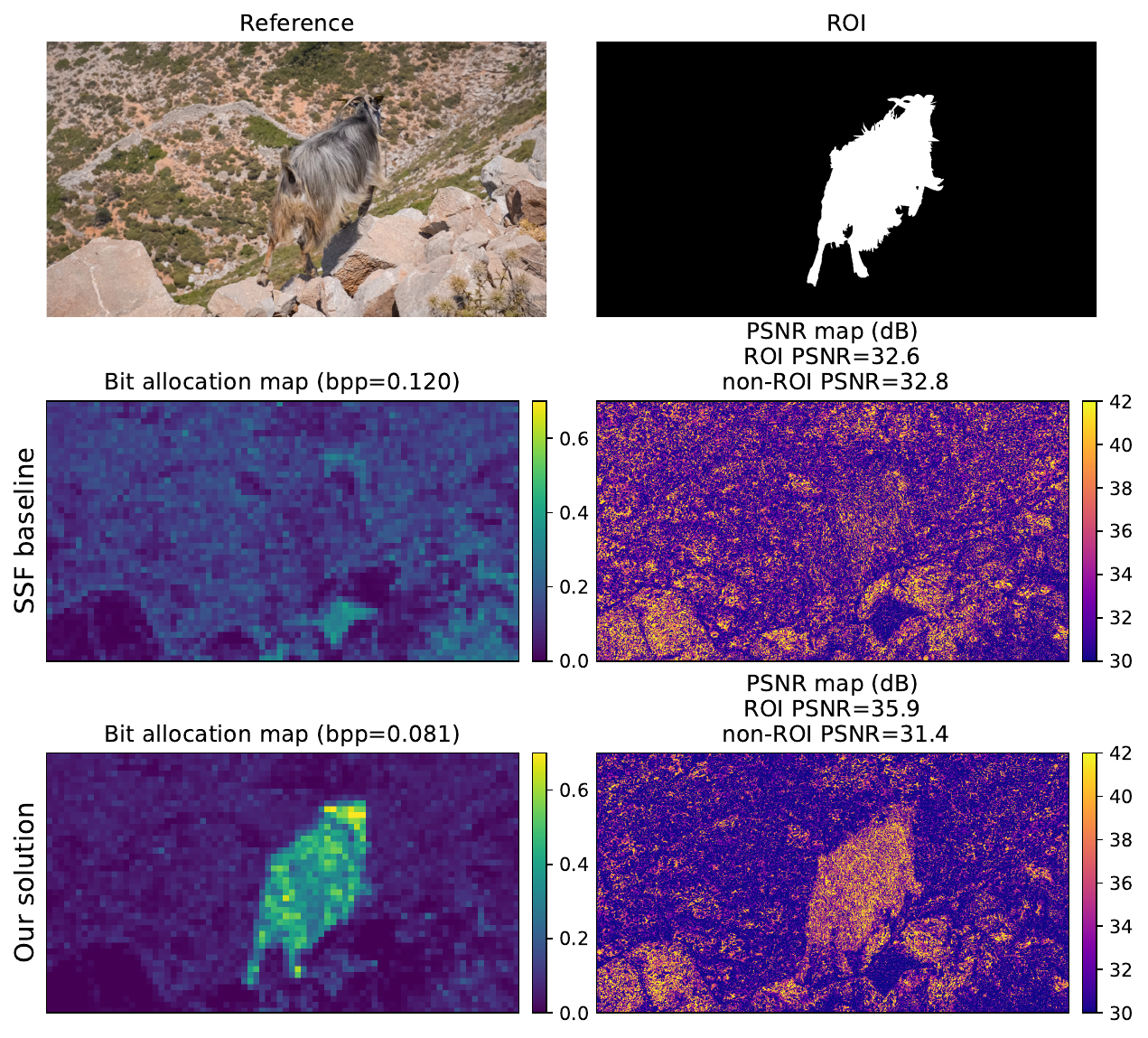}
\end{center}
\caption{Bitrate and PSNR allocation maps for SSF and our proposed ROI-based codec, latent-scaling ROI SSF. We hereby report frame 5 of DAVIS ``\texttt{goat}'' sequence.}
\label{fig:psnr_bit_allocation}
\end{wrapfigure}
For all compared models, solid lines and dashed lines correspond to RD curves in ROI and non-ROI regions respectively.
The figure shows several insights.
First, the plain SSF shows better compression results on non-ROI regions, that are seemingly easier to compress than ROI areas on DAVIS.
This result - that we hypothesize is due to the high degree of motion affecting foreground objects on the dataset - underlines that such a codec might be suboptimal.
The ROI-based baselines we consider, namely ROI-aware loss and OBIC SSF, succeed in delivering a better tradeoff for foreground regions.
Overall, their performances seem comparable across the rate spectrum. 
Interestingly, the separate hyperprior models envisioned by OBIC SSF for foreground and background barely outperforms a simple ROI-aware loss in our experiments.
Finally, the figure clearly shows the superiority of the proposed implicit and latent-scaling ROI SSF.
Indeed, their RD-curves performs on par with the mentioned baselines on background regions, while achieving a superior tradeoff for ROI regions.
In this respect, our latent-scaling based model seems to slightly outperform the implicit model in ROI areas, especially at higher bpps ($>0.1$).

Furthermore, we investigate the behavior of the proposed Latent Scaling ROI SSF codec in terms of spatial bit allocation and reconstruction quality.
Fig.~\ref{fig:psnr_bit_allocation} shows, on a reference validation frame from DAVIS, the pixel-wise bpp and PSNR as compared to the ones achieved by SSF.
For SSF, bit allocation and reconstruction quality are roughly uniformly distributed over the image.
Differently, Latent Scaling ROI SSF model focuses both bpp and PSNR on the region of interest.
Moreover, it is worth noting how, despite the fact latent scaling operates at the reduced resolution of the latents (resulting in block-wise bpp allocation), the PSNR of the reconstructed frame properly aligns with the ROI at pixel-level.
Finally, in Fig.~\ref{fig:qualitative} we shows a few qualitative compression results of our model, compared to SSF.
\bgroup
\setlength{\tabcolsep}{1pt}
\begin{figure*}[t]
\centering
\resizebox{0.95\textwidth}{!}{
\begin{tabular}{cccc}
SSF - 0.054 bpp & Implicit ROI SSF - 0.048 bpp & Latent Scaling ROI SSF - 0.048 bpp & Reference\\
\includegraphics[width=0.3\textwidth]{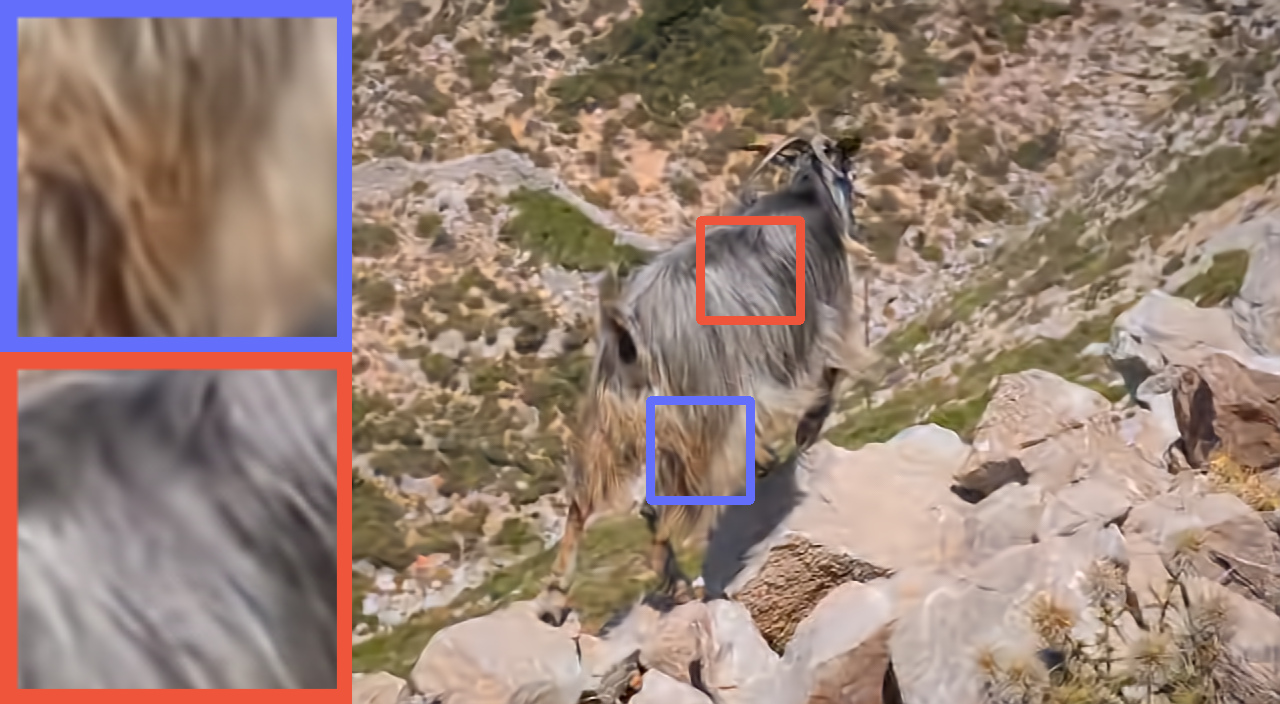}&  
\includegraphics[width=0.3\textwidth]{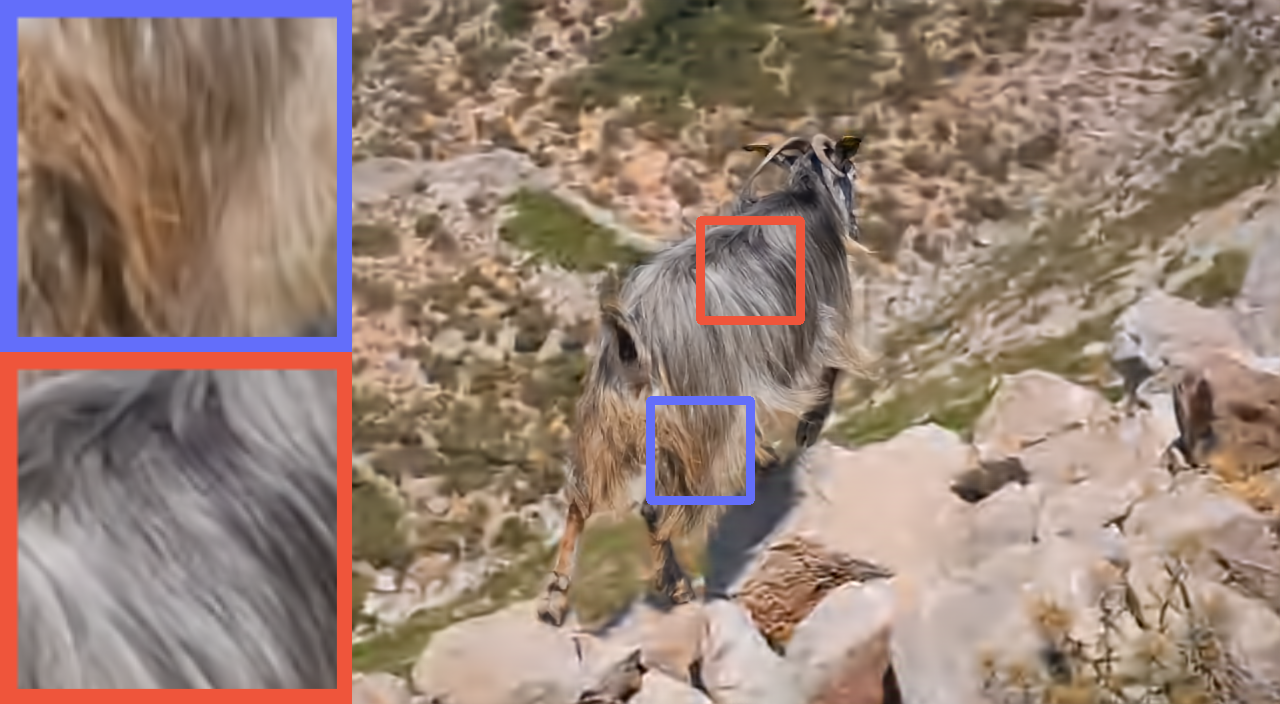}&
\includegraphics[width=0.3\textwidth]{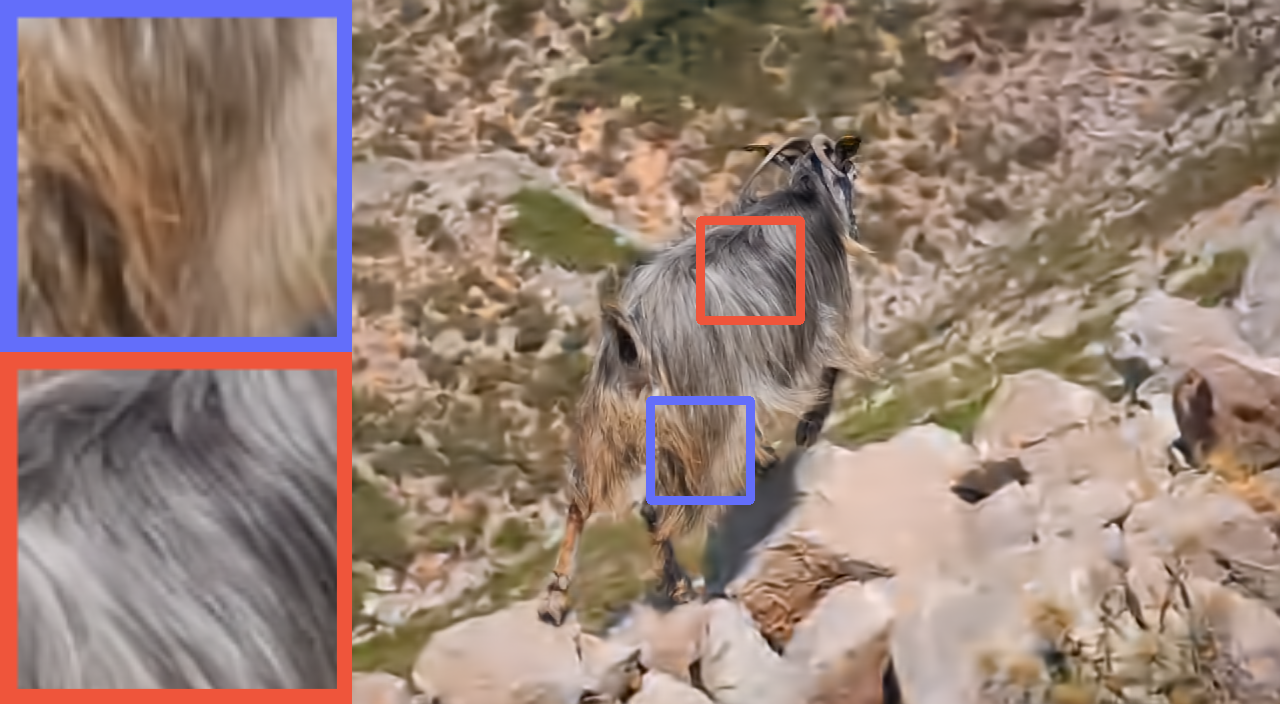}&
\includegraphics[width=0.3\textwidth]{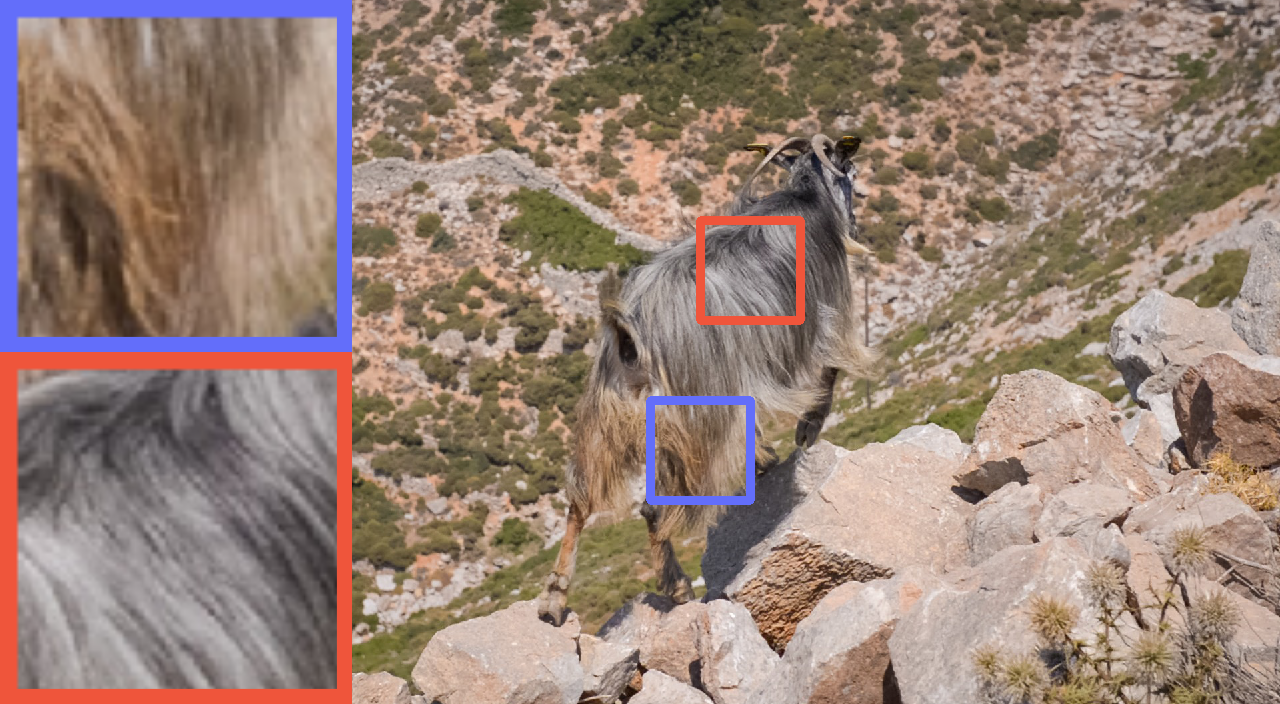}\\
\includegraphics[width=0.3\textwidth]{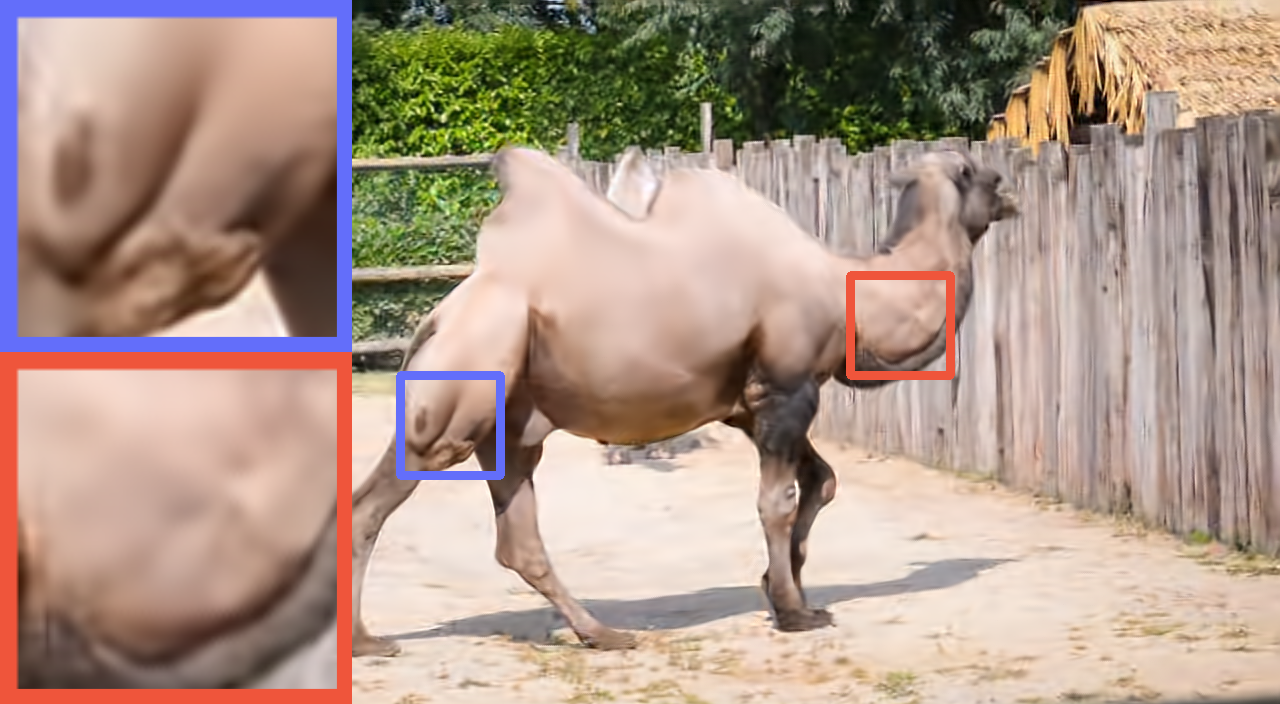}&  
\includegraphics[width=0.3\textwidth]{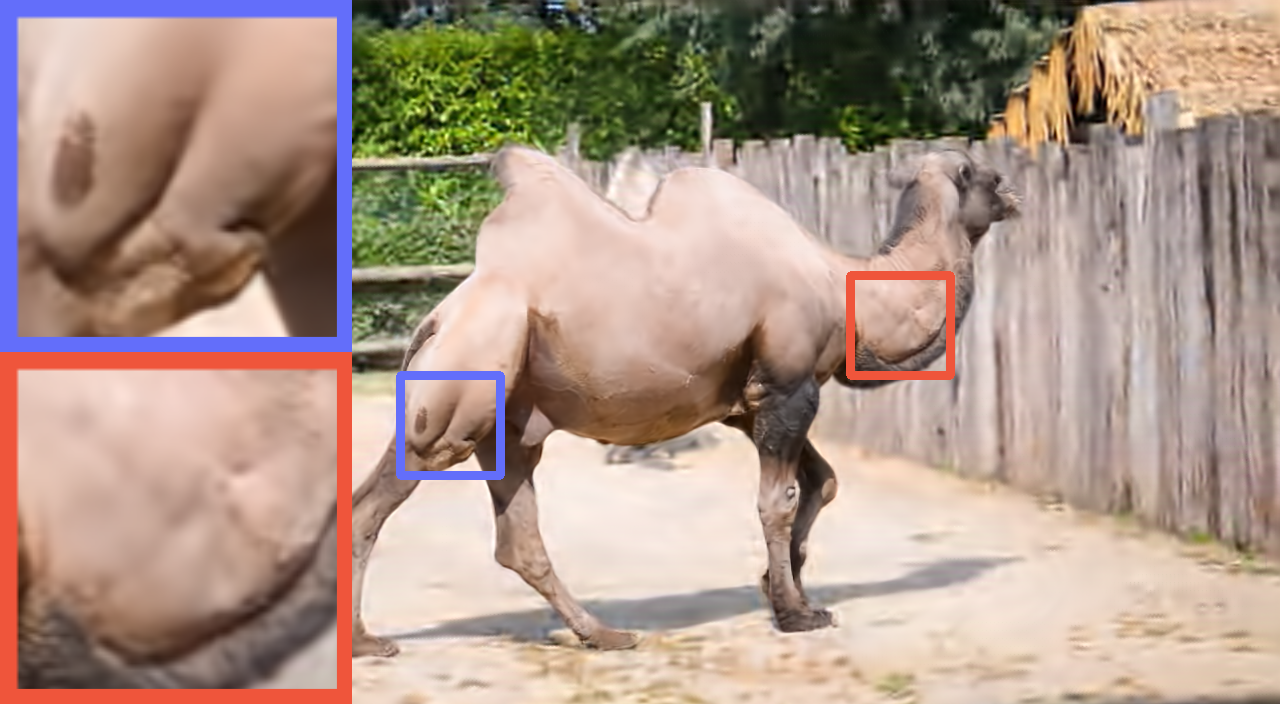}&
\includegraphics[width=0.3\textwidth]{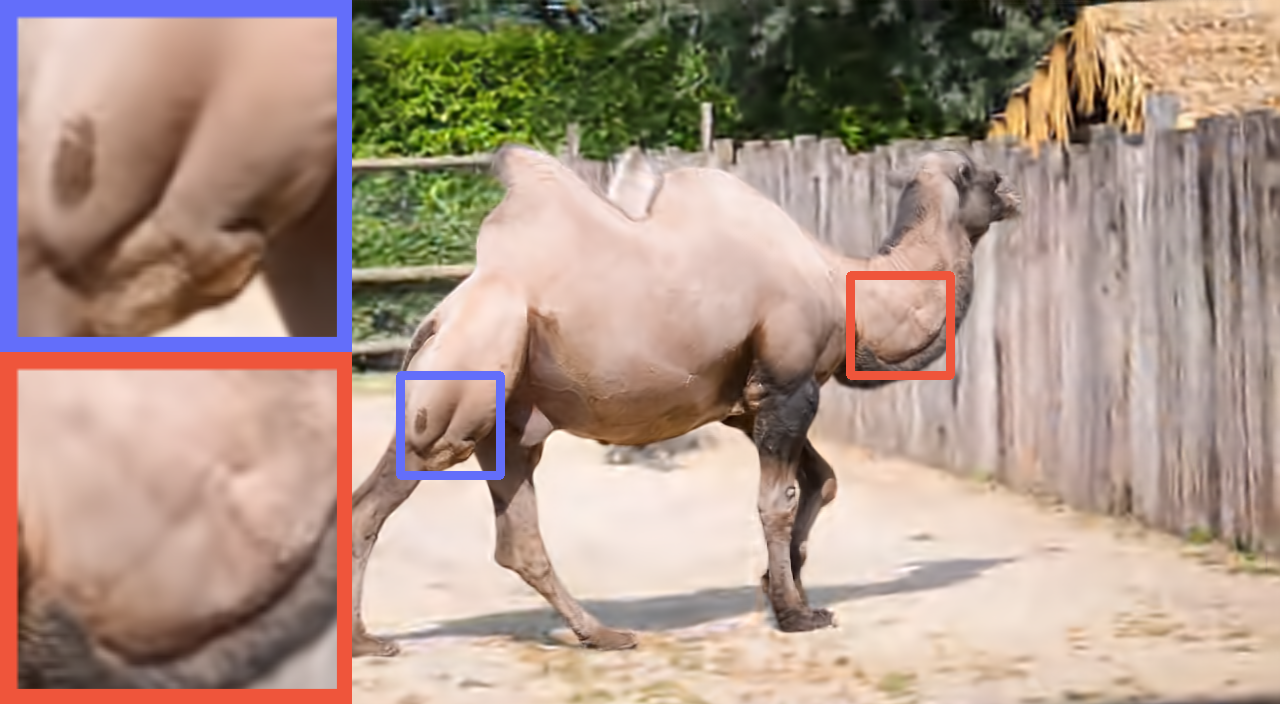}&
\includegraphics[width=0.3\textwidth]{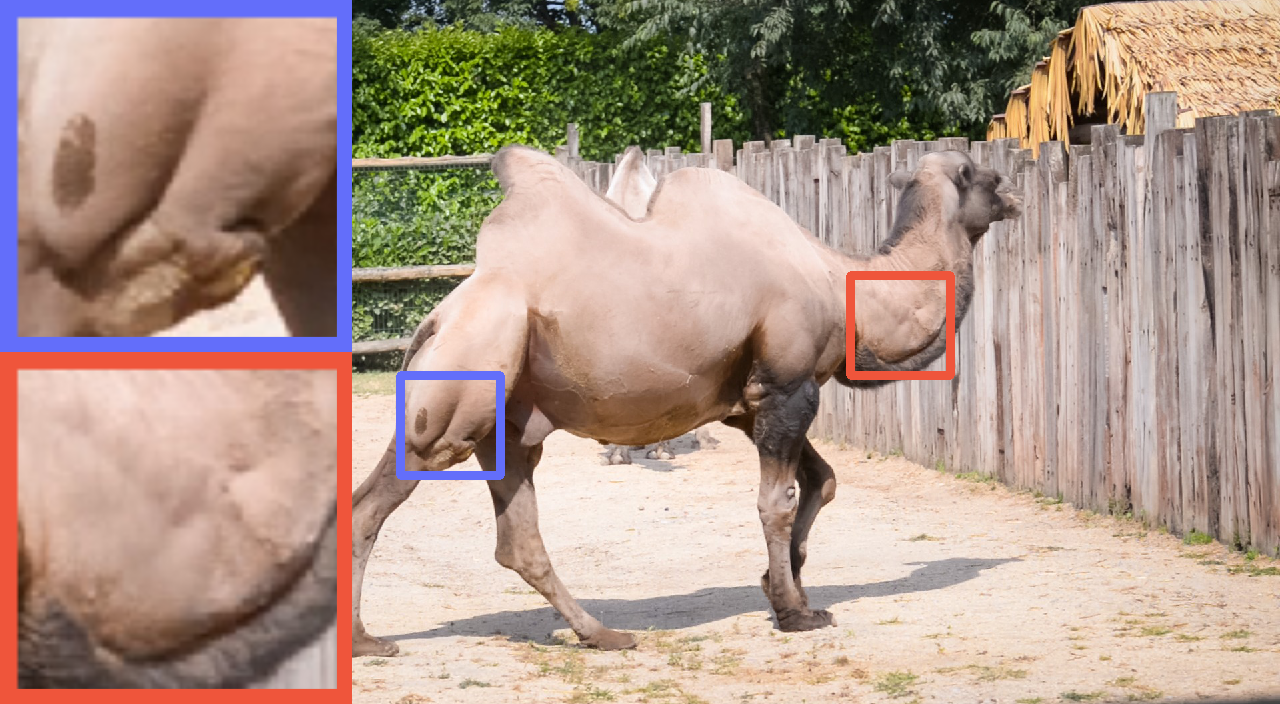}\\
\includegraphics[width=0.3\textwidth]{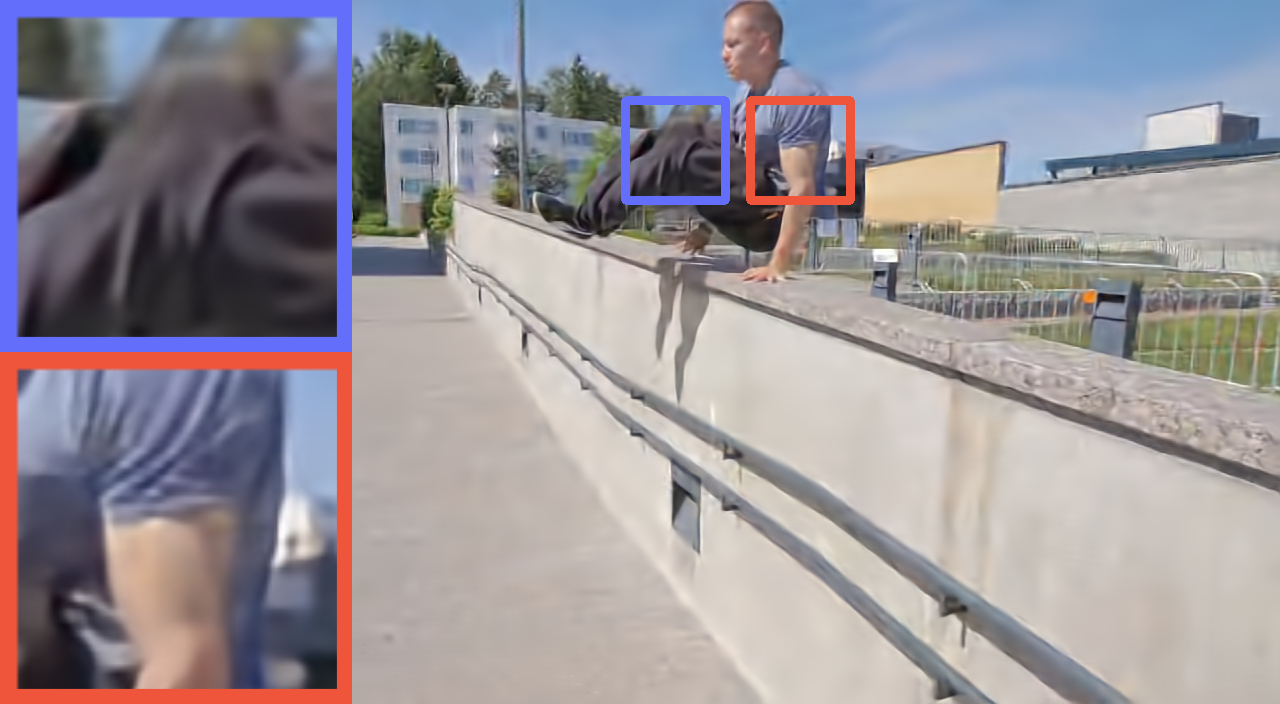}&  
\includegraphics[width=0.3\textwidth]{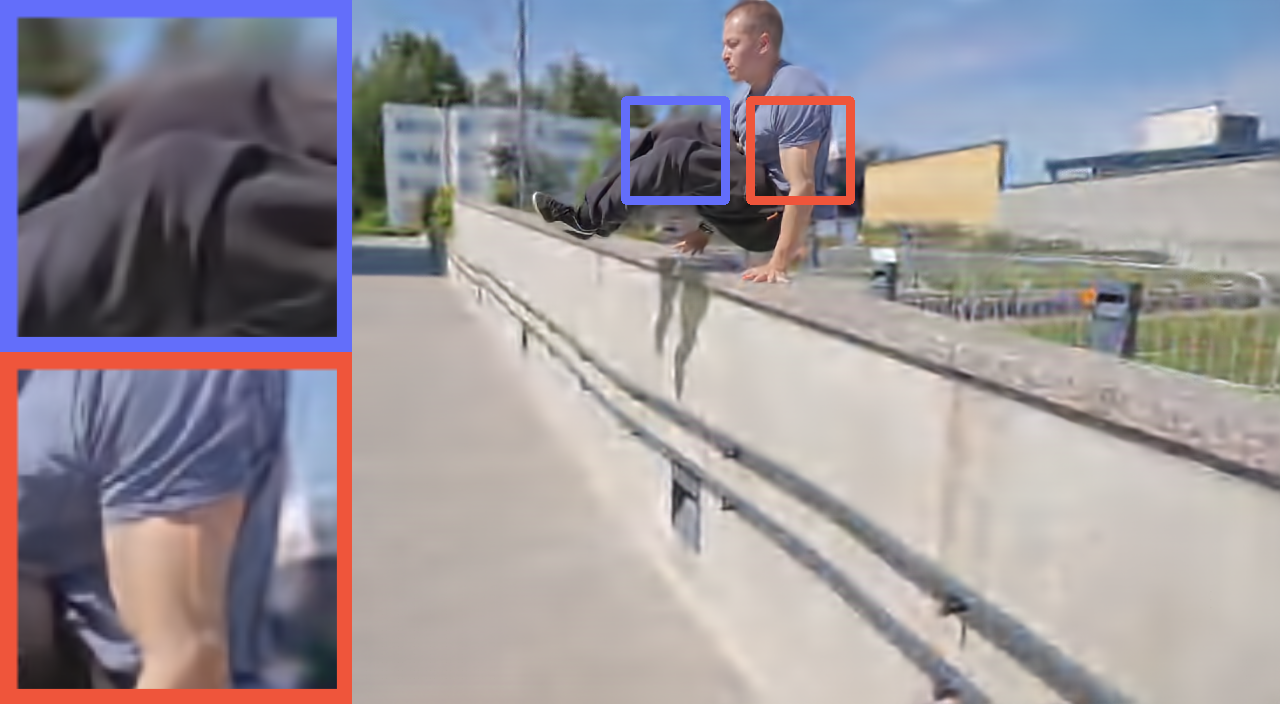}&
\includegraphics[width=0.3\textwidth]{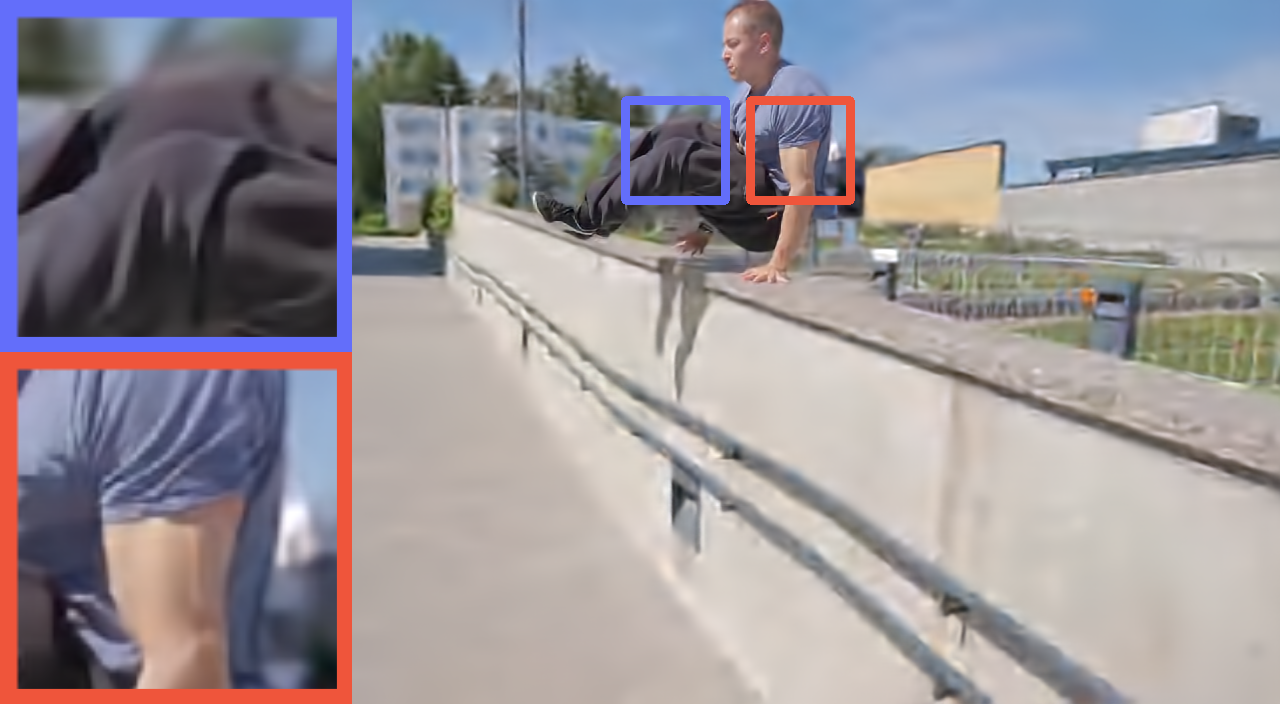}&
\includegraphics[width=0.3\textwidth]{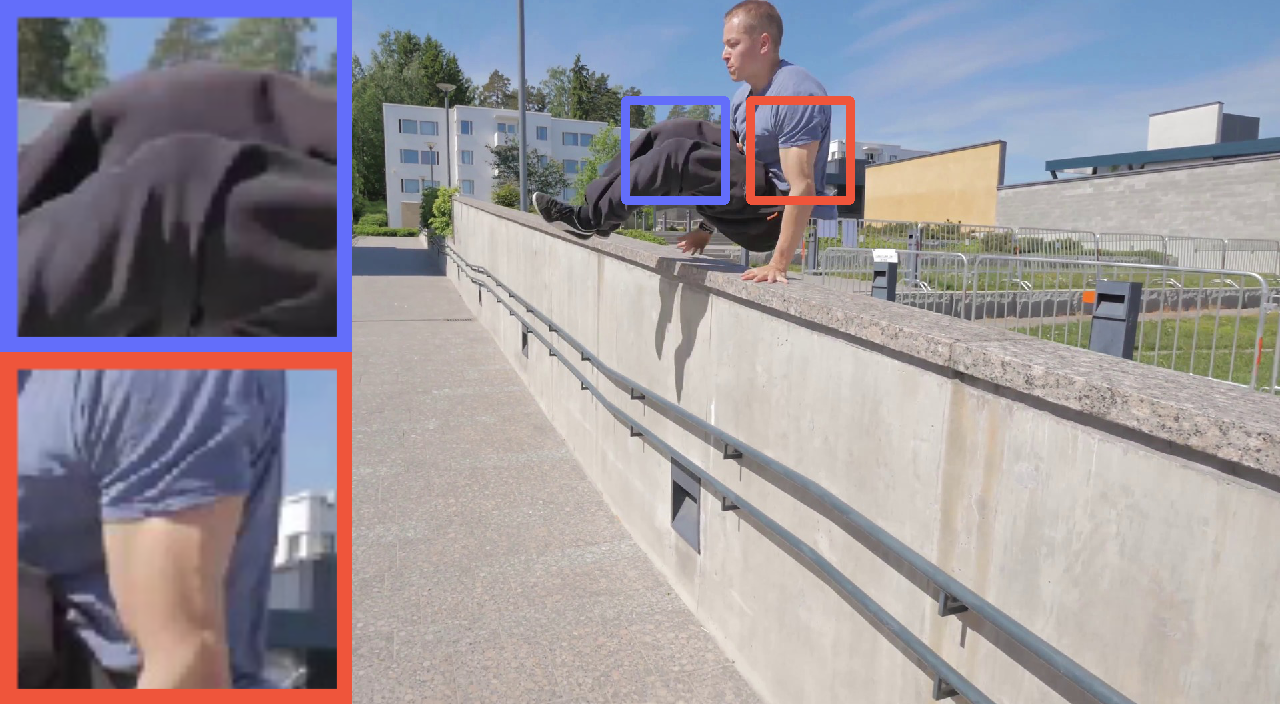}\\
\end{tabular}}
\caption{Qualitative results of SSF and our proposed ROI-based codecs, implicit ROI SSF and latent-scaling ROI SSF, on the sequences ``\texttt{goat}, \texttt{camel}, \texttt{parkour}'' sequence of DAVIS Val 2017. We hereby report frames 5, 11, 31 respectively.}
\label{fig:qualitative}
\end{figure*}
\egroup
\paragraph{Generalization} 
We investigate the generalization capability of our proposed latent-scaling ROI SSF model to different data and regions of interest.
To do so, train a model on DAVIS and measure its performance on Cityscapes.
We expect (at least) two main sources of generalization gap.
First, the videos in the two datasets depict very different content (\textit{data gap}), and differences in the acquisition settings may generate discrepancies in low-level image statistics and global motion\footnote{for instance, in Cityscapes the motion is dominated by the ego-motion of the camera, which is car-mounted.}.
Moreover, the ROI specification described above might impact training (\textit{ROI gap}).
To monitor both effects, we plot in Fig.~\ref{fig:cityscapes_and_perlin_noise}a the RD curves of our latent scaling model and plain SSF, trained either on DAVIS or on Cityscapes, and evaluated on Cityscapes.
By considering the gap between the SSF model (blue lines) trained on DAVIS and the one trained on Cityscapes, we notice how the former performs slightly worse than the latter, both for ROI and non-ROI areas.
This gives a sense of the severity of the data gap alone, as no ROI was employed during training whatsoever.
In order to assess the effect of the ROI gap, we examine the margin between the two trainings of Latent Scaling ROI SSF (pink lines).
Interestingly, we observe a similar edge as the one observed for plain SSF.
The fact that the performance gap does not increase significantly suggests that most of the discrepancy is still due to the data gap, and that our codec is barely susceptible to the nature of ROIs used during training.
Finally, we observe that, when evaluated on Cityscapes ROI areas, the ROI-based model trained on DAVIS outperforms the SSF model.
This observation suggests that, when interested in ROI-based compression on a target dataset, our codec trained on a different dataset might still be a better choice than its non ROI-based counterpart, even when the latter is trained on the target dataset itself.
\begin{figure*}[t]
\centering
\resizebox{0.95\columnwidth}{!}{
\begin{tabular}{cc}
\includegraphics[width=0.6\columnwidth]{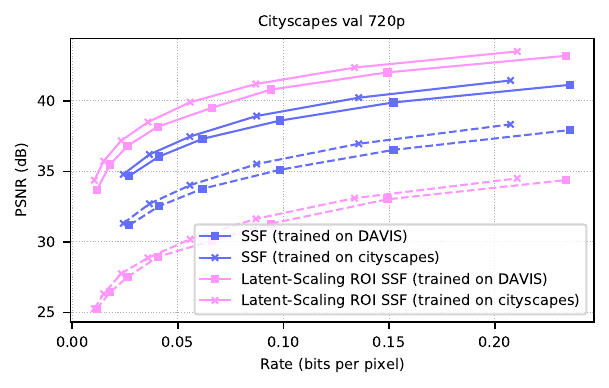}& 
\includegraphics[width=0.6\columnwidth]{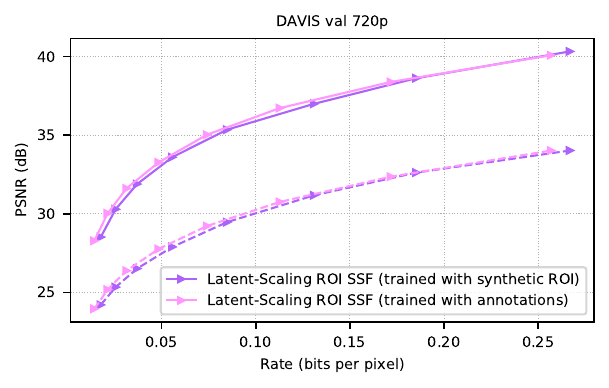}\\
(a) & (b)
\end{tabular}}
\caption{(a) Latent-scaling ROI SSF tested on Cityscapes. (b) Effect of training with synthetic ROI instead of ground-truth annotation for the binary ROI mask. Solid / dashed lines denote ROI/non-ROI PSNR respectively.}
\vspace{-4mm}
\label{fig:cityscapes_and_perlin_noise}
\end{figure*}
\paragraph{Synthetic ROI masks}
\label{sec:results_synthetic_roi}
In order to further investigate the sensitivity of our latent scaling based codec to the nature of ROIs used during training, we carry out an experiment where we train it using synthetically generated masks.
Specifically, we rely on the DAVIS dataset and we generate the ROI for every training clip randomly, by taking advantage of perlin noise~\cite{perlin_noise}.
The resulting masks are temporally smooth, but do not correlate with the content of the video itself.
In Fig.~\ref{fig:cityscapes_and_perlin_noise}b we plot the performance of such a model (in purple) against a model trained on regular semantic masks, obtained by manual annotation (in pink).
We emphasize that both models are tested, on the validation set, on regular semantic masks of ROI objects.
Thus, we expect the model trained on realistic ROI masks to trace an upper bound RD-curve for the model trained on synthetic.
Interestingly, results show that a gap exists between the two models, but it is almost negligible, confirming the intuition that our model is minimally affected by the nature of training ROIs.
The close performance represented in RD-curves suggests that, although in the case ROI masks are available at training time their use is worthwhile, their lack does not represent a serious impediment for optimizing the model, as the use of synthetic masks yields similar performance on realistic use cases.
\section{Limitations and societal impact}
Our main motivation for the Latent Scaling ROI Scale-Space Flow was to allow for inference-time single model multirate behavior for the largest rate model, without the need to re-train or to adapt the training scheme like in~\cite{cui2020g,elfvc} (similar to what was demonstrated in~\cite{plonq} for image compression).
This would make our ROI codec more practical to deploy by drastically reducing the number of parameters and allowing fine-grain control of the rate.
However, it does not allow for a fully multirate model (\textit{i.e.} a single model covering the whole rate spectrum), and it comes with an increase in implementation complexity with minor performance benefits over the simpler implicit ROI approach.

In addition, visual assessments highlighted how, in their current implementation, both ROI-based models can sometimes produce sharp quality transitions between ROI and non-ROI regions. The problem would probably be exacerbated if the ROI masks suffered both in terms of quality and in temporal consistency. Both of these issues may be overcome by using smooth masks during training and/or inference.

Finally, a user study would benefit the evaluation of quality of the compressed videos as quantitative quality metrics were shown to poorly correlate with human judgment~\cite{le2016vmaf}.
Such an analysis, based on subjective metrics such as Mean Opinion Scores (MOS), would further confirm that higher fidelity in the ROI at the cost of fidelity in the non-ROI can lead to a net boost in perceptual quality.

Concerning societal impact, we do not see immediate harmful applications of our method that might negatively affect any public. Note that because the ROI codecs depend on an ROI retrieval algorithm, the methods may suffer from (and potentially amplify) its biases and shortcomings.
\section{Conclusions}
In this paper, we introduced two methods for ROI-based neural video compression, capable of allocating more bits to pre-specified regions of interest in order to increase their fidelity. 
More specifically, we introduced an implicit model being fed with the ROI, as well as a latent scaling model explicitly controlling the quantization bitwidth of the latent variables in a spatial variant fashion.
Both models are optimized by a ROI-aware rate-distortion objective.
We showed that our methods outperform all baselines in terms of Rate-Distortion performance in the regions of interest, and that they can generalize to different datasets at inference time. 
Finally, they do not require expensive pixel-level annotations during training, as synthetic ROI masks can be used with little to no degradation in performance.
\begin{center}
\Large
\color{bmv@sectioncolor}
\textbf{
Supplementary Material
}
\end{center}
\appendix
\section{ROI creation}
\label{apd:roi_creation}
\bgroup
\setlength{\tabcolsep}{1pt}
\begin{figure}[b]
\centering
\resizebox{0.9\columnwidth}{!}{
\begin{tabular}{cc}
\includegraphics[width=0.5\columnwidth]{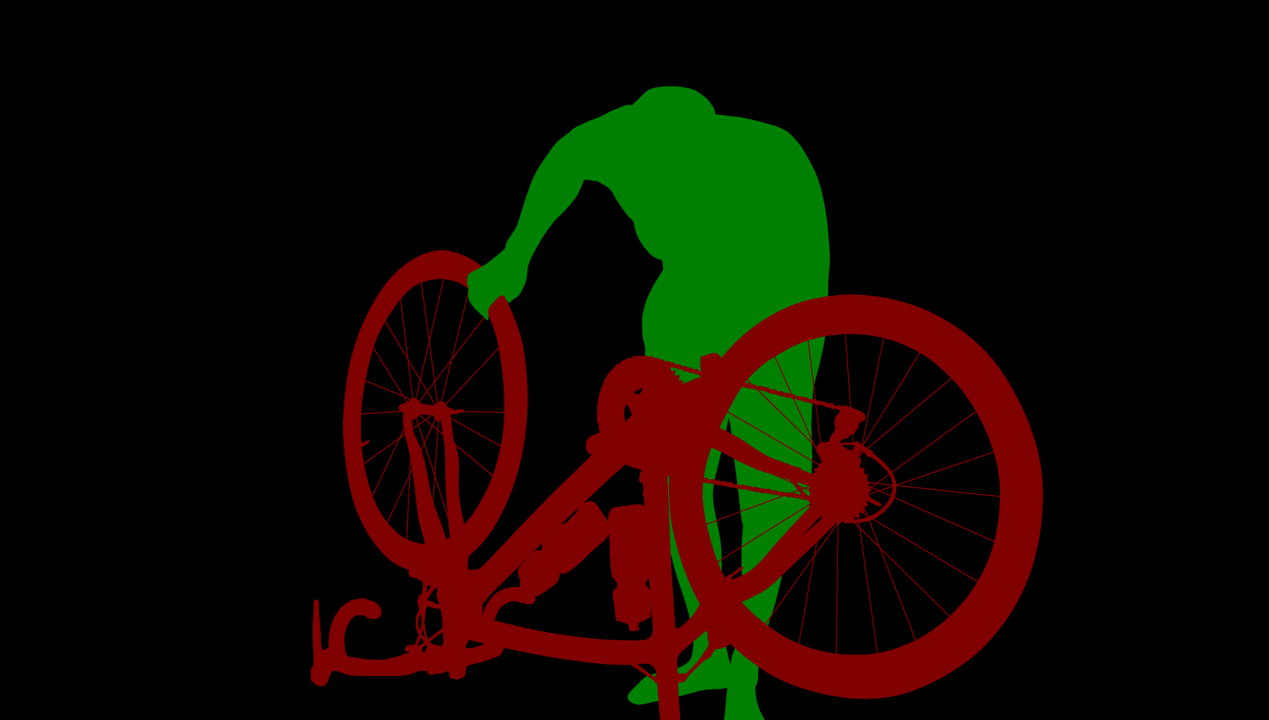}&
\includegraphics[width=0.5\columnwidth]{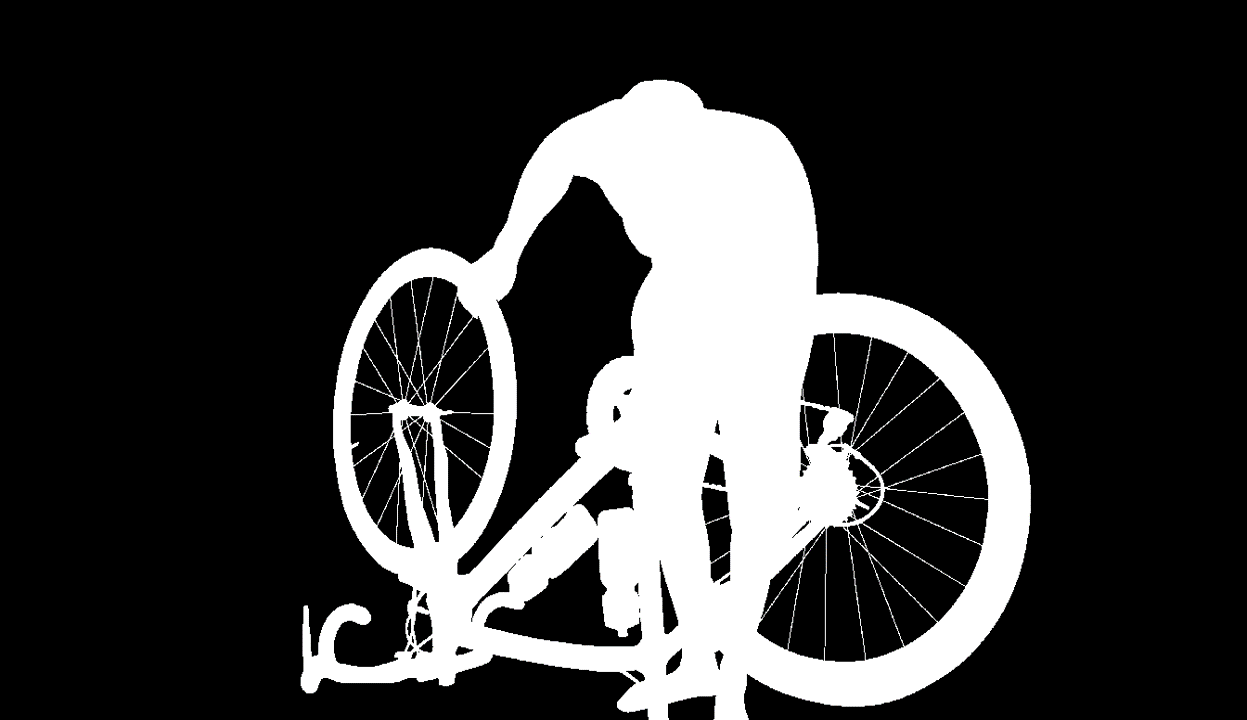}\\
DAVIS annotation & DAVIS ROI\\
\includegraphics[width=0.5\columnwidth]{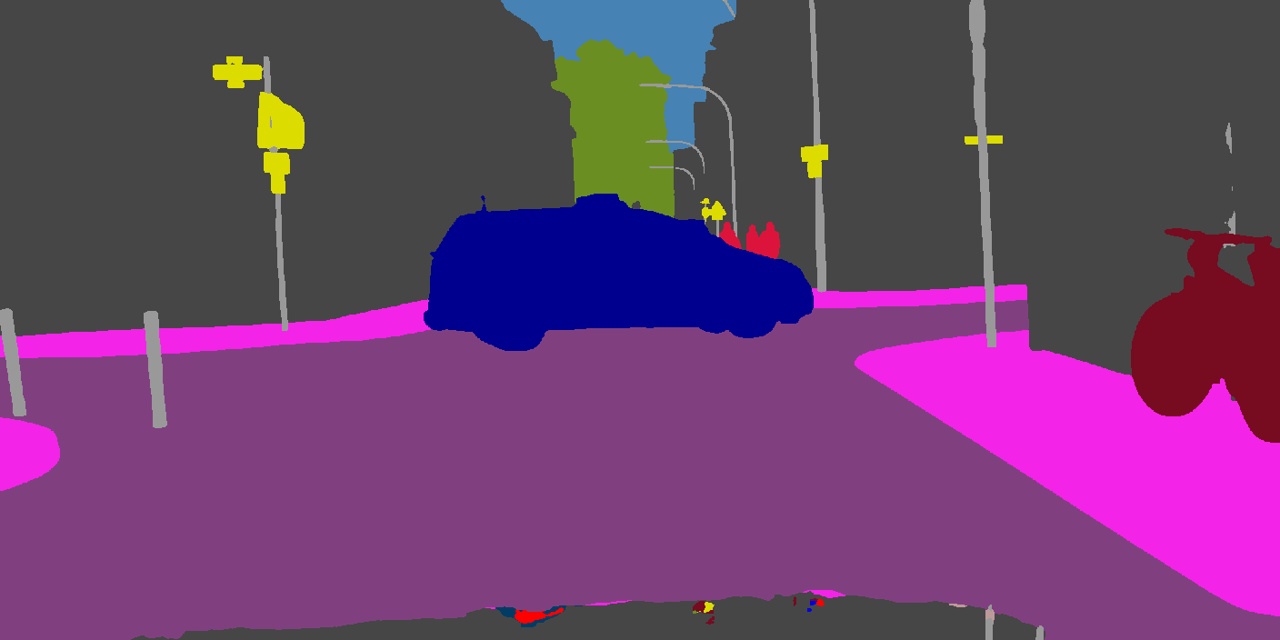}&
\includegraphics[width=0.5\columnwidth]{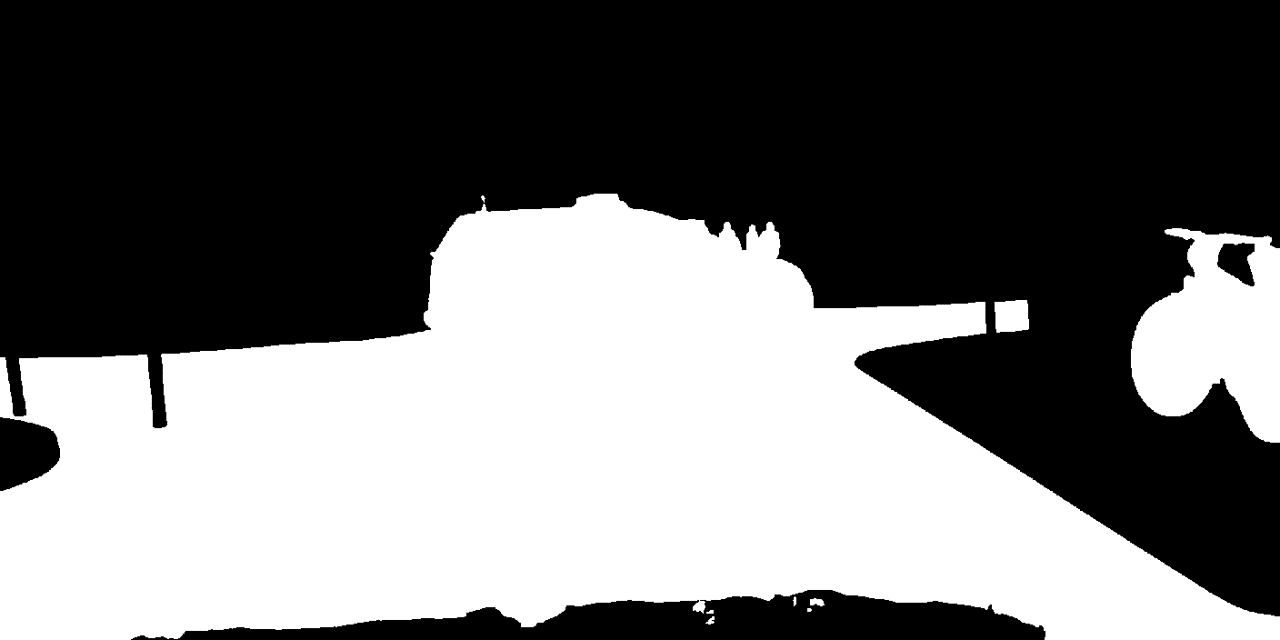}\\
Cityscapes annotation & Cityscapes ROI 
\end{tabular}}
\caption{Example of ROI creation for the DAVIS and Cityscapes datasets.}
\label{fig:roi_creation}
\end{figure}
\egroup
In Sec~4 (main paper) we explained how we created binary ROI mask from ground-truth annotations. In Fig.~\ref{fig:roi_creation} we show visual examples of this process for the DAVIS (top) and Cityscapes (bottom) datasets.
\section{Additional results} 
\label{apd:additional}
\subsection{Quantitative results} 
\label{apd:gammas}
During our research we tested two different penalty term $\gamma$ for non-ROI distortion, as defined in Eq.~5 (main paper), namely $\gamma=\{10, 30\}$. In Sec.~4 (main paper), all results are shared with $\gamma=30$ for ease of exposition.
In this section we provide additional results with $\gamma=10$. We allow side-by-side comparison for all experiments of Sec.~4 (main paper) for each penalty $\gamma$. Finally, we provide an additional multirate analysis.

\paragraph{ROI-based coding}
In Fig.~\ref{fig:apd_a_all_models} we show all ROI-based models trained with $\gamma=\{10, 30\}$ on DAVIS and evaluated on DAVIS val, with SSF as reference.
As expected from our loss formulation, a smaller penalty $\gamma$ results in a smaller performance gap between ROI and non-ROI across all ROI-based methods.
Interestingly, both the ROI-aware loss and OBIC SSF baselines which are blind to the ROI mask seem to only allow higher PSNR in the ROI than in the non-ROI at low bitrate, namely $\leq 0.15$ bpp.
For $\gamma=30$, the ROI PSNR is consistently better than non-ROI PSNR across the entire rate spectrum.
The two methods may perform similarly as they are both blind to the ROI mask, \ie the encoding operation does not get the ROI mask as input, although OBIC SSF foreground and background hypercodecs do get ROI information as their input is the ROI masked latent. We hypothesize that it may be insufficient for the hyper-codec network to implicitly learn to scale the prior parameters, and does not allow the encoder to scale the latent.
\bgroup
\setlength{\tabcolsep}{1pt}
\begin{figure*}[tbh]
\resizebox{\linewidth}{!}{
\begin{tabular}{cc}
\quad \quad $\gamma=10$ & \quad \quad $\gamma=30$\\
\includegraphics[width=0.5\textwidth]{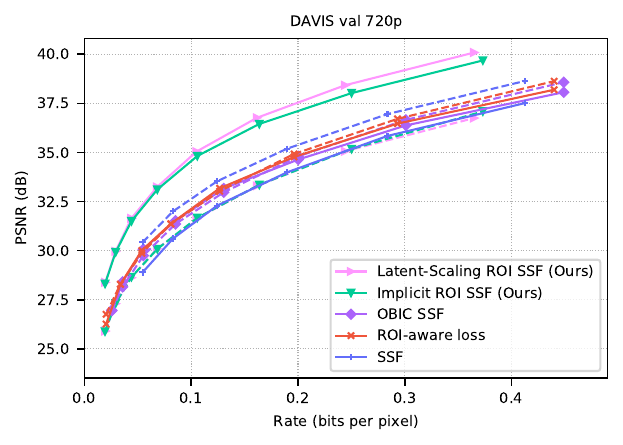}&  
\includegraphics[width=0.5\textwidth]{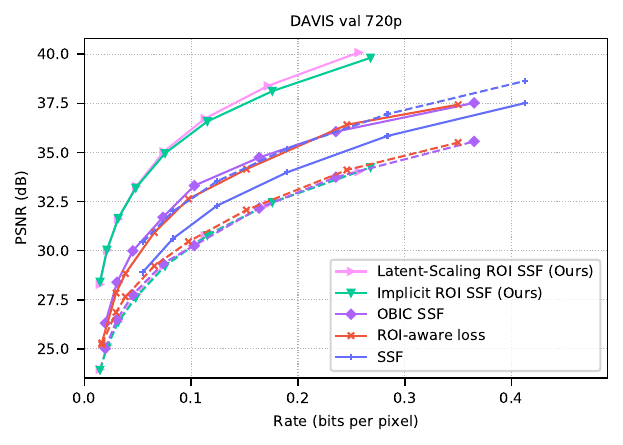}
\end{tabular}}
\caption{All ROI-based neural video compression approaches vs SSF, trained on DAVIS and evaluated on DAVIS val. ROI-based models are trained with $\gamma=\{10, 30\}$, left and right plot respectively. Right plot is Fig.~4 in the main text. Solid line denotes ROI-PSNR, while dashed non-ROI PSNR.}
\label{fig:apd_a_all_models}
\end{figure*}
\egroup

\paragraph{Generalization}
In Fig.~\ref{fig:apd_c_cityscapes} we show the SSF and latent-scaling ROI SSF models trained on either DAVIS or Cityscapes and evaluated on Cityscapes val for both values of $\gamma = \{10, 30\}$. As expected from our loss formulation, for $\gamma=10$ latent-scaling ROI SSF exhibits a smaller gap between ROI PSNR and non-ROI PSNR than with $\gamma=30$. Yet, irrespective of $\gamma$, the same observation can be made: the ROI PSNR of latent-scaling ROI SSF trained on DAVIS is higher than SSF trained on Cityscapes. This indicates that when interested in ROI-based compression on a target dataset, our codec trained on a different dataset might still be a better choice than its non ROI-based counterpart, even when the latter is trained on the target dataset itself.
\bgroup
\setlength{\tabcolsep}{1pt}
\begin{figure*}[tbh]
\resizebox{\textwidth}{!}{
\begin{tabular}{cc}
$\gamma=10$ & $\gamma=30$\\
\includegraphics[width=0.5\textwidth]{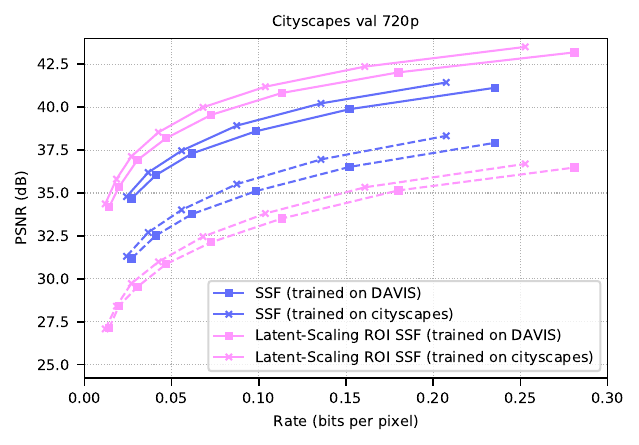}&  
\includegraphics[width=0.5\textwidth]{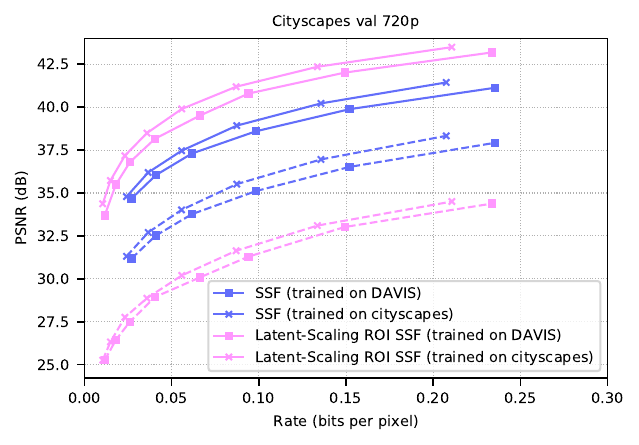}
\end{tabular}}
\caption{SSF and Latent-scaling ROI SSF trained on either DAVIS or Cityscapes with ground-truth annotations and evaluated on Cityscapes val. Our models are trained with $\gamma=\{10, 30\}$, left and right plot respectively. Right plot is Fig.~7a in the main text. Solid line denotes ROI-PSNR, while dashed non-ROI PSNR.}
\label{fig:apd_c_cityscapes}
\end{figure*}
\egroup
\paragraph{Synthetic ROI masks}
\label{apd:perlin_noise}
In Fig.~\ref{fig:apd_perlin_noise1} we show the effect of using synthetic ROI mask during training instead of ground-truth annotations, for $\gamma=\{10, 30\}$.
In addition to the experiment in the main text, we not only show latent-scaling ROI SSF but also implicit ROI SSF.
We find that for each of our proposed models, training with synthetically generated masks results only in a minor performance drop, albeit slightly larger for the implicit model. Since the performance of our proposed ROI-based models seem to be minimally affected by the type of ROI masks used during training, one could train them without requiring expensive pixel-wise annotations. This allows training on a target dataset of interest which may be different from dataset with available annotations like DAVIS. Consider, for instance, cartoons instead of natural videos.
\bgroup
\setlength{\tabcolsep}{1pt}
\begin{figure*}[tbh]
\resizebox{\textwidth}{!}{
\begin{tabular}{cc}
\quad \quad  $\gamma=10$ & \quad \quad $\gamma=30$\\
\includegraphics[width=0.5\textwidth]{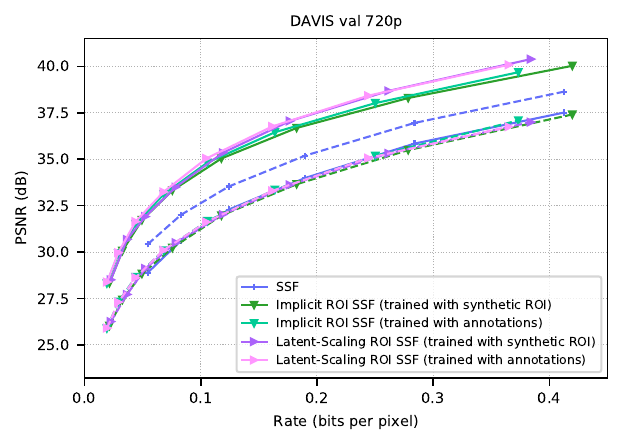}&  
\includegraphics[width=0.5\textwidth]{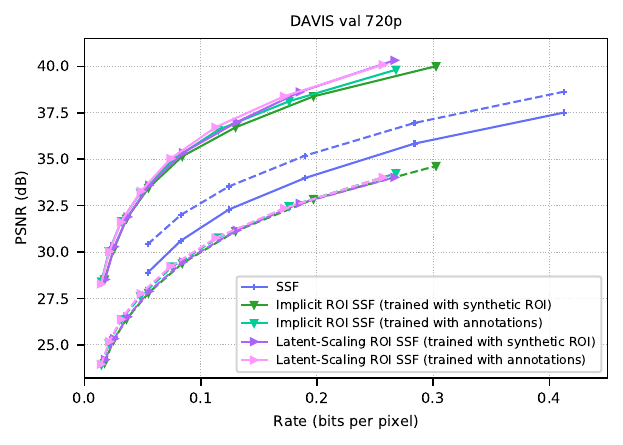}
\end{tabular}}
\caption{Effect of training with synthetic ROI masks instead of ground-truth annotations on ROI PSNR R-D performance for DAVIS val dataset. We show the implicit and latent-scaling ROI SSF versus the original SSF. Our models are trained with $\gamma=\{10, 30\}$, left and right plot respectively. Right plot is a modified version of Fig.~7b in the main text. Solid line denotes ROI-PSNR, while dashed non-ROI PSNR.}
\label{fig:apd_perlin_noise1}
\end{figure*}
\egroup
\paragraph{Inference time ROI selection}
We herby evaluate the capability of our model to adapt to different ROI specifications in front of the same video to be compressed.
We remark that this trait is appealing as it would elect our model as general purpose, as the same trained model could be deployed for ROI-based compression in disparate use cases.
We also notice how this feature lacks in current works for neural codecs~\cite{habibian2019video,agustsson2019extreme}, as they typically commit to specific semantic classes during optimization and are trained such that their encoder would implicitly recognize and favor important regions.
On the contrary, our model is explicitly fed with a mask specifying the desired (non-)ROI areas, allowing to compress the same video differently, depending on the desired ROI specifications.

\begin{figure*}
\centering
\resizebox{\textwidth}{!}{
\begin{tabular}{ccc}
\includegraphics[height=5cm]{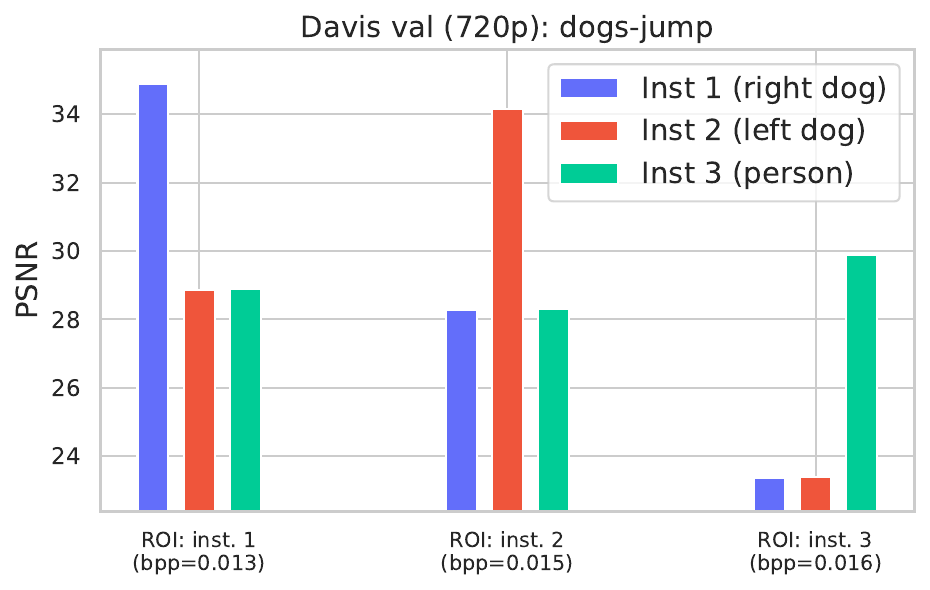}&
\includegraphics[height=5cm]{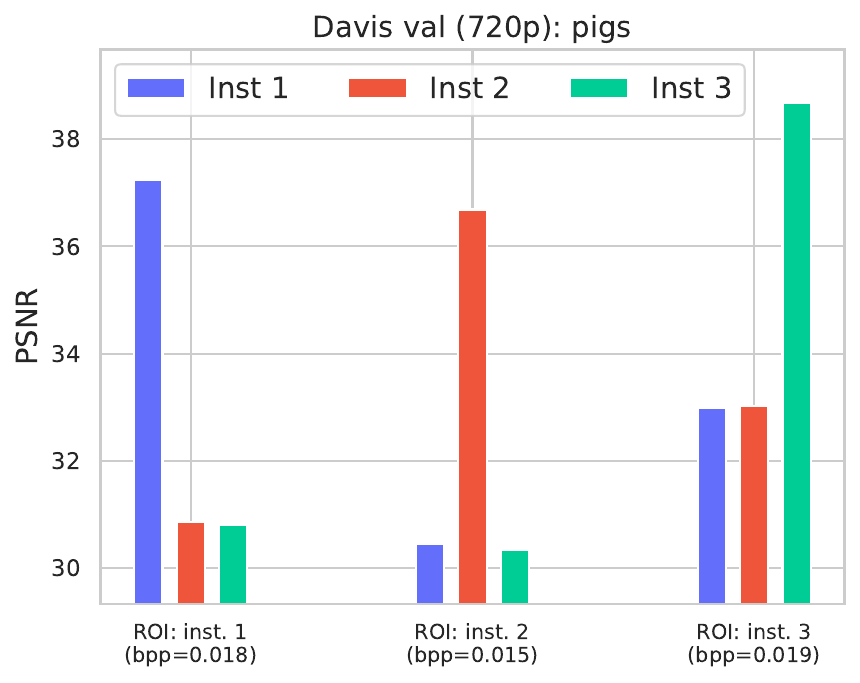}&
\includegraphics[height=5cm]{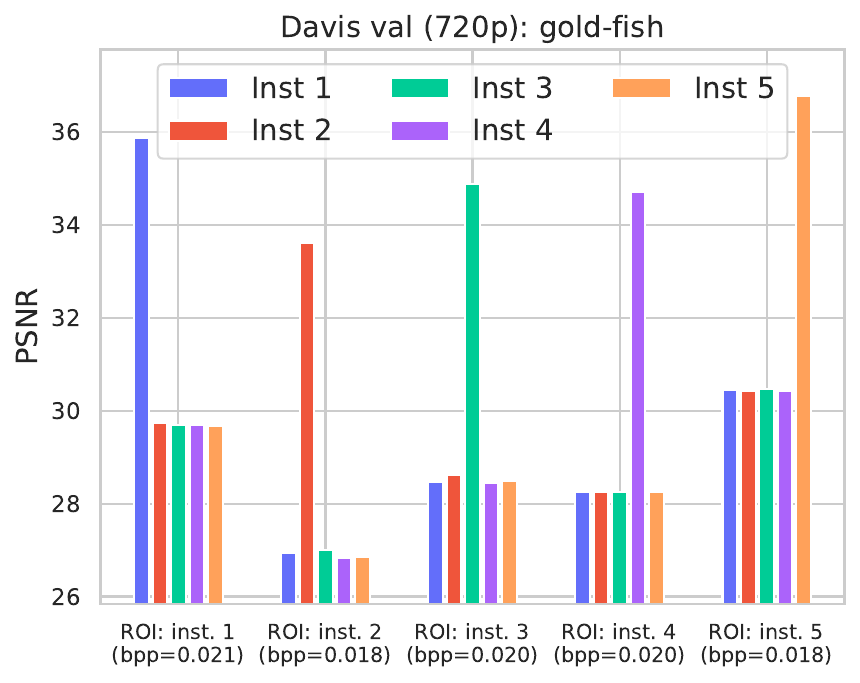}
\end{tabular}}
\caption{PSNR of each instance when ROI-coding the  different foreground instances in the ``\texttt{dogs-jump}``, ``\texttt{pigs}`` and ``\texttt{gold-fish}`` sequences in the DAVIS validation set.}
\label{fig:custom_foreground_barplots}
\end{figure*}

We select several sequences from the DAVIS validation set (\texttt{dogs-jump}, \texttt{pigs} and \texttt{gold-fish}), being labeled with more than one instance.
Instead of merging all instances into a single ROI mask (as we do in all other experiments), we compress the video multiple times, by considering different instances as ROI in different runs.
We consistently monitor PSNR on all instances, and observe it is consistently higher in the region considered as ROI.
We represent these results color-coded in the barplots in Fig.~\ref{fig:custom_foreground_barplots}. 
In all videos being considered, the instance considered as ROI benefits a boost of 5dB or more in PSNR.
This result clearly shows that our codec can be used, at approximately the same bitrate, to improve reconstruction quality in any ROI of choice.
A qualitative representation of such a feature is represented, for the \texttt{dogs-jump} sequence, in Fig.~\ref{fig:custom_foreground_qualitative}.
\bgroup
\begin{figure}
\begin{center}
\resizebox{0.9\columnwidth}{!}{
\begin{tabular}{cc}
\includegraphics[width=0.5\columnwidth]{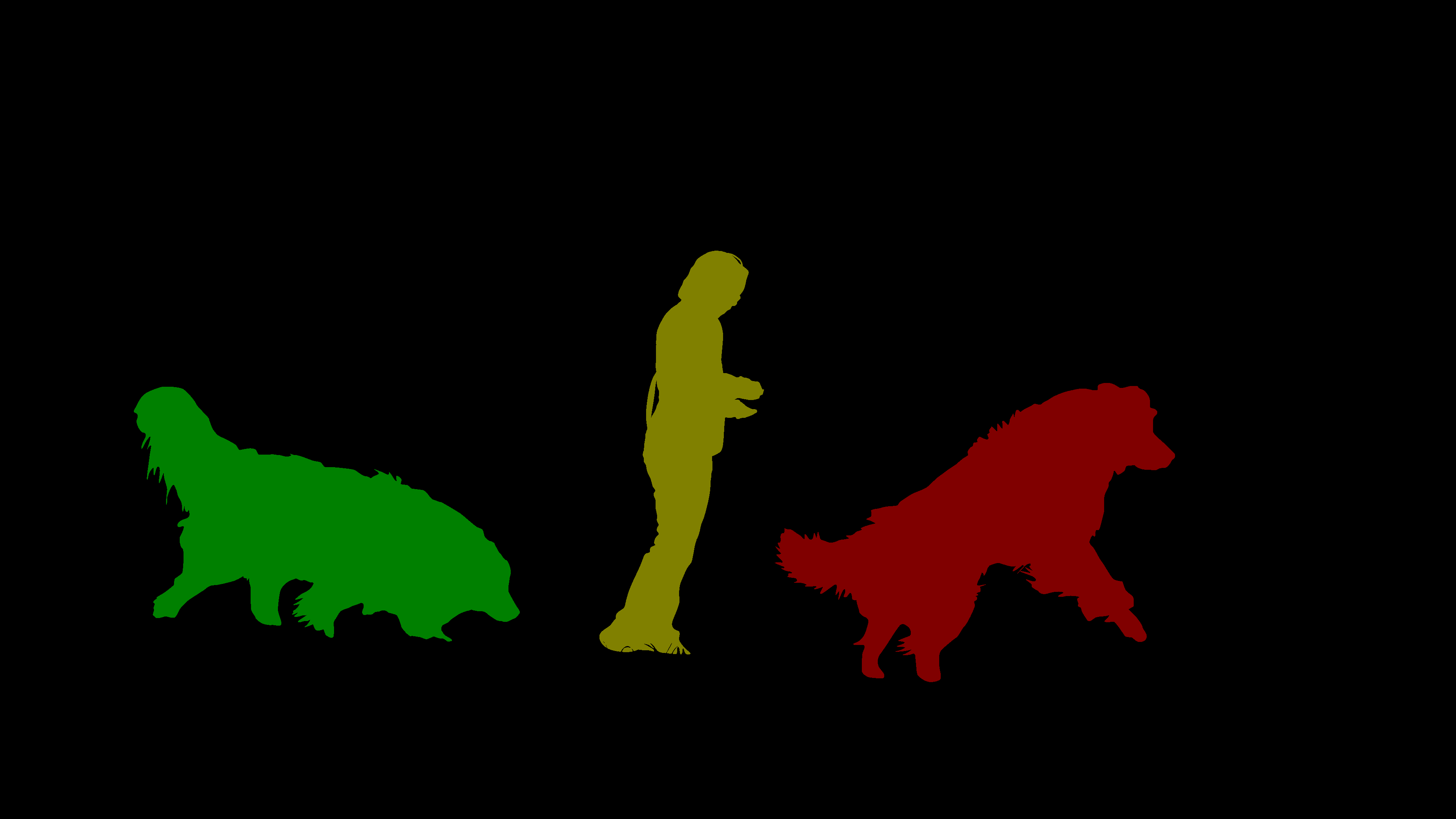}&  
\includegraphics[width=0.5\columnwidth]{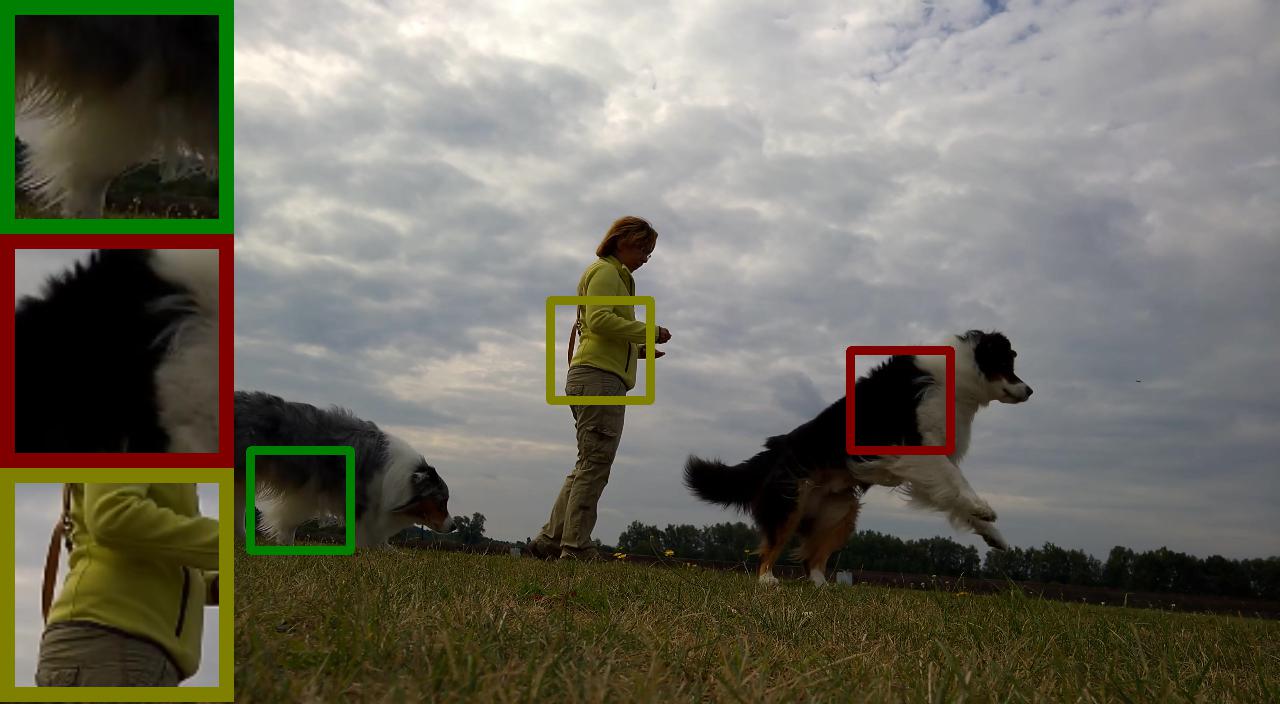}\\
Semantic instances & Reference\\
\includegraphics[width=0.5\columnwidth]{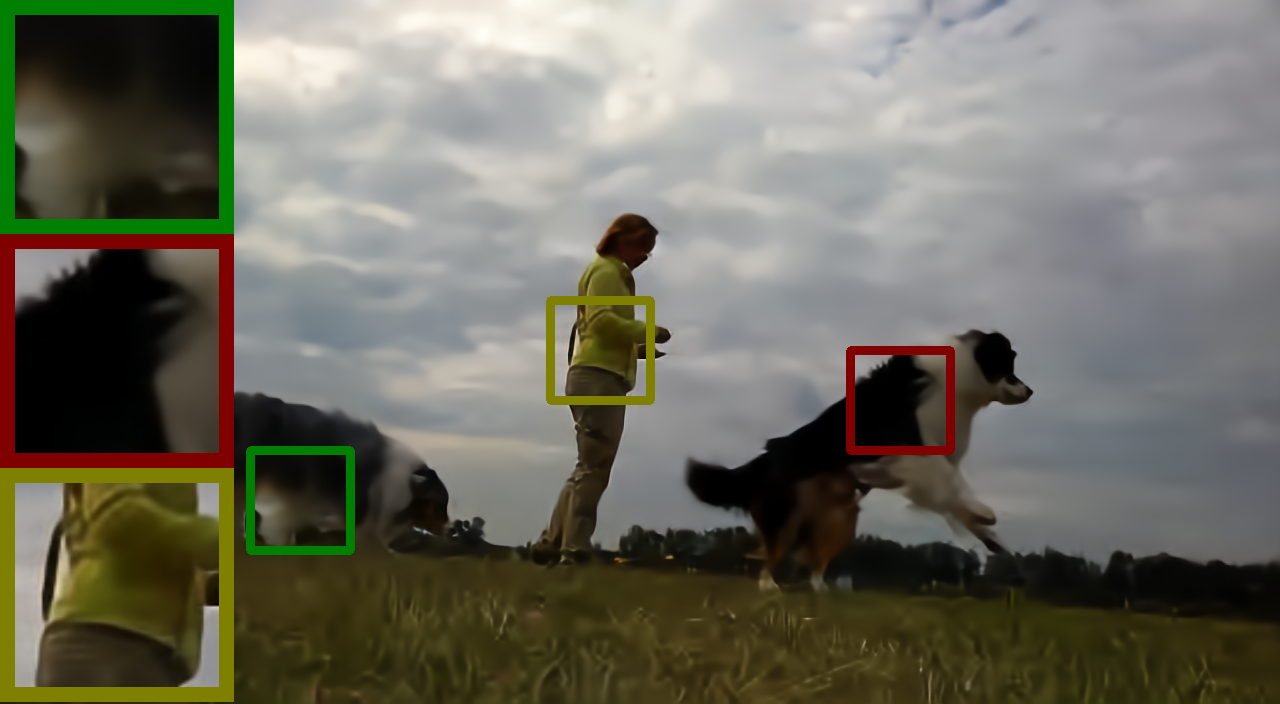}&
\includegraphics[width=0.5\columnwidth]{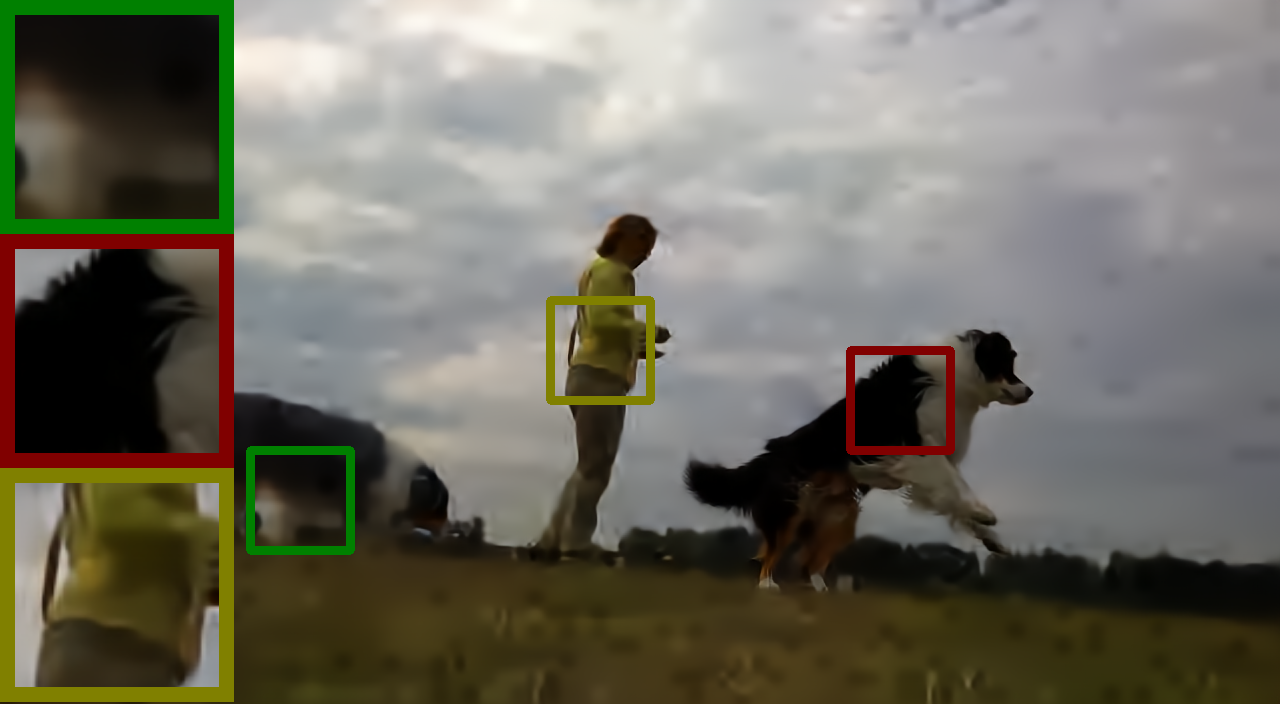}\\
SSF - 0.02 bpp & LS-SSF (red) - 0.01 bpp\\
\includegraphics[width=0.5\columnwidth]{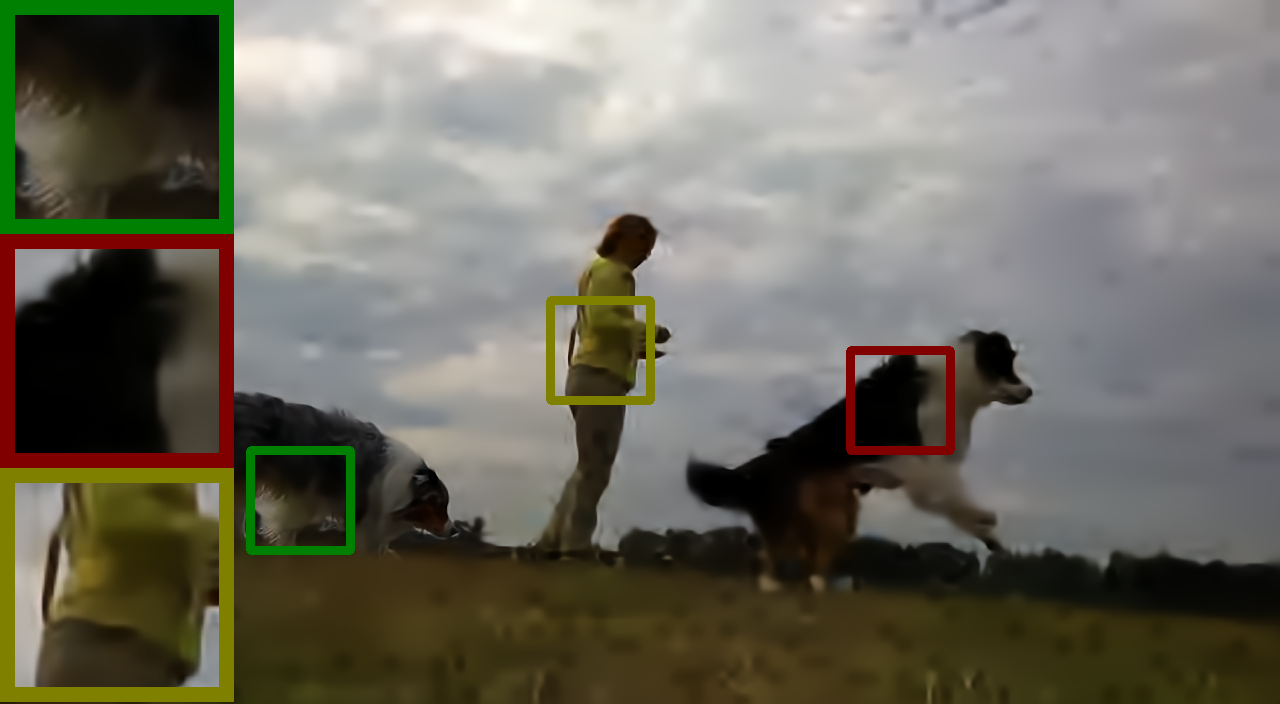}& \includegraphics[width=0.5\columnwidth]{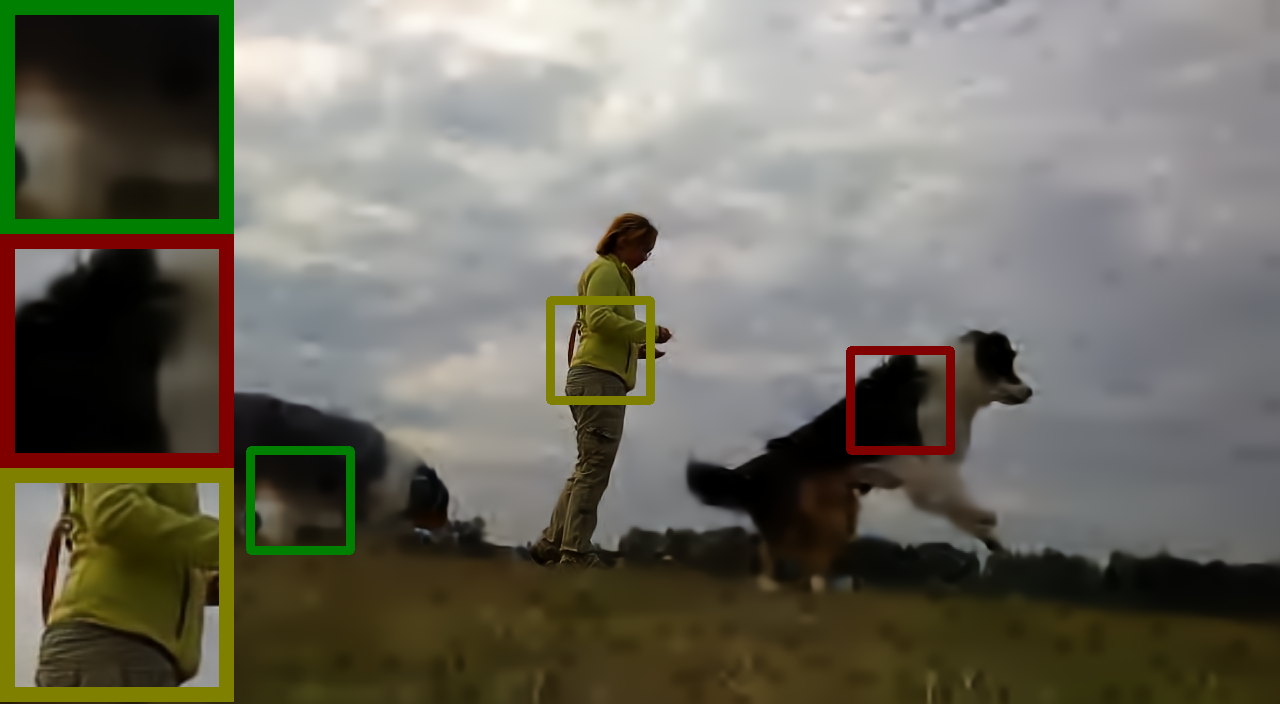}\\
LS-SSF (green) - 0.02 bpp & LS-SSF (yellow) - 0.02 bpp
\end{tabular}}
\end{center}
\caption{ROI-coding of different foreground instances (red, green and yellow) in the 37-th frame of the ``\texttt{dogs-jump}'' sequence in the DAVIS validation set. The same pretrained latent-scaling ROI SSF model can be conditioned to achieve a higher ROI PSNR on different ROIs at eval time.}
\label{fig:custom_foreground_qualitative}
\end{figure}
\egroup
\paragraph{Multirate capabilities}\label{apd:additional:multirate}
We experimented with the ``naive'' latent-scaling technique described in Lu \etal~\cite{plonq}. With the use of a gain amplifier $\mathrm{ga}$, it allows navigating different R-D tradeoffs with a single trained model during evaluation. The gain variable $h$ output by the gain hyperprior AE is transformed using
\begin{equation}
\label{eq:gain_amplifier}
\tilde{h} = (h-1) \cdot \mathrm{ga} + 1
\end{equation}
before being used to scale the prior parameters and latent code, see Sec~3 (main paper) for details.
Note that the higher the $\mathrm{ga}$ value, the coarser the quantization grid becomes, which in return is cheaper to encode.

In Fig.~\ref{fig:apd_b_ls} we show the latent-scaling ROI SSF for different rate regularization coefficients $\beta$ with gain amplifier $ga=1$ in pink. In addition
we select three trained models ($\beta=\{0.0001, 0.0008, 0.0064\}$) and sweep the gain amplifier $\mathrm{ga} \in \{1,2,4,8,16,32,64\}$; such curves are represented in red, purple and brown, and marked as ``MR'' (multirate) in the plot.
The figure shows how, in general, the multirate curves can follow the baseline curve for several values of the gain amplifier, before falling below it. 
This allows to cover the target bpp range with 3 trained models instead of the 8 originally achieved by separate trainings.
More specifically, for high bpps ($\beta=0.0001$) we observe favorable performance for low values of the gain amplifier, with a severe drop as $ga$ increases.
We however appreciate that for higher compression rates ($\beta=\{0.0008, 0.0064\}$) the MR curves closely resemble the one achieved by separate trainings.
This shows promise for training a single model to support multiple bitrate by following training schemes as proposed in Cui \etal~\cite{cui2020g}.
\bgroup
\setlength{\tabcolsep}{1pt}
\begin{figure*}[tbh]
\resizebox{\textwidth}{!}{
\begin{tabular}{cc}
\quad \quad $\gamma=10$ & \quad \quad $\gamma=30$\\
\includegraphics[width=0.5\textwidth]{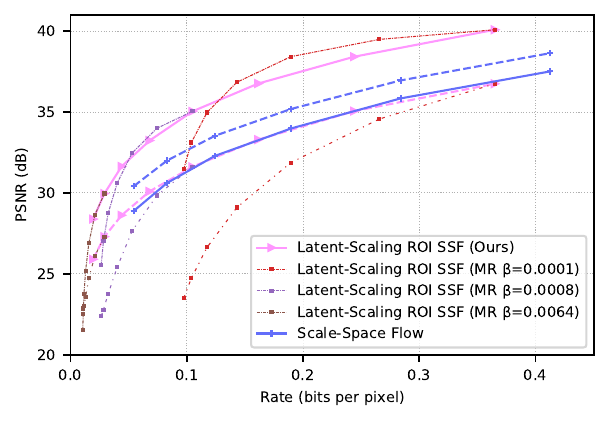}&  
\includegraphics[width=0.5\textwidth]{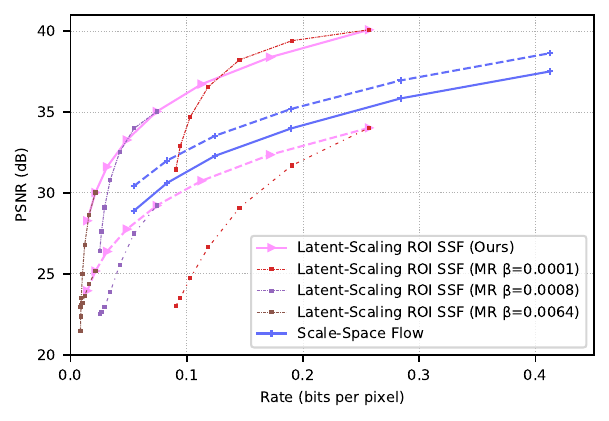}
\end{tabular}}
\caption{Latent-scaling ROI SSF trained on DAVIS with ground-truth ROI masks evaluated on DAVIS val. ROI-based models are trained with $\gamma=\{10, 30\}$, left and right plot respectively. Right plot is a modified version of Fig.~1 in the main text, with additional curves obtained by multirate (MR). Solid line denotes ROI-PSNR, while dashed non-ROI PSNR. The dashed-dotted lines, marked with "MR" in the legend, are obtained using a model trained for a single $\beta$, and then varying gain amplifier $ga$ as outlined in Eq.~\ref{eq:gain_amplifier}.}
\label{fig:apd_b_ls}
\end{figure*}
\egroup
\subsection{Qualitative results} \label{apd:qualitative}
In this section, we provide additional visual results for several variants of the proposed ROI-based methods.
\paragraph{Different background penalty}
In Fig.~\ref{fig:qualitative_gammas} we report for frames from the DAVIS validation set the ROI-based encodings achieved by Implicit ROI-SSF and Latent Scaling ROI-SSF at different values of the background penalty $\gamma$ (Eq.~5 in main text).
Such an hyperparameter controls to which extent background distortion can be de-emphasized to achieve (under rate constraints) a better quality in ROI regions.
\paragraph{Training on synthetic ROI masks}
As validated in Fig.~7b (main paper) and Fig.~\ref{fig:apd_perlin_noise1}, our models can be trained even in the absence of pixel-level ROI masks, as synthetically generated ones can be used instead, with similar validation performances. 
In Fig.~\ref{fig:qualitative_random_maps} we report some examples of encodings for comparable models, when trained either on synthetic or groundtruth masks.
The visual quality of the resulting encoded frames appears comparable, confirming quantitative measurements.
\bgroup
\setlength{\tabcolsep}{1pt}
\begin{figure*}
\resizebox{\textwidth}{!}{
\begin{tabular}{ccccc}
\includegraphics[width=0.4\textwidth]{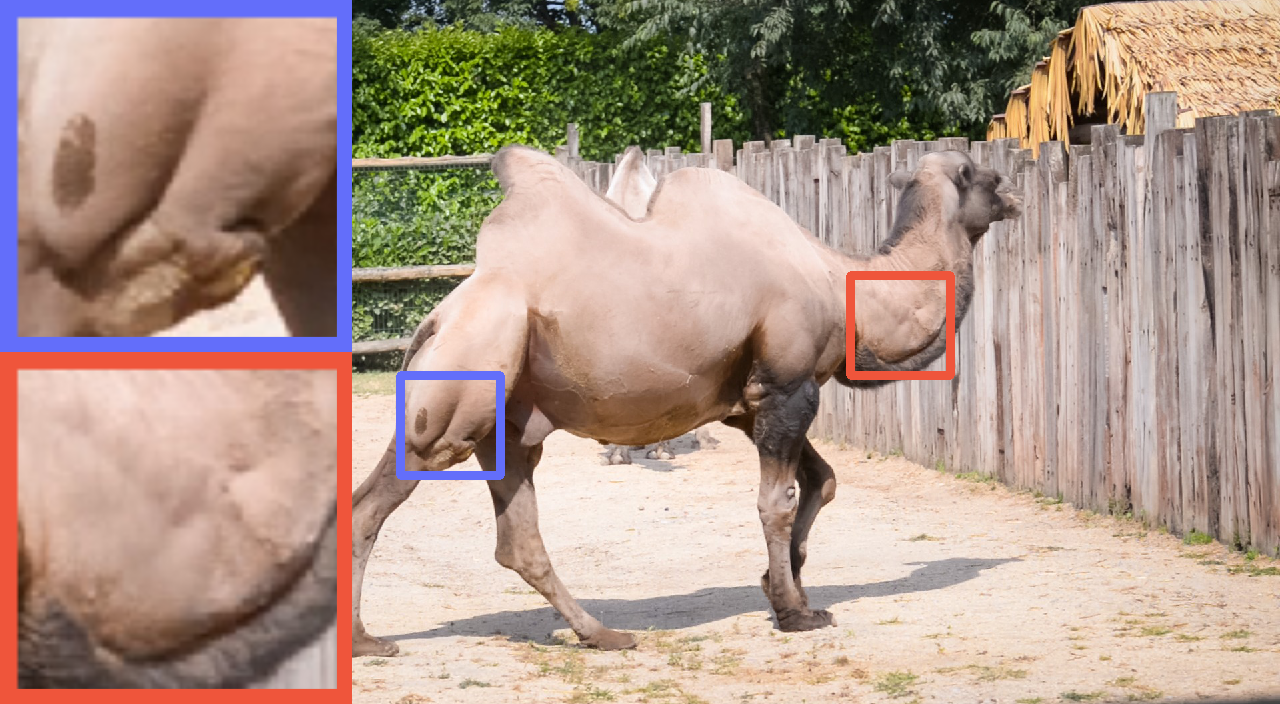}&
\includegraphics[width=0.4\textwidth]{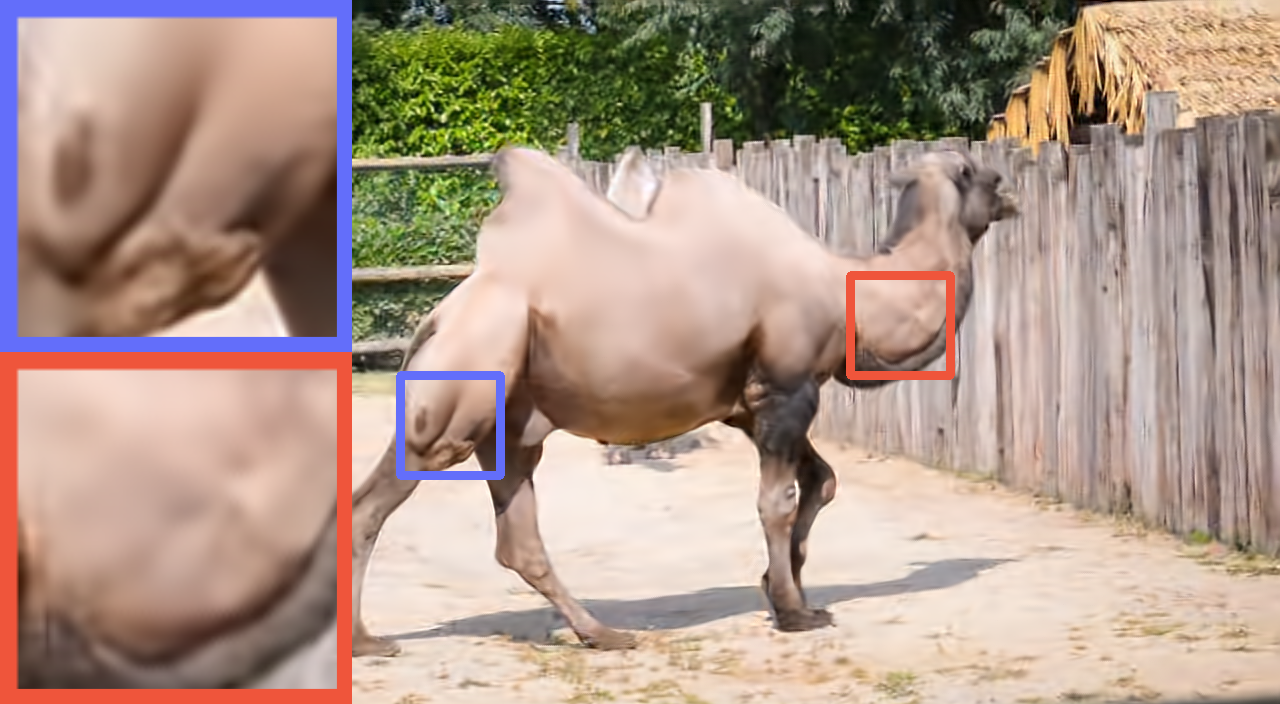}&&
\includegraphics[width=0.4\textwidth]{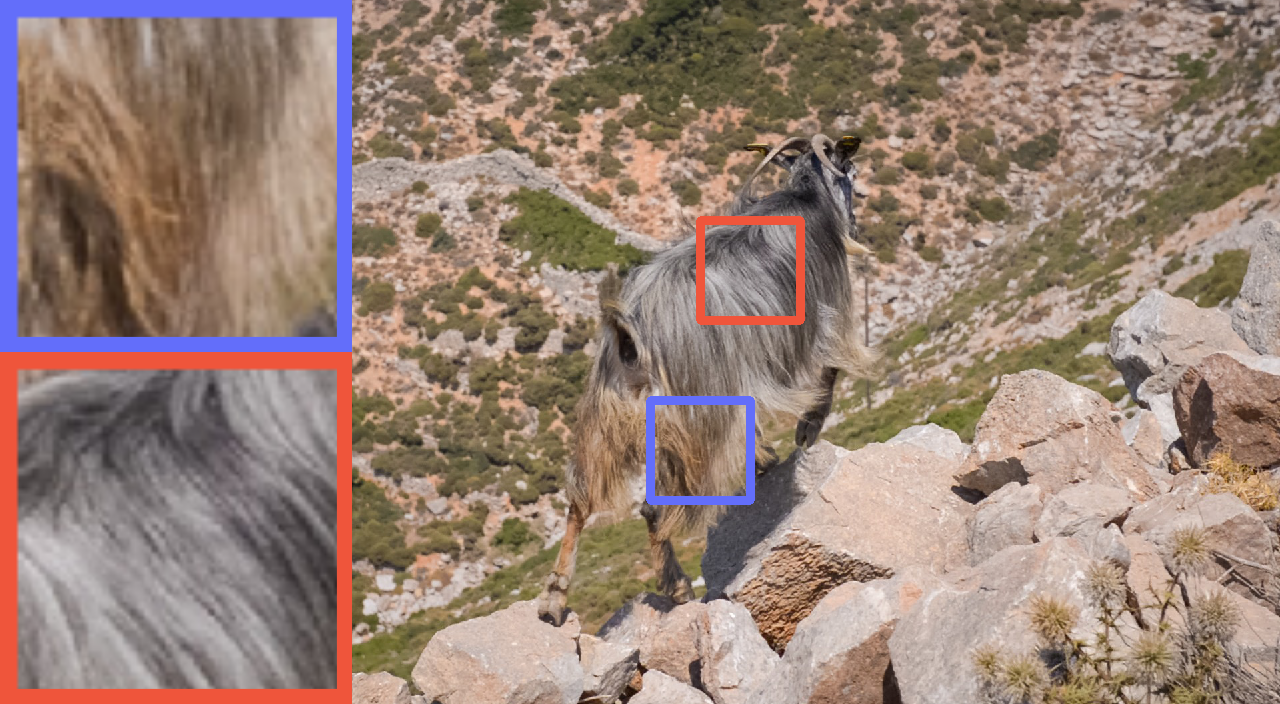}&  
\includegraphics[width=0.4\textwidth]{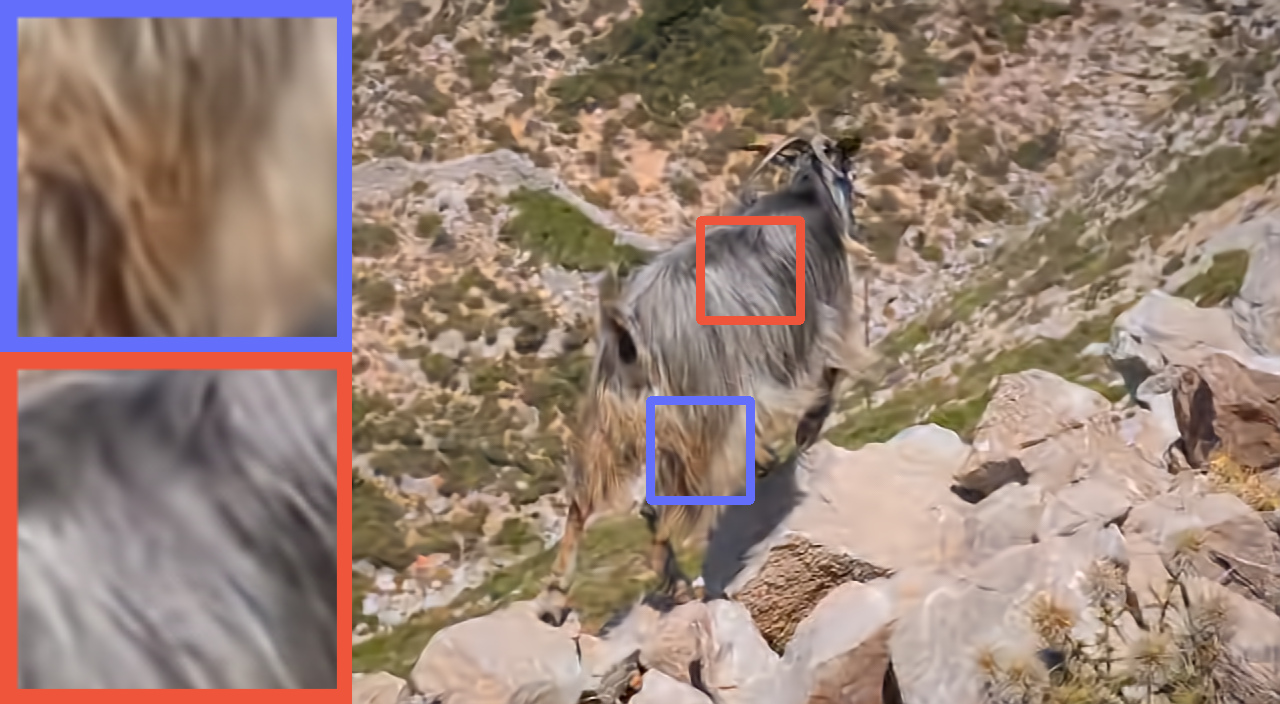}\\
Reference & SSF (0.054 bpp) && Reference & SSF (0.054 bpp) \\
\includegraphics[width=0.4\textwidth]{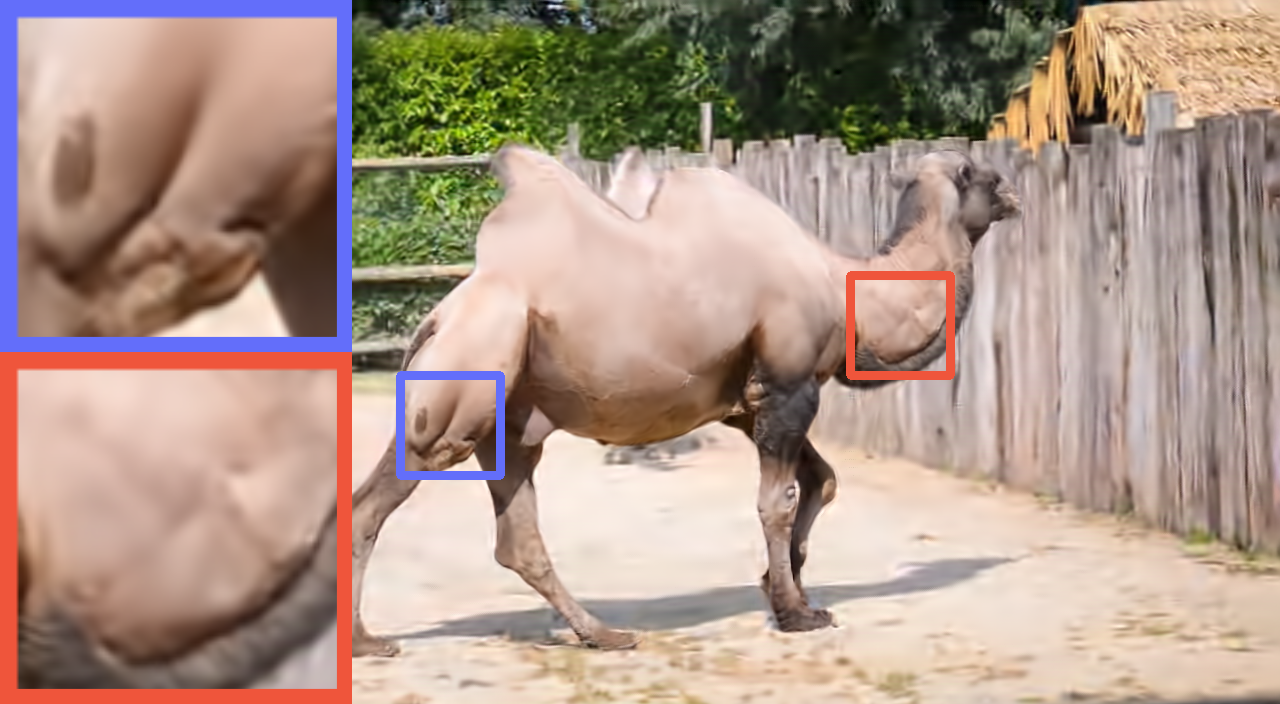}&
\includegraphics[width=0.4\textwidth]{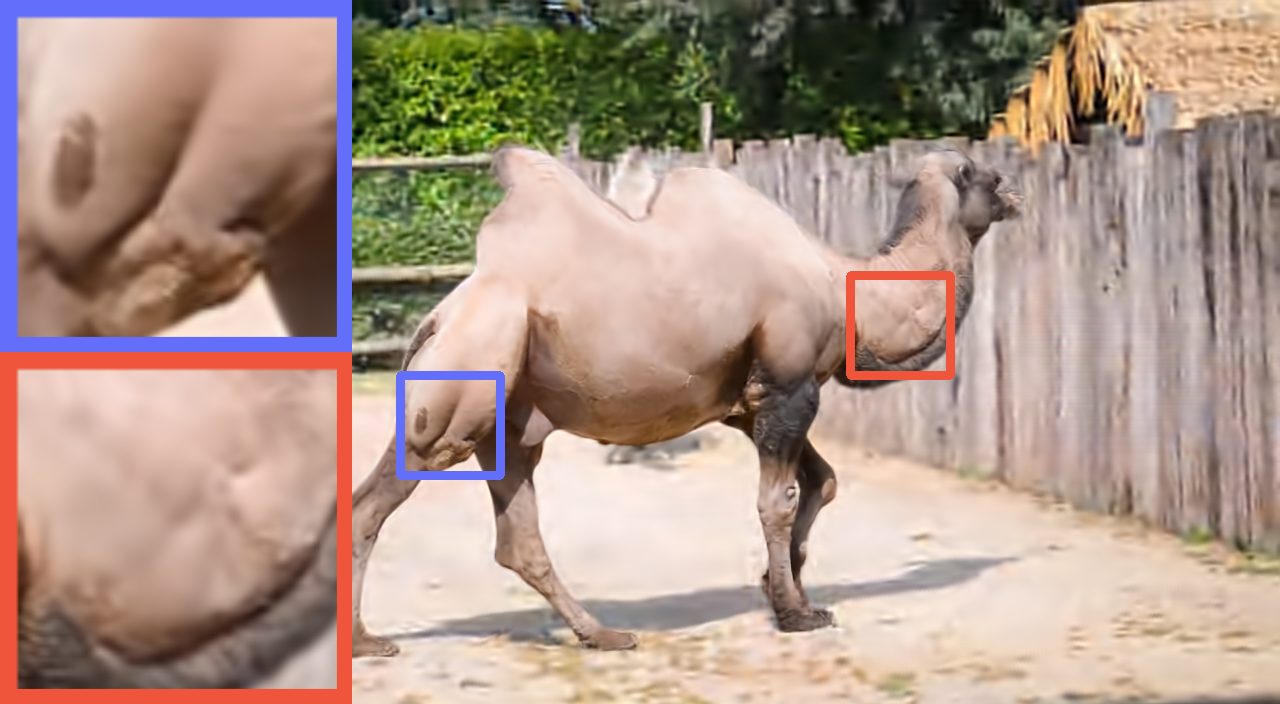}&&
\includegraphics[width=0.4\textwidth]{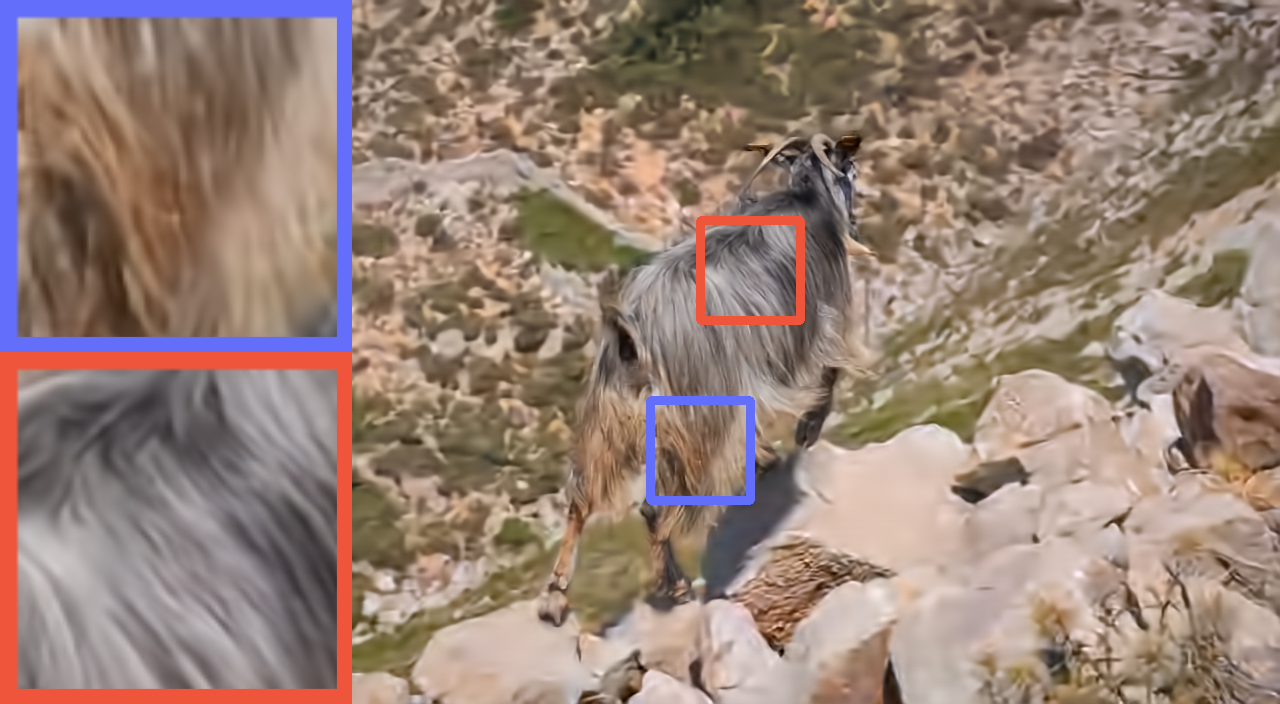}&
\includegraphics[width=0.4\textwidth]{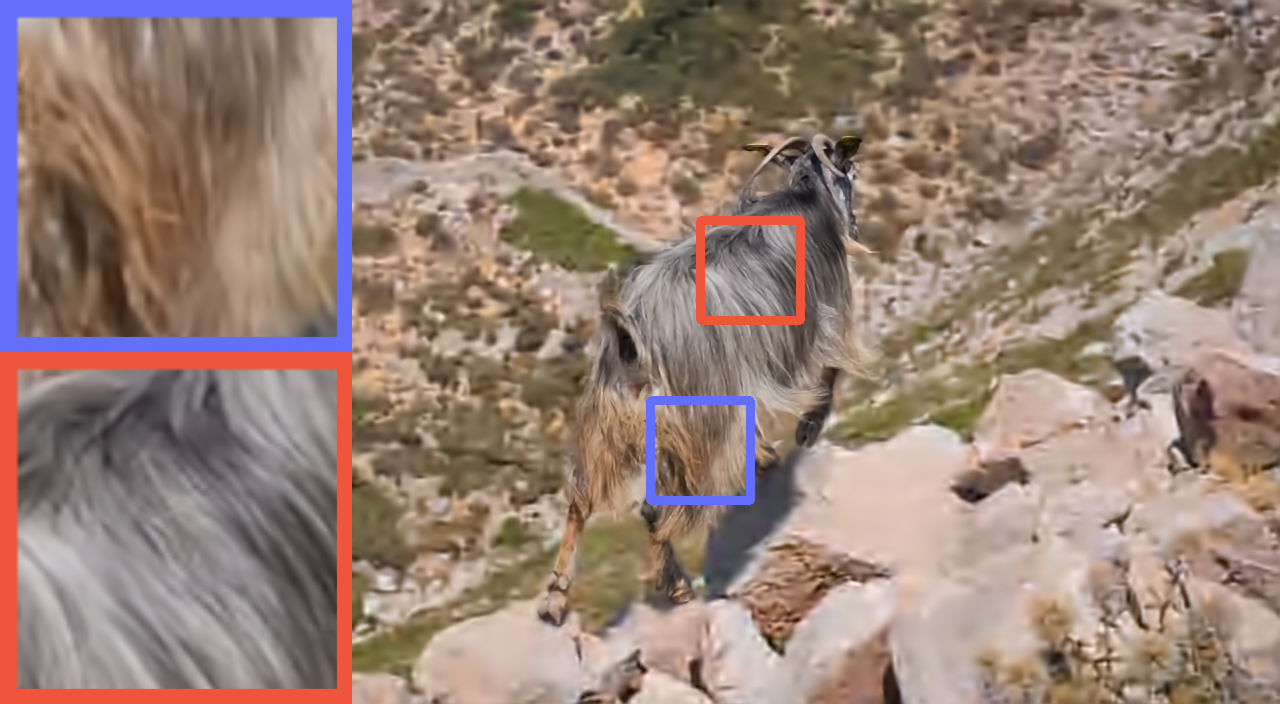}\\
Implicit ROI SSF ($\gamma=10$, 0.044 bpp) & Implicit ROI SSF ($\gamma=30$, 0.049 bpp) && Implicit ROI SSF ($\gamma=10$, 0.044 bpp) & Implicit ROI SSF ($\gamma=30$, 0.049 bpp)\\
\includegraphics[width=0.4\textwidth]{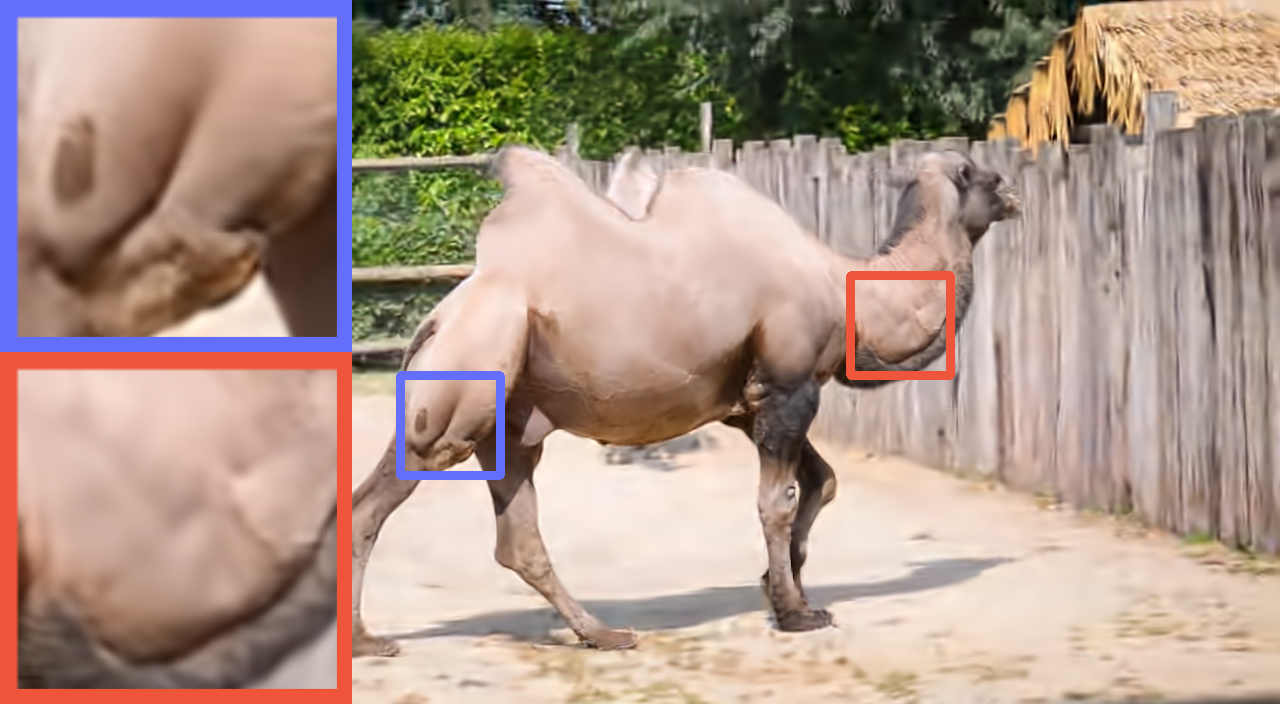}&
\includegraphics[width=0.4\textwidth]{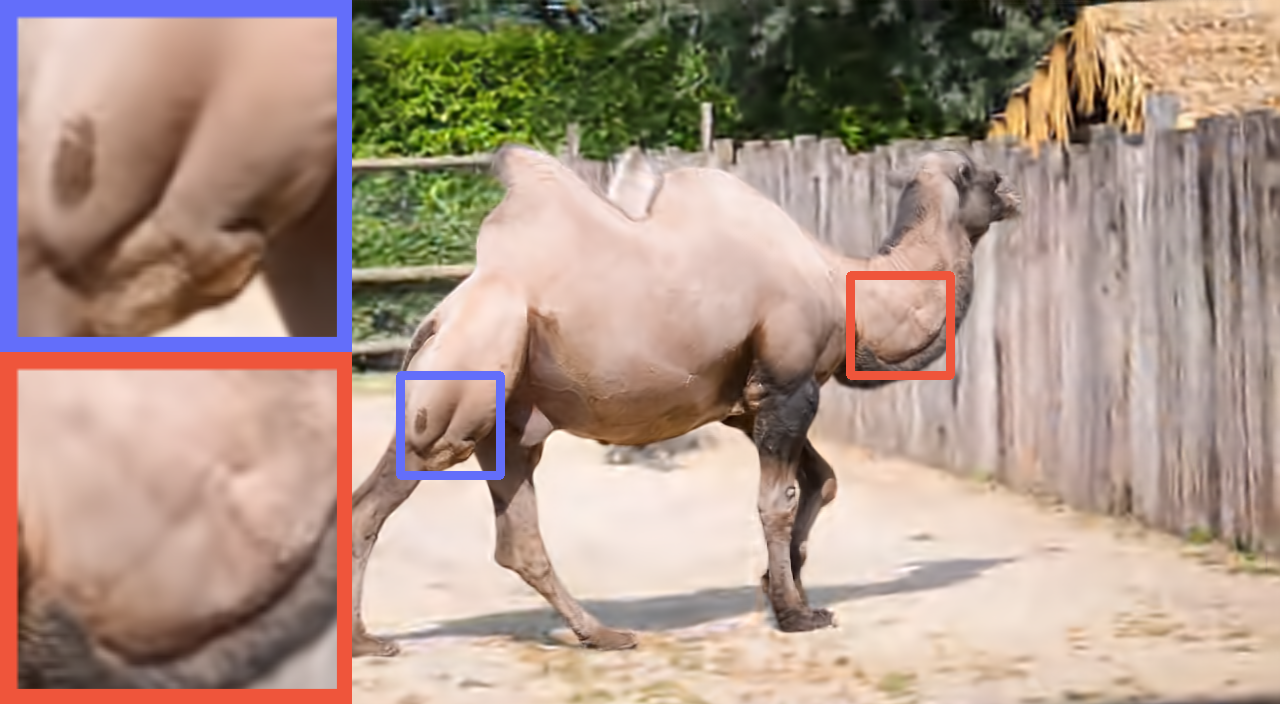}&&
\includegraphics[width=0.4\textwidth]{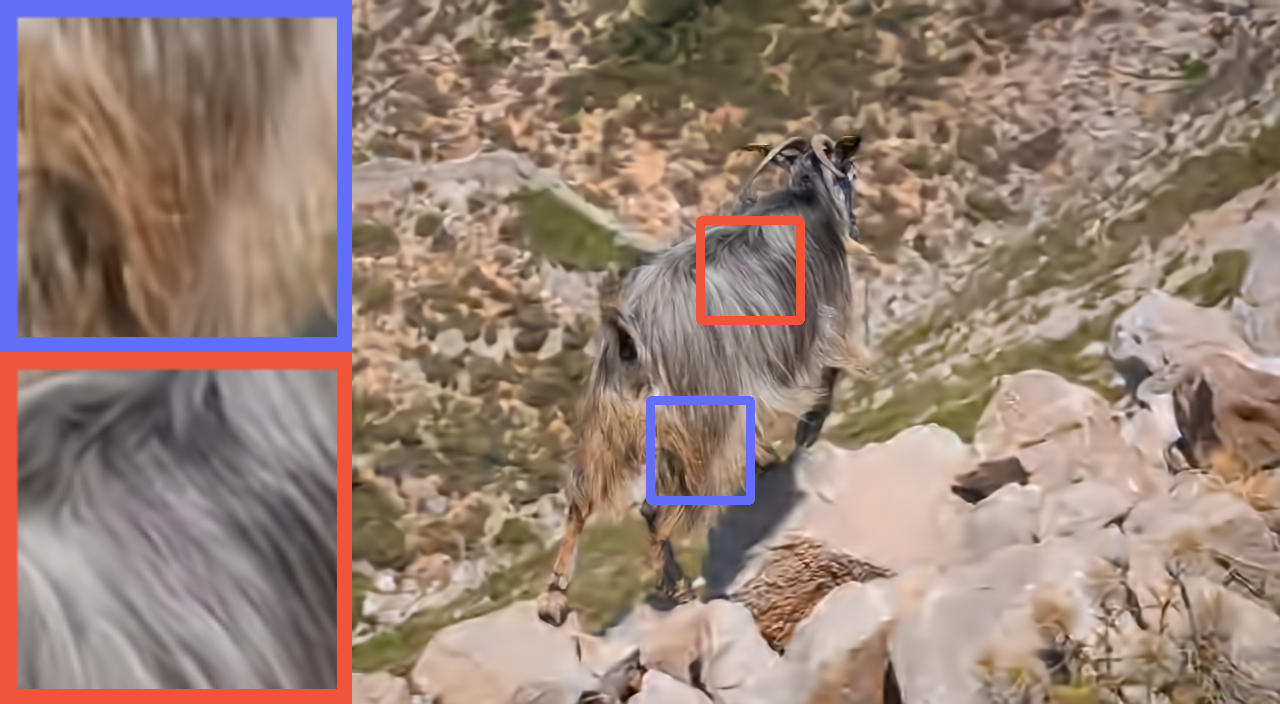}&
\includegraphics[width=0.4\textwidth]{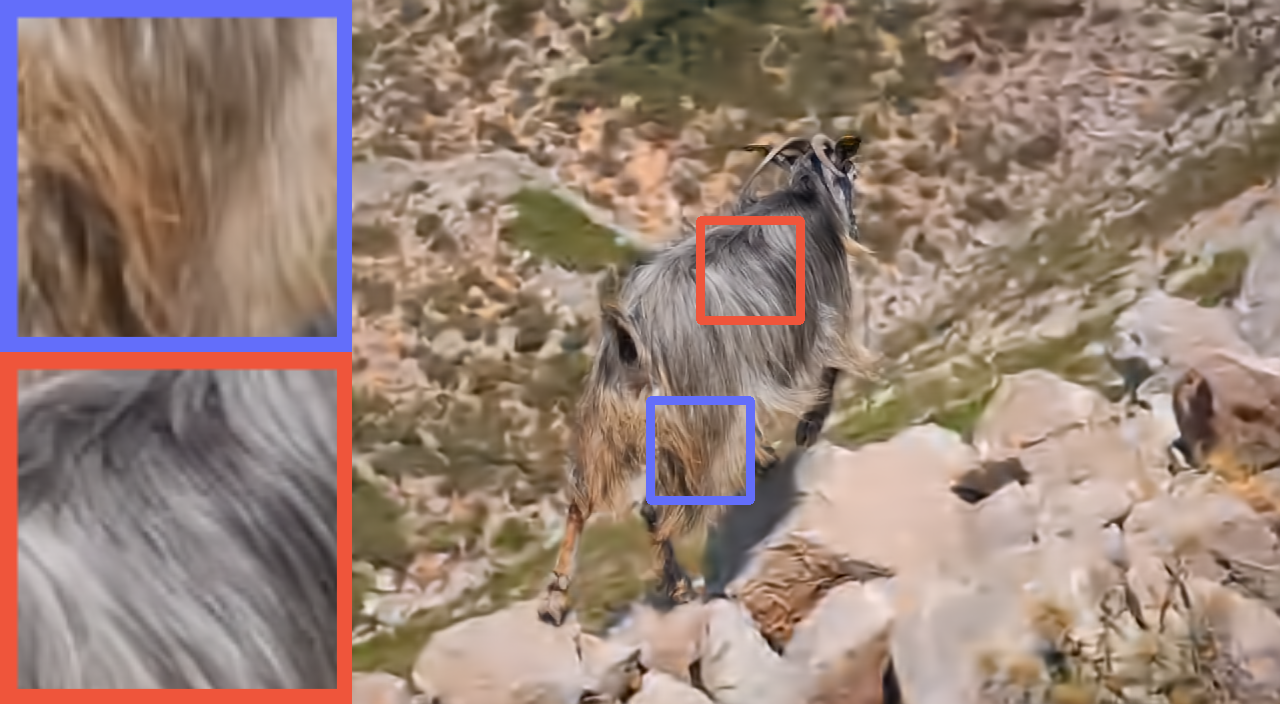}\\
Latent Scaling ROI SSF ($\gamma=10$, 0.044bpp) & Latent Scaling ROI SSF ($\gamma=30$, 0.048 bpp) && Latent Scaling ROI SSF ($\gamma=10$, 0.044bpp) & Latent Scaling ROI SSF ($\gamma=30$, 0.048 bpp)
\end{tabular}}
\caption{Qualitative results of the Implicit and Latent-Scaling ROI SSF for $\gamma=\{10, 30\}$. Benchmarked against the SSF and the reference frame. We use the ``\texttt{camel}'' and ``\texttt{goat}'' sequences from the DAVIS validation set, at frames 11 and 5 respectively.}
\label{fig:qualitative_gammas}
\end{figure*}
\egroup
\bgroup
\setlength{\tabcolsep}{1pt}
\begin{figure*}
\resizebox{\textwidth}{!}{
\begin{tabular}{ccccc}
\includegraphics[width=0.4\textwidth]{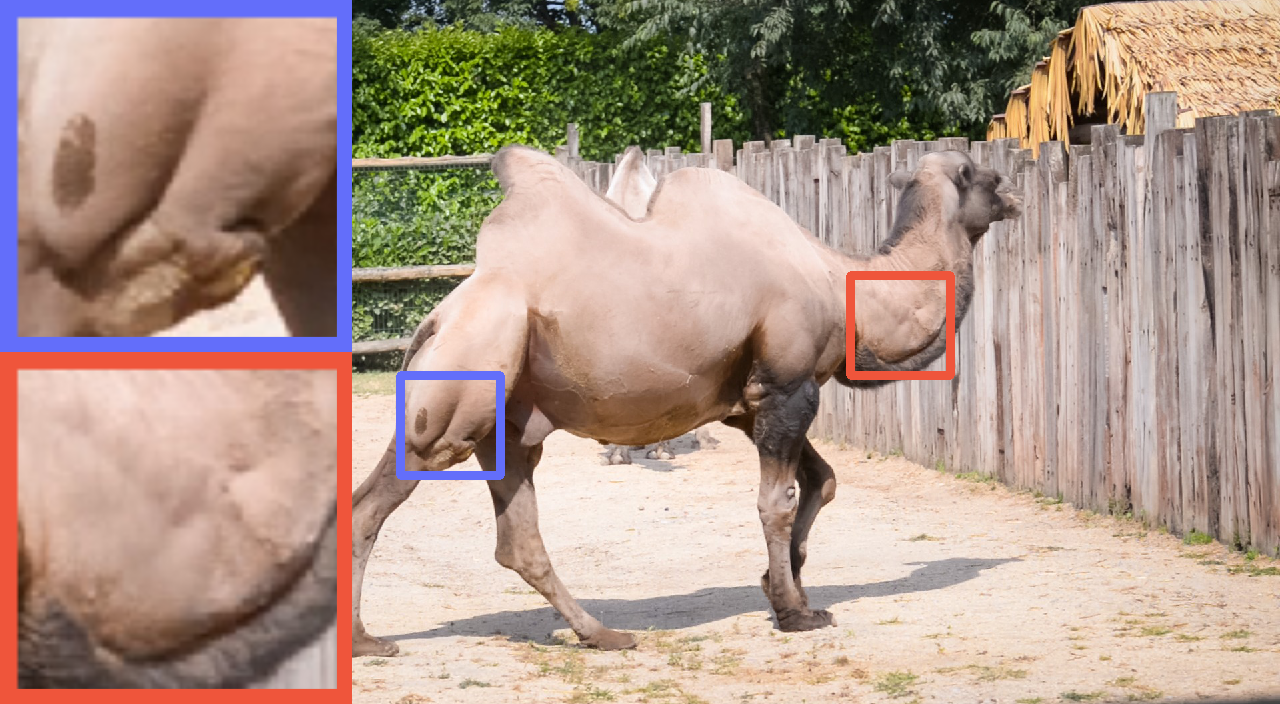}&
\includegraphics[width=0.4\textwidth]{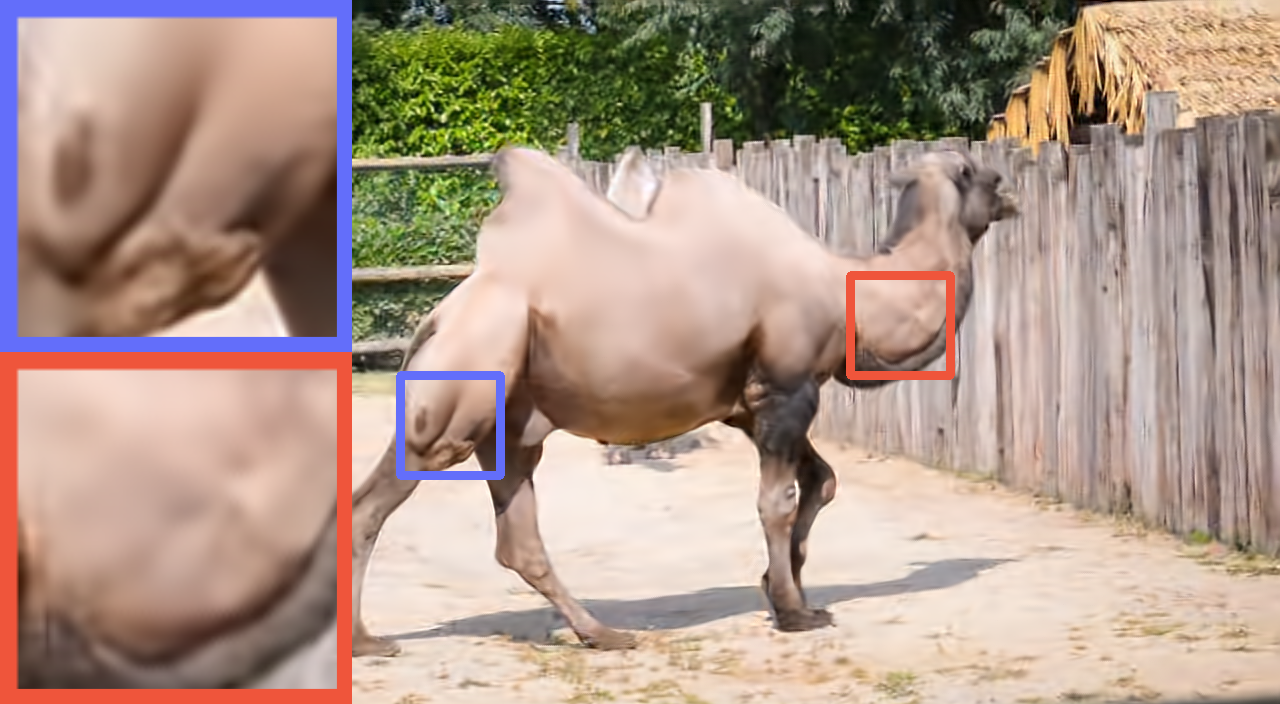}&&
\includegraphics[width=0.4\textwidth]{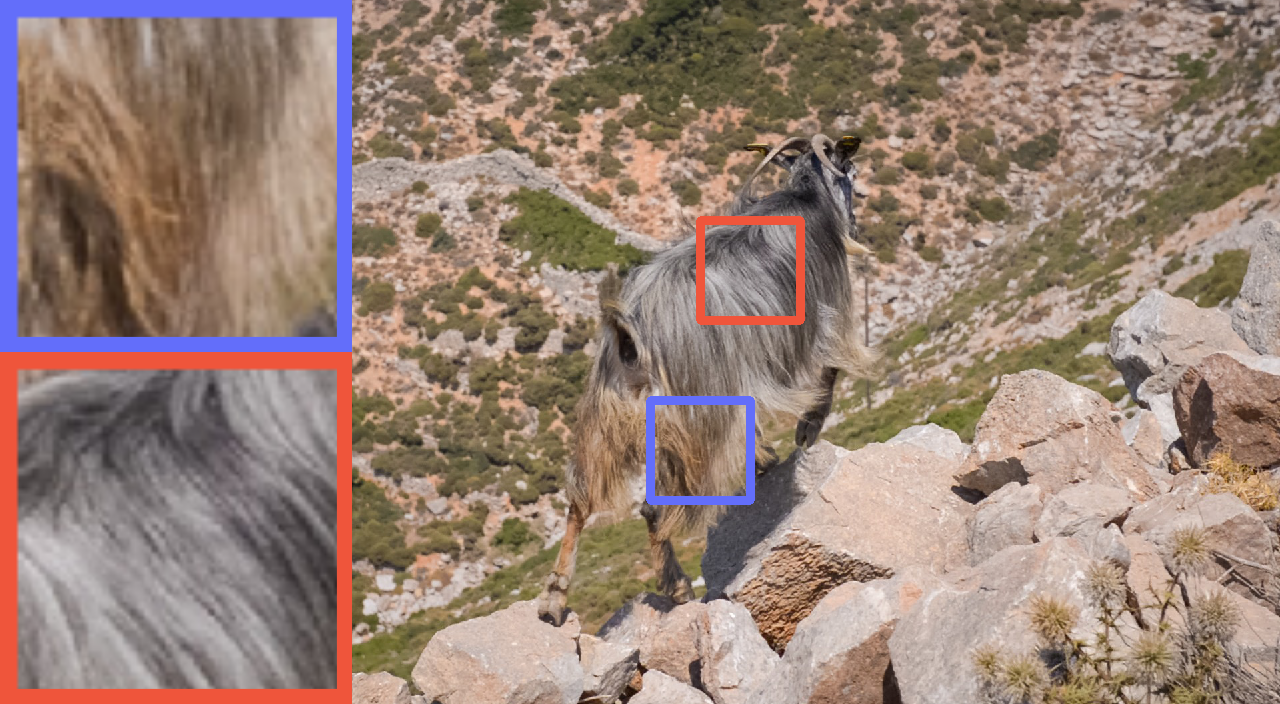}&  
\includegraphics[width=0.4\textwidth]{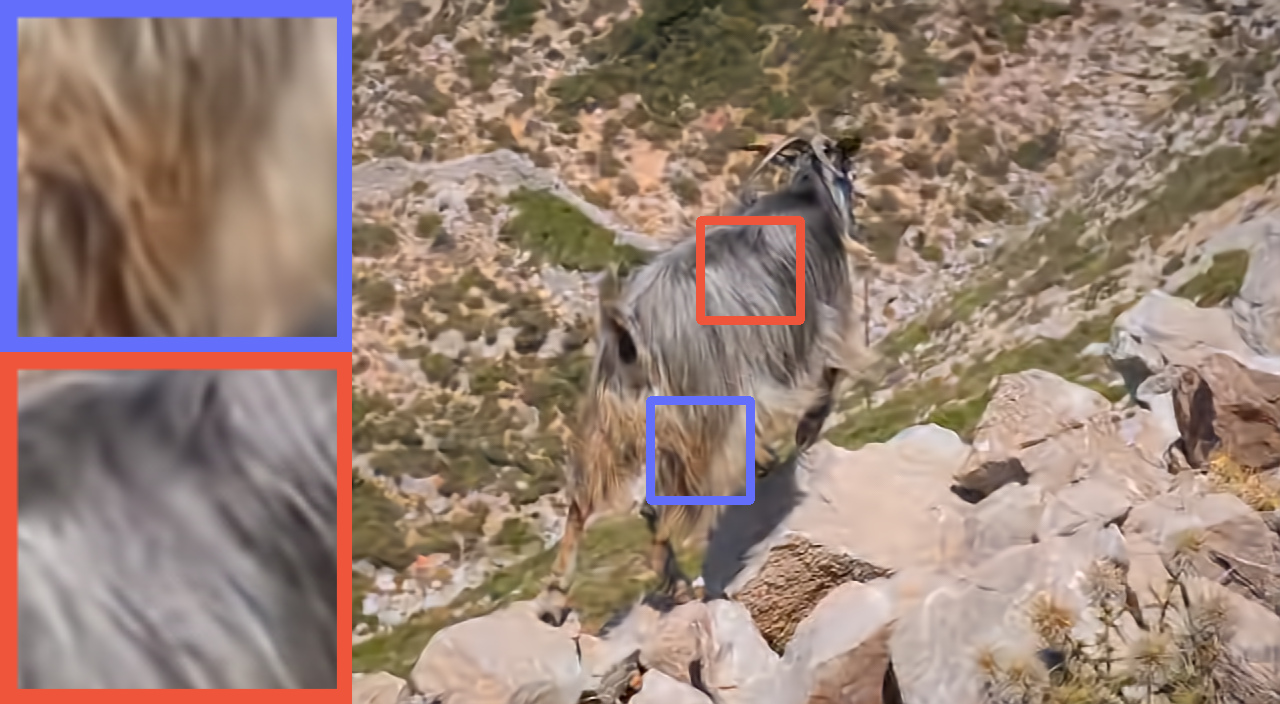}\\
Reference & SSF && Reference & SSF \\
\includegraphics[width=0.4\textwidth]{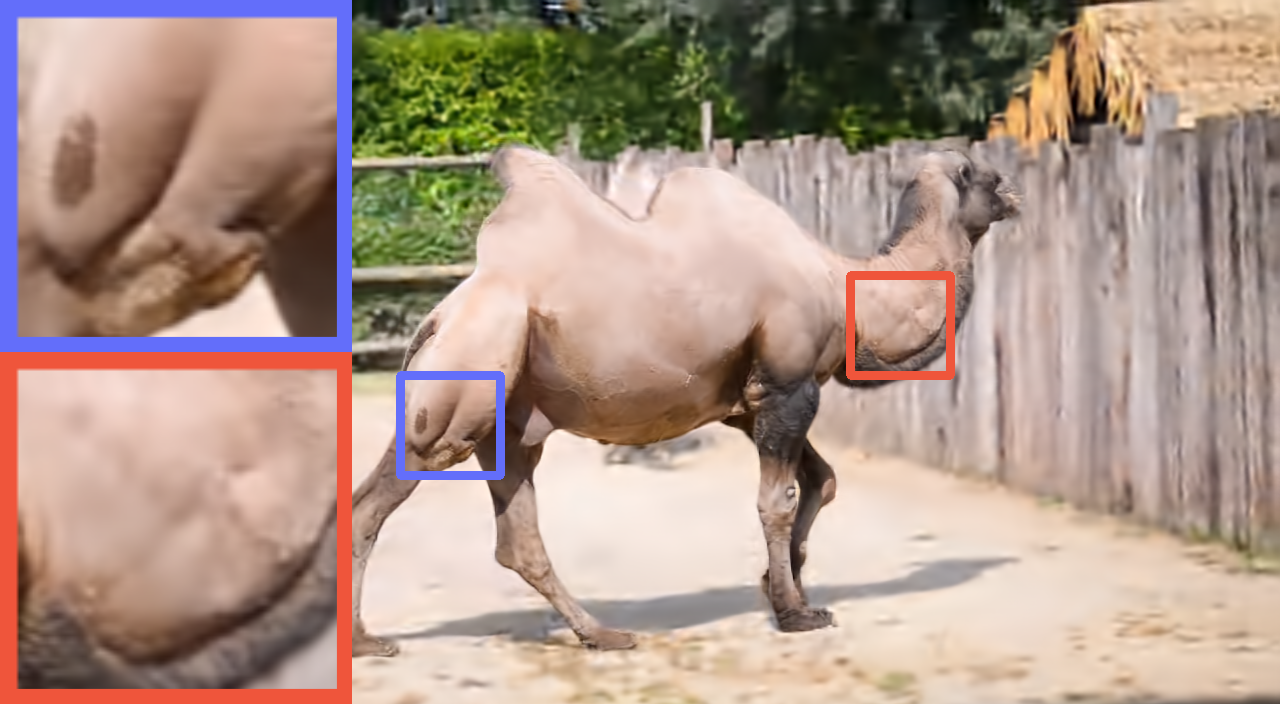}&
\includegraphics[width=0.4\textwidth]{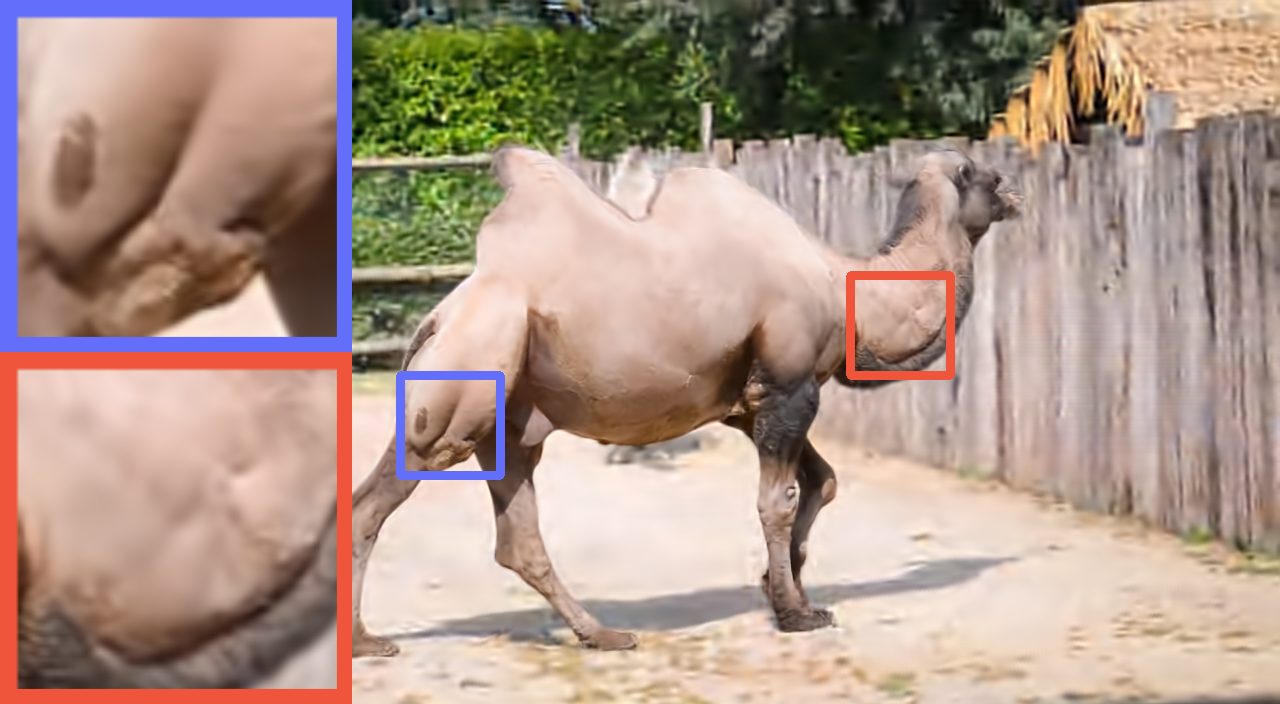}&&
\includegraphics[width=0.4\textwidth]{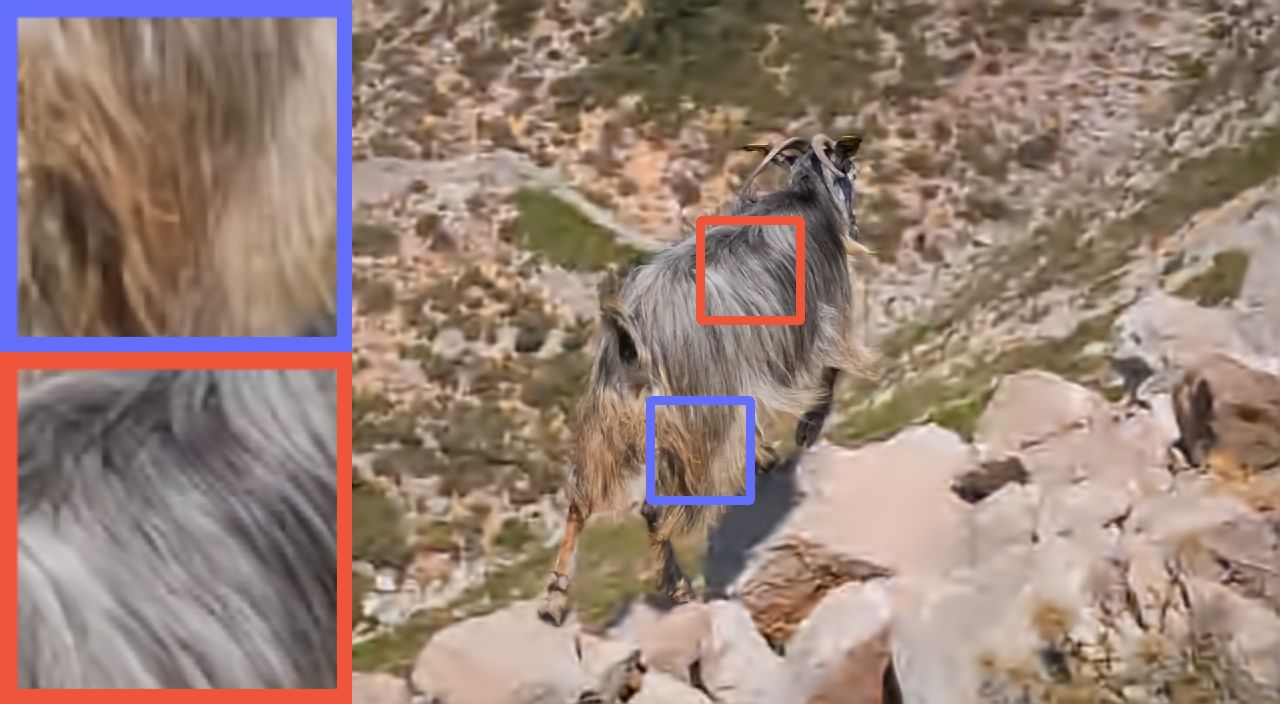}&
\includegraphics[width=0.4\textwidth]{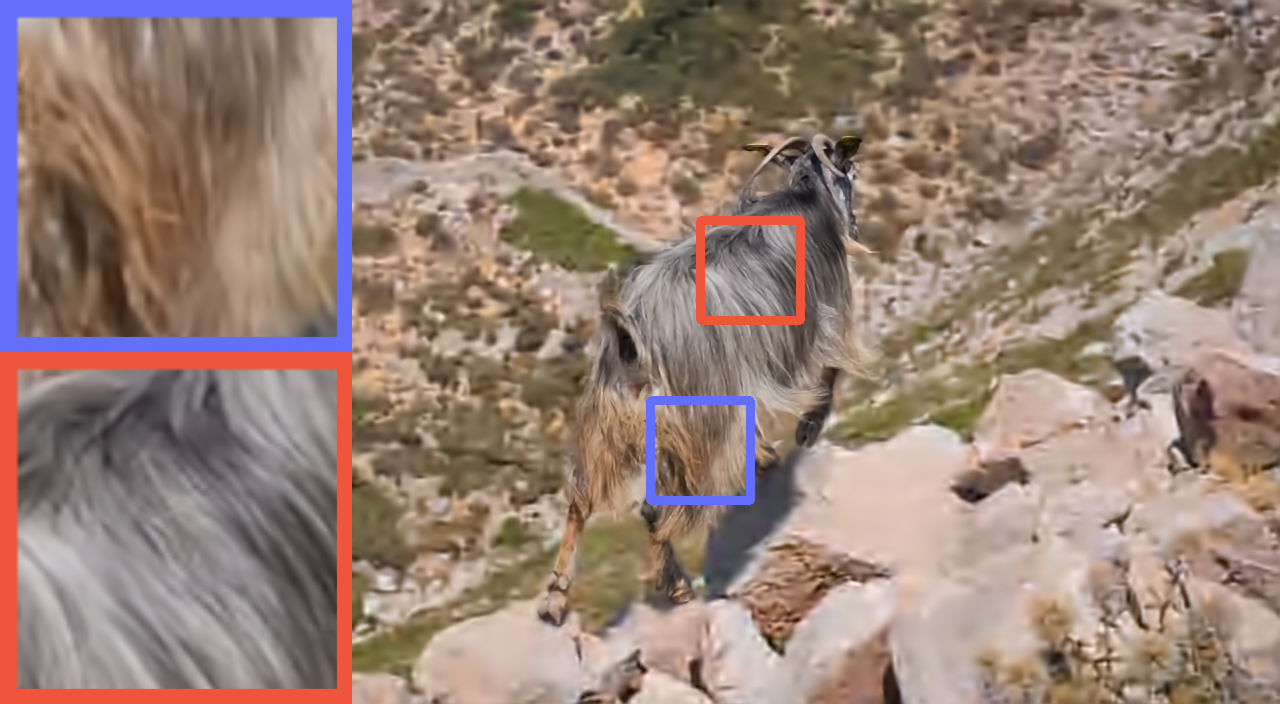}\\
Implicit ROI SSF (syn, 0.055 bpp) & Implicit ROI SSF (gtt, 0.049 bpp) && Implicit ROI SSF (syn, 0.055 bpp) & Implicit ROI SSF (gtt, 0.049 bpp)\\
\includegraphics[width=0.4\textwidth]{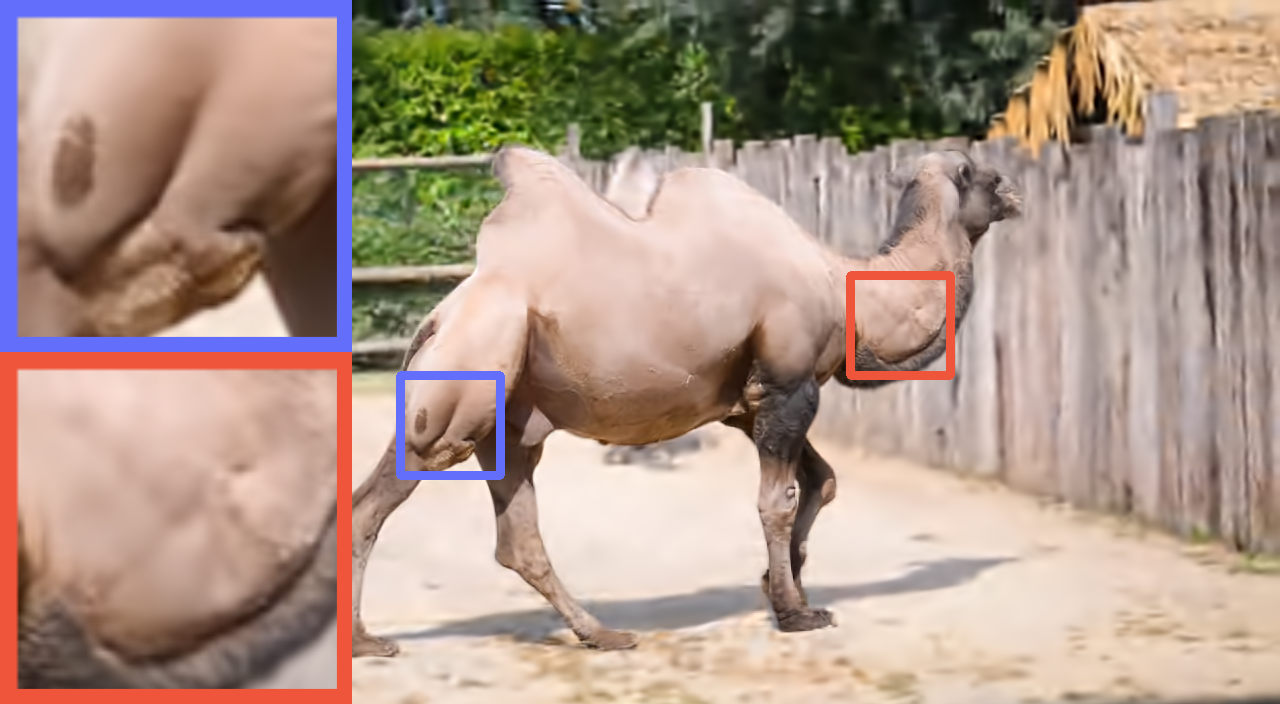}&
\includegraphics[width=0.4\textwidth]{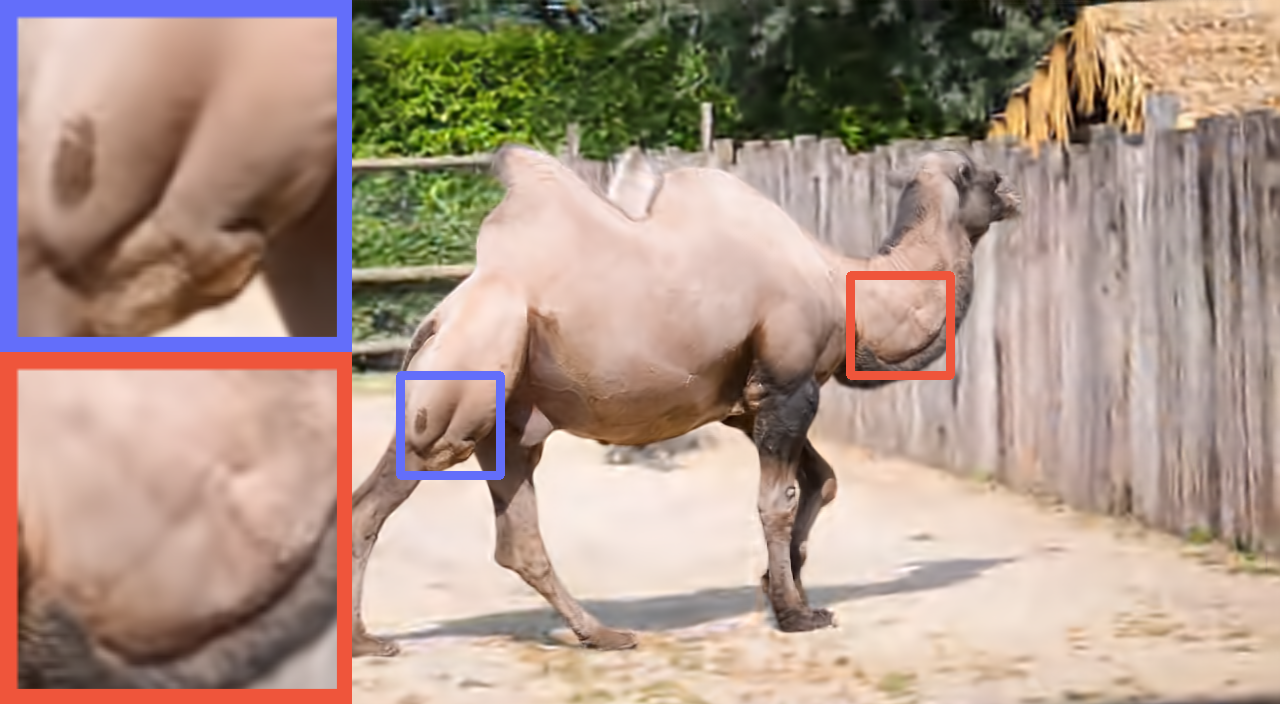}&&
\includegraphics[width=0.4\textwidth]{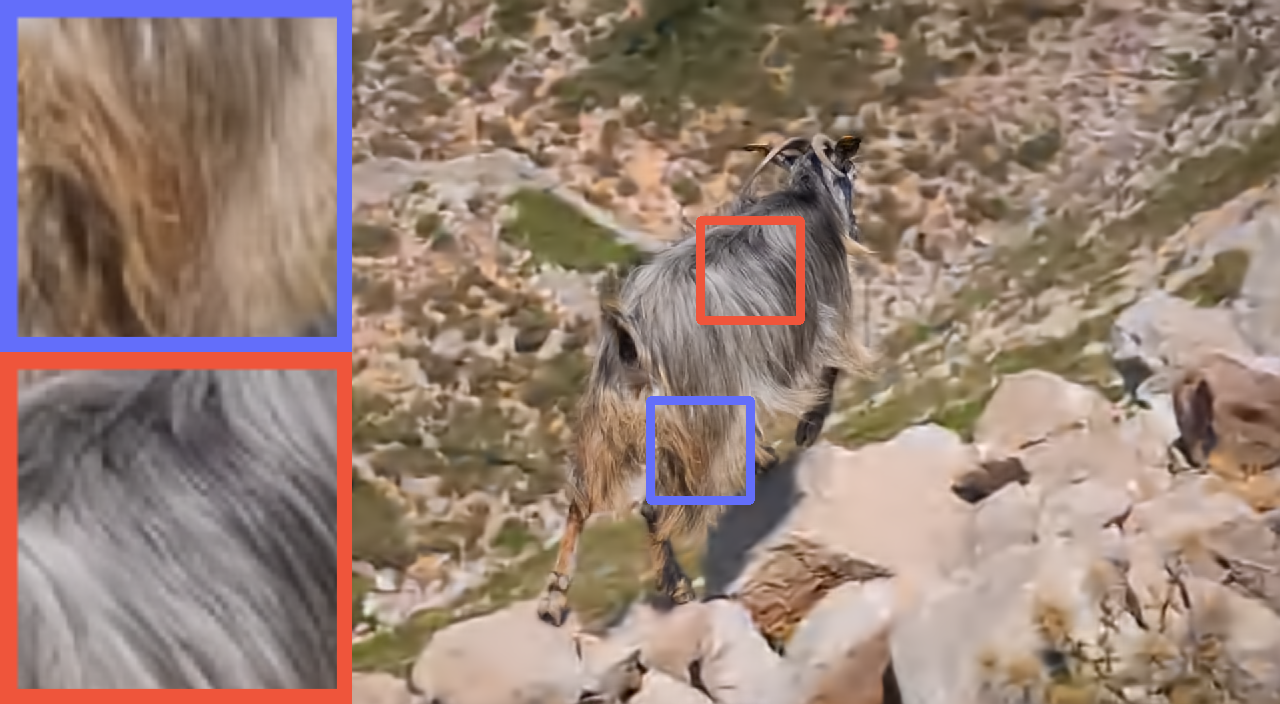}&
\includegraphics[width=0.4\textwidth]{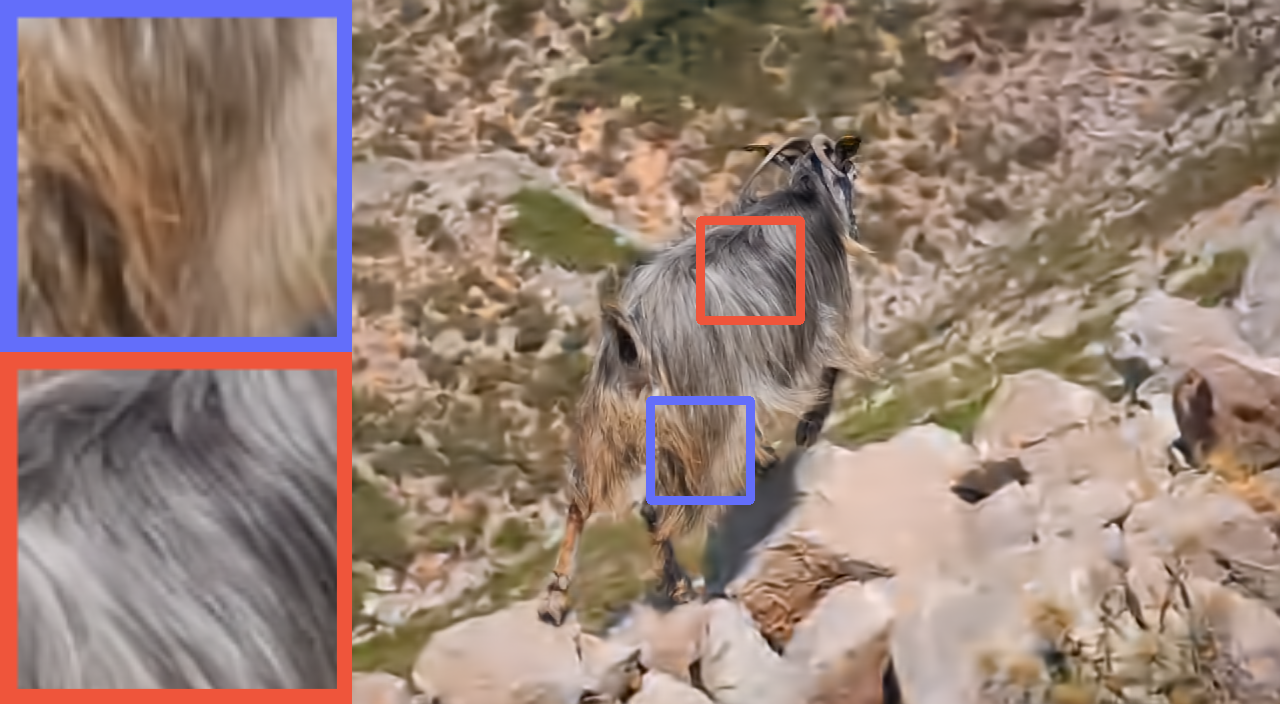}\\
Latent Scaling ROI SSF (syn, 0.055 bpp) & Latent Scaling ROI SSF (gtt, 0.048 bpp) && Latent Scaling ROI SSF (syn, 0.055 bpp) & Latent Scaling ROI SSF (gtt, 0.048 bpp)
\end{tabular}}
\caption{Qualitative results of the Implicit and Latent-Scaling ROI SSF when trained using synthetic ROI maps (syn) or ground-truth ROI maps (gtt). We use the ``\texttt{camel}'' and ``\texttt{goat}'' sequences from the DAVIS validation set, at frames 11 and 5 respectively.}
\label{fig:qualitative_random_maps}
\end{figure*}
\egroup

\subsection{Runtime performance} \label{apd:complexity}

In table \ref{table:timings}, we benchmarked the runtime of SSF and Latent-Scaling ROI SSF on HD 720p inputs on an NVIDIA Tesla V100 and Intel CPU E5-1620 v4 @ 3.50GHz. We show timings in frames-per-second (FPS) for encode and decode operations: neural-network execution only, and together with entropy coding on CPU including data transfer, for I-frame and P-frame codec separately.

Note that the computational complexity of the Implicit ROI SSF is negligibly higher than that of the original SSF, as it only adds an input channel to each autoencoder.

\bgroup
\begin{table*}[tbh]
\centering
\resizebox{\columnwidth}{!}{
\begin{tabular}{lccccccccc}
\toprule
           &          & \multicolumn{2}{c}{Encode} & \multicolumn{2}{c}{Decode} & \multicolumn{2}{c}{Encode (no EC)} & \multicolumn{2}{c}{Decode (no EC)} \\ 
           &          & I-frame      & P-frame     & I-frame      & P-frame     & I-frame          & P-frame         & I-frame          & P-frame         \\ \midrule
SSF        &    FPS      & 3.5          & 1.7         & 3.8          & 1.8         & 378              & 192             & 682              & 340             \\ \midrule
\multirow{2}{*}{LS ROI SSF} & FPS & 2.9          & 1.5         & 3.2          & 1.7         & 247              & 156             & 410              & 259             \\
           & FPS drop & -17\%        & -12\%       & -16\%        & -6\%        & -35\%            & -19\%           & -40\%            & -24\%          
           \\\bottomrule
\end{tabular}
}
\caption{Comparison of runtime (FPS) for 720p inputs of SSF and LS ROI SSF I/P-frame codecs on NVIDIA V100.
}
\label{table:timings}
\end{table*}
\egroup
\section{Architecture Details}

\begin{figure}
\centering
\resizebox{\columnwidth}{!}{
\begin{tabular}{cc}
\includegraphics[width=0.6\columnwidth]{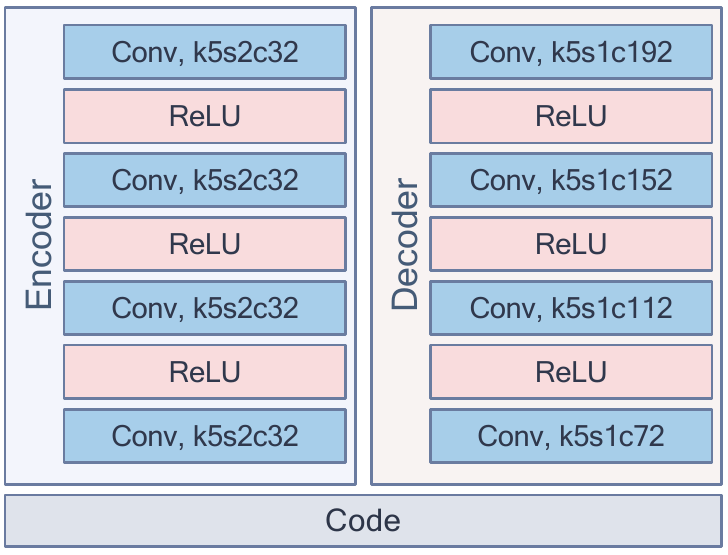}&
\includegraphics[width=0.6\columnwidth]{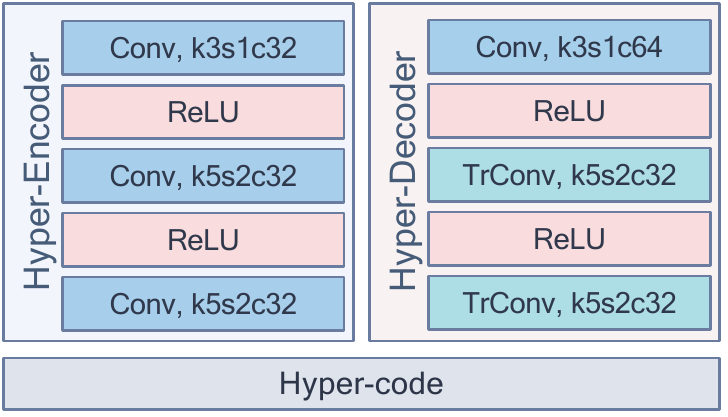}
\end{tabular}}
\caption{(a) Gain hyperprior codec details. $k$, $s$, and $c$ denote kernel
size, stride and the number of output channels, respectively. (b) Gain hyperprior hyper-codec details. $k$, $s$, and $c$ denote kernel size, stride and the number of output channels, respectively.} 
\label{fig:gain_hyperprior}
\end{figure}

We use the same SSF architecture as described in Pourreza~\etal~\cite{pourreza2021extending}, Appendix A.1, except we share the hyper decoder for mean and scale, and last layer outputs twice as many channels. Our gain hyperprior autoencoder follows a similar architecture, except for the codec decoder which does not upsample and replaces transpose convolutions with regular convolutions with stride 1, see details in  Fig.~\ref{fig:gain_hyperprior}a for the codec and Fig~\ref{fig:gain_hyperprior}b for the hyper-codec.

We adopt the quantization strategy in Guo et al.~\cite{guo2021soft}. 
Calling $y$ the latent, we apply additive uniform noise ($\tilde{y} = y + u$ with $u\sim \mathcal{U}(-0.5, 0.5)$) and rounding with straight-through gradient estimation ($\bar{y} = \lfloor y \rceil $). 
During training, we use the noisy $\tilde{y}$ for the entropy computation in the prior, whereas we feed the decoder with the rounded latent $\bar{y}$. The same strategy holds for the hyper-latents.

\bibliography{references}
\end{document}